\documentclass[useAMS,usenatbib]{mn2e}

\usepackage{graphicx,color}
\usepackage{amssymb,amsmath}
\usepackage{amsmath} %Alinhamento das equacoes
\usepackage{multicol}
\usepackage{latexsym}
\usepackage{amsfonts}
\usepackage{amssymb}
\usepackage{wasysym}
\usepackage{textcomp}
\usepackage{longtable,lscape}
\usepackage{rotating}
\usepackage{mathtools}
\usepackage{aas_macros}
\usepackage[dvipsnames]{xcolor}
\usepackage{morefloats}
\usepackage{enumitem}
\usepackage{cleveref}

\definecolor{amethyst}{rgb}{0.6, 0.4, 0.8}   %see more in https://latexcolor.com/

\usepackage{natbib}
\usepackage{amssymb,amsmath}
\title[Exploring OCs dynamical evolution from observations]{Exploring the dynamical state of Galactic open clusters using Gaia DR3 and observational parameters}

\author[M. S. Angelo et al.]{M. S. Angelo$^{1,2}$\thanks{E-mail:
mateusangelo@cefetmg.br}, J. F. C. Santos Jr.$^{3}$, W. J. B. Corradi$^{2,3}$ and F. F. S. Maia$^{4}$   \\ 
\noindent
$^1$Centro Federal de Educa\c{c}\~ao Tecnol\'ogica de Minas Gerais, Av. Amazonas, 7675, 30510-000 Belo Horizonte, MG, Brazil\\
$^2$Laborat\'orio Nacional de Astrof\'isica, R. Estados Unidos 154, 37530-000 Itajub\'a, MG, Brazil\\
$^3$Departamento de F\'isica, ICEx, Universidade Federal de Minas Gerais, Av. Ant\^onio Carlos 6627, 31270-901 Belo Horizonte, MG, Brazil\\
$^4$Universidade Federal do Rio de Janeiro, Instituto de F\'isica, 21941-972, Brazil\\
}

\begin{document}

\date{Accepted XXX. Received XXX; in original form XXX}

\pagerange{\pageref{firstpage}--\pageref{lastpage}} \pubyear{XXXX}

\maketitle

\label{firstpage}

\begin{abstract}

%(manter maÂx de 250 palavras)

Galactic open clusters (OCs) are subject to internal and external destructive effects that gradually deplete their stellar content, leaving imprints on their structure. To investigate their dynamical state from an observational perspective, we employed Gaia DR3 data to perform a comprehensive analysis of 174 OCs ($\sim$10\% of \citeauthor{Dias:2021a}'s\,\,\citeyear{Dias:2021a} catalogue). We employed radial density profiles and astrometrically decontaminated colour-magnitude diagrams to derive structural parameters, distance, mass and time-related quantities. We explored the parameters space and searched for connections relating the clusters' structure with the internal evolutionary state and the external Galactic tidal field. Correlations were verified after segregating the sample according to the Galactocentric distance and half-light to Jacobi radius ratio ($r_h/R_J$). This tidal filling ratio decreases with both the cluster mass and dynamical age. At a given evolutionary stage, OCs with larger $r_h/R_J$ tend to present larger fractions of mass loss due to dynamical effects. Regarding the impact of the external conditions, we identified different evaporation regimes: for ambient densities ($\rho_{\textrm{amb}}$) larger than $\sim$0.1\,$M_{\odot}/$pc$^3$, clusters tend to be more tidally filled as they are subject to weaker tidal stresses. For $\rho_{\textrm{amb}}\lesssim0.1\,M_{\odot}/$pc$^3$, the opposite occurs: $R_J$ increases for smaller $\rho_{\textrm{amb}}$, causing $r_h/R_J$ to decrease. In turn, two-body relaxation tends to compact the cluster core, which is less sensitive to variations of the external potential. The higher the degree of central concentration, the larger the number of relaxation times a cluster takes until its dissolution.

\end{abstract}


\section{Introduction}

After the early gas expulsion phase (\citeauthor{Hills:1980}\,\,\citeyear{Hills:1980}; \citeauthor{Geyer:2001}\,\,\citeyear{Geyer:2001}; \citeauthor{Goodwin:2006}\,\,\citeyear{Goodwin:2006}), Galactic open clusters (OCs) have their mass content gradually depleted due to dynamical interactions and stellar evolution (e.g., \citeauthor{de-La-Fuente-Marcos:1997}\,\,\citeyear{de-La-Fuente-Marcos:1997}; \citeauthor{Portegies-Zwart:2001}\,\,\citeyear{Portegies-Zwart:2001}; \citeauthor{Baumgardt:2003}\,\,\citeyear{Baumgardt:2003}; \citeauthor{Fukushige:2000}\,\,\citeyear{Fukushige:2000}). The mass loss process by dynamical effects results from the combination of internal two-body relaxation with the external tidal field: as lower mass stars progressively occupy the high-velocity tail of the velocity distribution at each relaxation time \citep{Spitzer:1987}, they can reach the cluster outskirts and eventually evaporate from the system.

Another disruptive effect is tidal stripping, which consists in the prompt removal of stars when they are located beyond the cluster Jacobi radius ($R_J$; e.g., \citeauthor{von-Hoerner:1957}\,\,\citeyear{von-Hoerner:1957}), which delimits the system gravitational influence amidst the external Galactic potential (Section~\ref{sec:analysis}). This effect can result from variations in the external tidal field in timescales shorter than the cluster crossing time, which may occur, for example, during collisions with molecular clouds (\citeauthor{Theuns:1991}\,\,\citeyear{Theuns:1991}; \citeauthor{Gieles:2006}\,\,\citeyear{Gieles:2006}) and passages through the Galactic disc (\citeauthor{Ostriker:1972}\,\,\citeyear{Ostriker:1972}; \citeauthor{Lamers:2006}\,\,\citeyear{Lamers:2006}). In the case of clusters describing more eccentric orbits, the shrinking of $R_J$ as they pass by the perigalacticon can cause energetic stars in the cluster outskirts to become unbound, therefore contributing to mass loss (\citeauthor{Portegies-Zwart:2010}\,\,\citeyear{Portegies-Zwart:2010}; \citeauthor{Webb:2014}\,\,\citeyear{Webb:2014}). The interplay between this set of destructive processes may affect the clusters' shape as they dynamically evolve and, in principle, their evolutionary stage can be verified from diagnostic plots involving structural parameters.

Since the first data release (DR1) of the \textit{Gaia} catalogue, a progressively more accurate characterization of the already known OCs population has made possible to better constrain the Milky Way (MW) structure (e.g., \citeauthor{Cantat-Gaudin:2020}\,\,\citeyear{Cantat-Gaudin:2020}; \citeauthor{Castro-Ginard:2021}\,\,\citeyear{Castro-Ginard:2021}; \citeauthor{Hao:2021}\,\,\citeyear{Hao:2021}; \citeauthor{Joshi:2023}\,\,\citeyear{Joshi:2023}), besides the discovery of new groups (e.g., \citeauthor{Castro-Ginard:2018}\,\,\citeyear{Castro-Ginard:2018}; \citeauthor{Castro-Ginard:2019}\,\,\citeyear{Castro-Ginard:2019}; \citeauthor{Castro-Ginard:2020}\,\,\citeyear{Castro-Ginard:2020}; \citeauthor{Ferreira:2019}\,\,\citeyear{Ferreira:2019}; \citeauthor{Ferreira:2020}\,\,\citeyear{Ferreira:2020}; \citeauthor{Ferreira:2021}\,\,\citeyear{Ferreira:2021}; \citeauthor{Liu:2019}\,\,\citeyear{Liu:2019}; \citeauthor{Qin:2023}\,\,\citeyear{Qin:2023}). The exquisite precision reached on the photometric and astrometric data\footnote[1]{For the \textit{Gaia} E+DR3 catalogue, the median uncertainties on parallaxes and proper motion components are, respectively: 0.01$-$0.02\,mas and 0.02$-$0.03\,mas.yr$^{-1}$ for $G$$\,<\,$15\,mag; 0.05\,mas and 0.07\,mas.yr$^{-1}$ at $G$$\sim$17\,mag; 0.4\,mas and 0.5\,mas.yr$^{-1}$ at $G$$\sim$20\,mag \citep{Fabricius:2021}. For the photometric data, uncertainties on the $G$, $G_{\textrm{BP}}$ and $G_{\textrm{RP}}$-band data are smaller than $\sim$0.9\,mmag for $G$$\,<\,$13\,mag, smaller than $\sim$12\,mmag at $G$$\sim$17\,mag and smaller than $\sim$108\,mmag at $G$$\sim$20\,mag \citep{Riello:2021}.} allows a proper disentanglement of cluster and field populations, therefore improving the member star lists and the determination of astrophysical parameters. 

\cite{Zhong:2022}, analysed the structure of 256 OCs and, based on the correlations found among the derived structural parameters, they proposed some scaling relations. Besides, they concluded that mass loss has lead to a slight decrease on the clusters' size for ages greater than $\sim$30\,Myr. In turn, \cite{Tarricq:2022} revisited the membership lists of 389 local OCs ($d\lesssim1.5\,$kpc), detecting vast coronae around most of the investigated sample, some of them presenting tidal tails. On average, they also found that the core radii tend to be smaller and less dispersed for old clusters (log\,$t$ $\gtrsim9$) in comparison to younger ones. None of these works, however, investigated possible correlations with internal dynamical timescales or position within the Galaxy.

\cite{Pang:2021} employed a clustering algorithm \citep{Yuan:2018} and performed a detailed analysis of the morphology and kinematics of 13 selected OCs. They also carried out a set of $N$-body simulations with different sets of initial conditions and gas expulsion regimes and explored the compatibility of the models predictions with the observed data. Their sample was restricted to the solar neighborhood ($d\lesssim 500\,$pc).

More recently, \cite{Hunt:2024} have improved the open cluster census within the MW after establishing objective criteria to distinguish between genuine OCs and gravitationally unbound moving groups, which undergo different disruption processes. The observed mass functions of the investigated OCs proved to be compatible with \citeauthor{Kroupa:2001}'s\,\,(\citeyear{Kroupa:2001}) initial mass function and the more centrally concentrated clusters presented, on average, larger total mass in comparison to sparser ones. In turn, for a given total mass, unbound moving groups are generally larger than OCs, which makes them more prone to disruptive effects. However, possible impacts of the external tidal field on the clusters dynamical state have not been deeply explored. 

Other worth mentioning works using \textit{Gaia} data, devoted to the characterization of OCs and their evolution, have: (i) investigated the clusters elongation and verified possible trends of the degree of deformation with the cluster age \citep{Hu:2021}, (ii) employed high-resolution spectra of giant/red clump stars in a sample of OCs, combined with Gaia DR2 astrometry, for the precise determination of their ages and chemical composition \citep{Zhang:2021}, (iii) employed deep infrared photometry in order to extend the membership analysis to the bottom of the main sequence (Pe\~na Ram\'irez et al.\,\,\citeyear{Pena-Ramirez:2021}, \citeyear{Pena-Ramirez:2022}), (iv) targeted clusters of different ages within all the main MW components and compiled a uniform high-resolution spectroscopic dataset (the \textit{Gaia-}ESO Suvey; \citeauthor{Bragaglia:2022}\,\,\citeyear{Bragaglia:2022}), (v) detected hints of a correlation between the fraction of binary stars and the central density of the host cluster, (vi) evaluated the role of the Galactocentric distance and initial mass on the longevity of old OCs \citep{Alvarez-Baena:2024}, among others (e.g., \citeauthor{Perren:2020}\,\,\citeyear{Perren:2020}; \citeauthor{Perren:2022}\,\,\citeyear{Perren:2022}; \citeauthor{Ding:2021}\,\,\citeyear{Ding:2021}; \citeauthor{Karatas:2023}\,\,\citeyear{Karatas:2023}; \citeauthor{Rangwal:2023}\,\,\citeyear{Rangwal:2023}; \citeauthor{Maurya:2023}\,\,\citeyear{Maurya:2023}; \citeauthor{Vaher:2023}\,\,\citeyear{Vaher:2023}; \citeauthor{Viscasillas-Vazquez:2023}\,\,\citeyear{Viscasillas-Vazquez:2023}; see also the review by \citeauthor{Krumholz:2019}\,\,\citeyear{Krumholz:2019}).

From the above reference list (far from being complete), it becomes evident that the intricate processes which lead a stellar system to dissolution are multifactorial and a proper comprehension demands uniform analysis procedures and data sources. The mass loss processes due to the internal interactions, regulated by the external tidal field, and stellar evolution play a role in determining the cluster dynamical state at a given age. The superposition of this set of disruptive effects, each one taking place during different timescales, combined with different cluster formation conditions, mean that structural parameters (and relations among them) should not be considered simple functions of time.

%Verso original submetida
%The present paper is inserted in this context. Here we investigated a set of 114 OCs, which were combined with the outcomes from a previous work (\citeauthor{Angelo:2023}\,\,\citeyear{Angelo:2023}, hereafter Paper\,I) that employed the same procedures for the study of 60 OCs. Therefore, our total sample contains 174 objects, corresponding to $\sim10\%$ of the number of OCs in the \citeauthor{Dias:2021a}\,\,(\citeyear{Dias:2021a}; hereafter DMML21) catalogue. We focused on non-embedded clusters containing reasonably large catalogued number of member stars ($N\gtrsim100$), which allowed us to derive structural parameters and effectively apply a decontamination technique (Section~\ref{sec:methods}) to derive optimized member star lists. The complete sample of 174 OCs spans different ages (7\,$\lesssim$\,log\,($t.\textrm{yr}^{-1}$)\,$\lesssim$\,10, therefore comprising different evolutionary stages) and Galactocentric distances (6 $\lesssim$ $R_\textrm{G}$ (kpc) $\lesssim$ 12, making it possible to sample different external environments). Possible imprints of the evolutionary process on the observed structural parameters were then outlined after separating the complete sample according to the clusters physical properties and location within the Galaxy. Their degree of mass segregation was also evaluated and discussed (Section~\ref{sec:discussion}).  

The present paper is inserted in this context. Here we investigated a set of 114 OCs (see Section\,2), which were combined with the outcomes from a previous work (\citeauthor{Angelo:2023}\,\,\citeyear{Angelo:2023}, hereafter Paper\,I) that employed the same procedures for the study of 60 OCs. Therefore, our total sample contains 174 objects, corresponding to $\sim10\%$ of the number of OCs in the \citeauthor{Dias:2021a}\,\,(\citeyear{Dias:2021a}; hereafter DMML21) catalogue. This investigation is part of a recent effort (e.g., \citeauthor{Angelo:2018}\,\,\citeyear{Angelo:2018}; Piatti, Angelo \& Dias\,\,\citeyear{Piatti:2019}; Paper\,I) devoted to provide a list of observational parameters, derived from uniform analysis procedures and databases, which can provide useful  constraints to models intended to describe the clusters evolution from, e.g., N-body simulations (e.g., Rossi et al. 2016; Pfeffer et al. 2018).

We derived structural parameters and effectively applied a decontamination technique (Section~\ref{sec:methods}) to obtain optimized member star lists. The complete sample of 174 OCs spans different ages (7\,$\lesssim$\,log\,($t.\textrm{yr}^{-1}$)\,$\lesssim$\,10, therefore comprising different evolutionary stages) and Galactocentric distances (6 $\lesssim$ $R_\textrm{G}$ (kpc) $\lesssim$ 12, making it possible to sample different external environments). Possible imprints of the evolutionary process on the observed structural parameters were then outlined after separating the complete sample according to the clusters physical properties and location within the Galaxy. Their degree of mass segregation was also evaluated and discussed (Section~\ref{sec:discussion}).

This paper is organized as follows: in Section~\ref{sec:sample_data} we present the collected data and the investigated sample. Section~\ref{sec:methods} presents the methodology used to investigate the clusters' structure, establish membership probabilities and determine astrophysical parameters. Our results are analysed in Section~\ref{sec:analysis} and  discussed in Section~\ref{sec:discussion}. Section~\ref{sec:conclusions} is devoted to our main conclusions.

%%%%%%%%MEUS INPUTS%%%%%%%%%
  %%%%%%%%%%%%%%%%%%%
\section{Sample and data collected}
\label{sec:sample_data}
%%%%%%%%%%%%%%%%%%%

We searched the DMML21 catalogue looking for clusters with reasonably large number of members ($N\,\gtrsim\,100$) and presenting low-to-moderate interstellar extinction ($E(B-V)$ typically smaller than $\sim0.6$\,mag), therefore avoiding embedded  stellar groups, severely affected by differential reddening. Only clusters with log\,$t > 7$ were selected. We focused on clusters presenting well-defined contrast in relation to the general Galactic field, as inferred from visual inspection of DSS\footnote[2]{https://archive.stsci.edu/cgi-bin/dss\_form} images and through preliminary analysis of their radial density profiles (Section~\ref{sec:struct_params}). In this search, we excluded those clusters tagged as non-physical groups or doubtful cases according to \cite{Cantat-Gaudin:2020a} and \cite{Cantat-Gaudin:2020}. Our final sample consists in a set of 114 OCs. 

We downloaded photometric, astrometric and spectroscopic data from the \textit{Gaia} DR3 catalogue ({\fontfamily{ptm}\selectfont \textit{gaiadr3.gaia\_source}} and {\fontfamily{ptm}\selectfont \textit{gaiadr3.astrophysical\_parameters}} tables) using dedicated Astronomical Data Query Language (ADQL) scripts run on the \textit{Gaia} Archive\footnote[3]{https://gea.esac.esa.int/archive/}. For each investigated OC, the extraction radius (typically, $r\gtrsim2^{\circ}$) is larger than $\sim5\times$ the cluster radius listed in DMML21, centred on the catalogued coordinates. 

Data collected from \textit{Gaia}'s main table ({\fontfamily{ptm}\selectfont \textit{gaiadr3.gaia\_source}}) were corrected according to the prescriptions and scripts\footnote[4]{https://www.cosmos.esa.int/web/gaia/dr3-software-tools} available in the online documentation\footnote[5]{https://gea.esac.esa.int/archive/documentation/GDR3/\\index.html}, namely: (i) corrections to the parallax zero-point \citep{Lindegren:2021}, (ii) corrections to the flux excess factor \citep{Riello:2021} and to the (iii) radial velocity ($V_{\textrm{rad}}$) for hot \citep{Blomme:2022} and cold stars \citep{Katz:2022}. Z\textit{Gaia}'s main table also incorporates atmospheric parameters (effective temperature, $T_{\textrm{eff}}$, surface gravity, log\,$g$, and iron abundance, $[Fe/H]$) derived from the {\fontfamily{ptm}\selectfont GSP-Phot} algorithm run on low-resolution BP/RP spectra \citep{Andrae:2022}, as part of the astrophysical parameters inference system (Apsis; \citeauthor{Creevey:2022}\,\,\citeyear{Creevey:2022} and \citeauthor{Fouesneau:2022}\,\,\citeyear{Fouesneau:2022}). 

Spectroscopic parameters available in the {\fontfamily{ptm}\selectfont \textit{gaiadr3.astrophysical\_parameters}} table were derived from the analysis methods implemented in the {\fontfamily{ptm}\selectfont GSP-Spec} algorithm, within \textit{Gaia}'s Apsis pipeline run on higher resolution RVS spectra \citep{Cropper:2018}. The set of atmospheric parameters\footnote[6]{The $[Fe/H]$ metallicity is not provided directly in the {\fontfamily{ptm}\selectfont  \textit{gaiadr3.astrophysical\_parameters}} table; instead, the global metallicity, $[M/H]$, and abundance of neutral iron, $[Fe/M]$, are available. Therefore, we employed the relation $[Fe/H]$ = $[Fe/M]$ + $[M/H]$, with the correspoding uncertainty $\Delta[Fe/H]$=$\sqrt{(\Delta[M/H])^2+(\Delta[Fe/M])^2}$.} available in this additional table were recalibrated according to the prescriptions outlined in \cite{Recio-Blanco:2022}. In the present paper, the uncertainties of the polynomial coefficients in the calibration equations (their tables~3, 4 and E.1) have been properly considered together with the catalogued parameter uncertainties. 

Complementarly, we also downloaded $T_{\textrm{eff}}$ and log\,$g$ values for stars analysed by the {\fontfamily{ptm}\selectfont ESP-HS} module \citep{Fremat:2023}, which deals specifically with hot stars ($T_{\textrm{eff}}\gtrsim7\,500\,$K) by assuming solar composition. Our final database contains the complete set of parameters obtained from the  \textit{Gaia} tables mentioned above and cross-matched via the stars' unique source identifier ({\fontfamily{ptm}\selectfont source\_id}). Finally, in order to avoid sources with problematic astrometry and/or photometry, we restricted our sample to stars with $G$\,$\le$\,19\,mag, which corresponds to the nominal completeness limit of the \textit{Gaia} catalogue (section 2 of \citeauthor{Fabricius:2021}\,\,\citeyear{Fabricius:2021}), and applied the following quality filters:

\begin{flalign}
    & RUWE < 1.4,   &   \\
    &  \vert C^{*} \vert\,<\,5\,\sigma_{C^*}\,\,\, \textrm{(for $G\,>\,4\,$mag)},   &      
\end{flalign}

\noindent
where $RUWE$ is the \textit{renormalised unit weight error} parameter for astrometry \citep{Lindegren:2021a}, $\sigma_{C^*}$ is given in equation 18 of \cite{Riello:2021} and $C^{*}$ is the corrected flux excess factor  parameter for photometry ($E(\textrm{BP/RP})$; \citeauthor{Evans:2018}\,\,\citeyear{Evans:2018}).

Table~\ref{tab:investig_sample} presents the coordinates and some of the derived astrophysical parameters for the 114 OCs investigated in the present paper (see additional parameters in Table~\ref{tab:masses_and_other_params}). Other 60 OCs were previously characterized in Paper\,I using the same procedures, thus totalizing a sample of 174 clusters.

  %%%%%%%%%%%%%%
\section{Methods}
\label{sec:methods}
%%%%%%%%%%%%%%

%%%%%%%%%%%%%%%%%%
\subsection{Preliminary analysis}
\label{sec:pre_analysis}
%%%%%%%%%%%%%%%%%%

\begin{figure*}
\begin{center}

\parbox[c]{1.00\textwidth}
  {
   \begin{center}
       \includegraphics[width=0.330\textwidth]{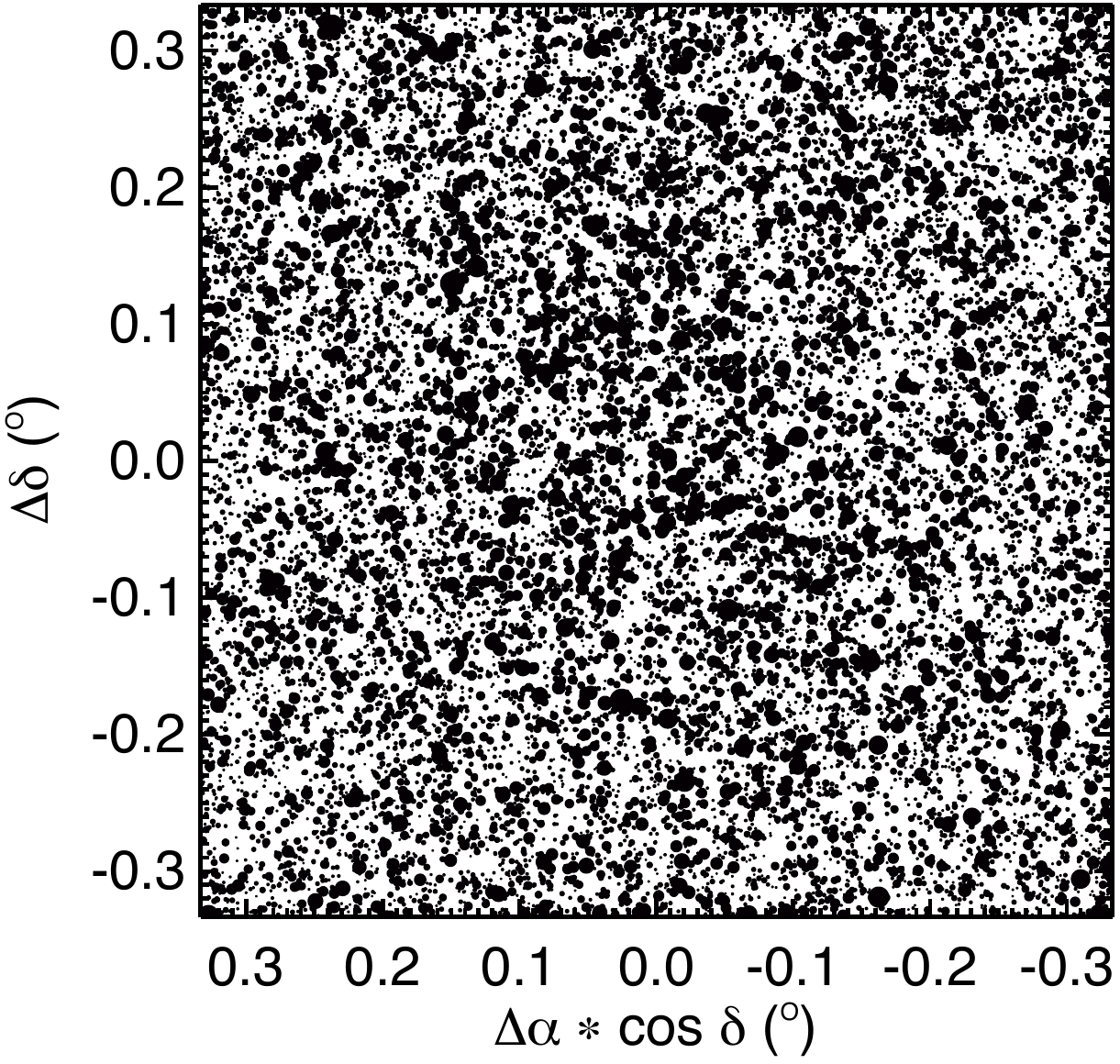}
       \includegraphics[width=0.325\textwidth]{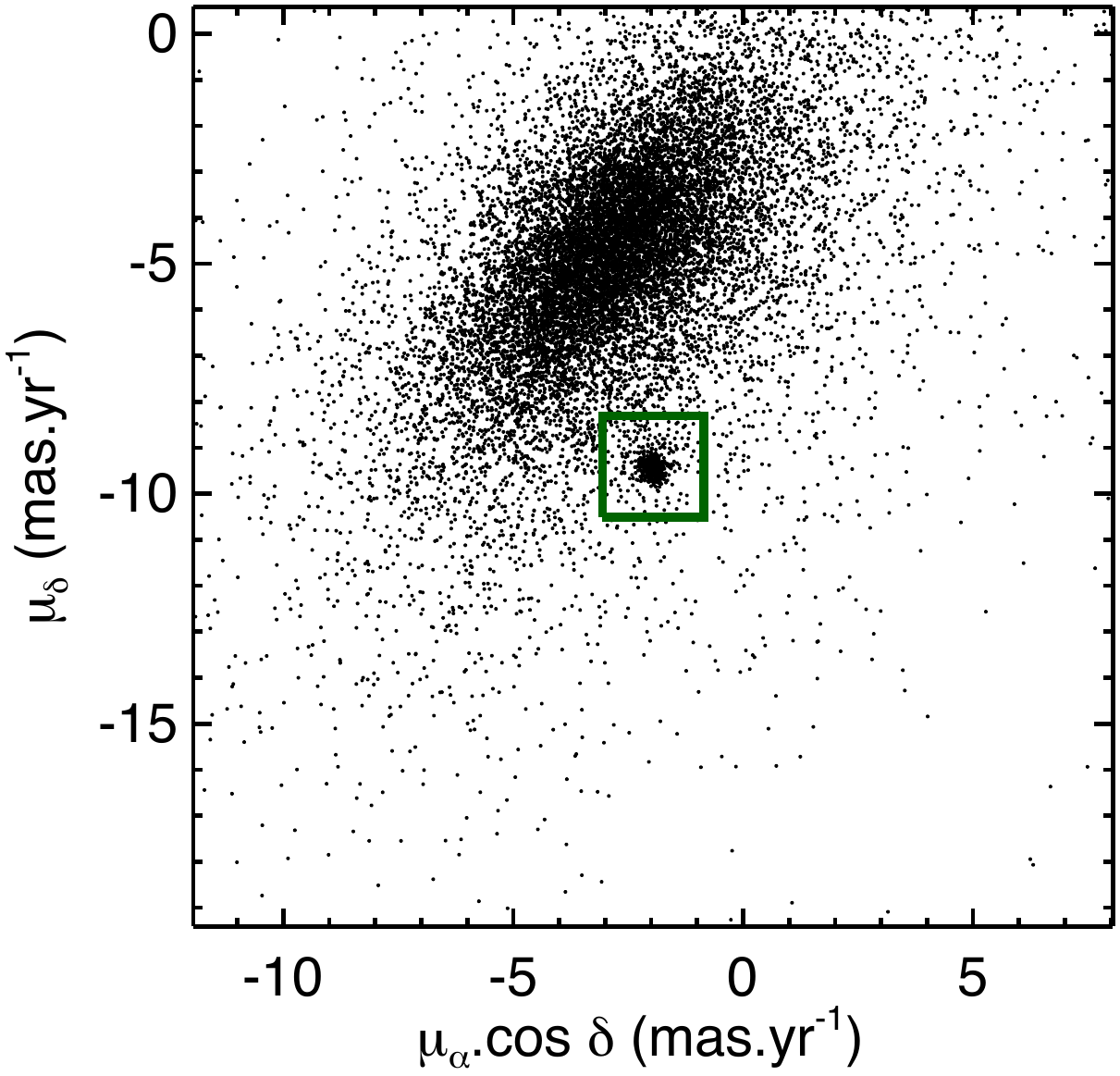}   
       \includegraphics[width=0.330\textwidth]{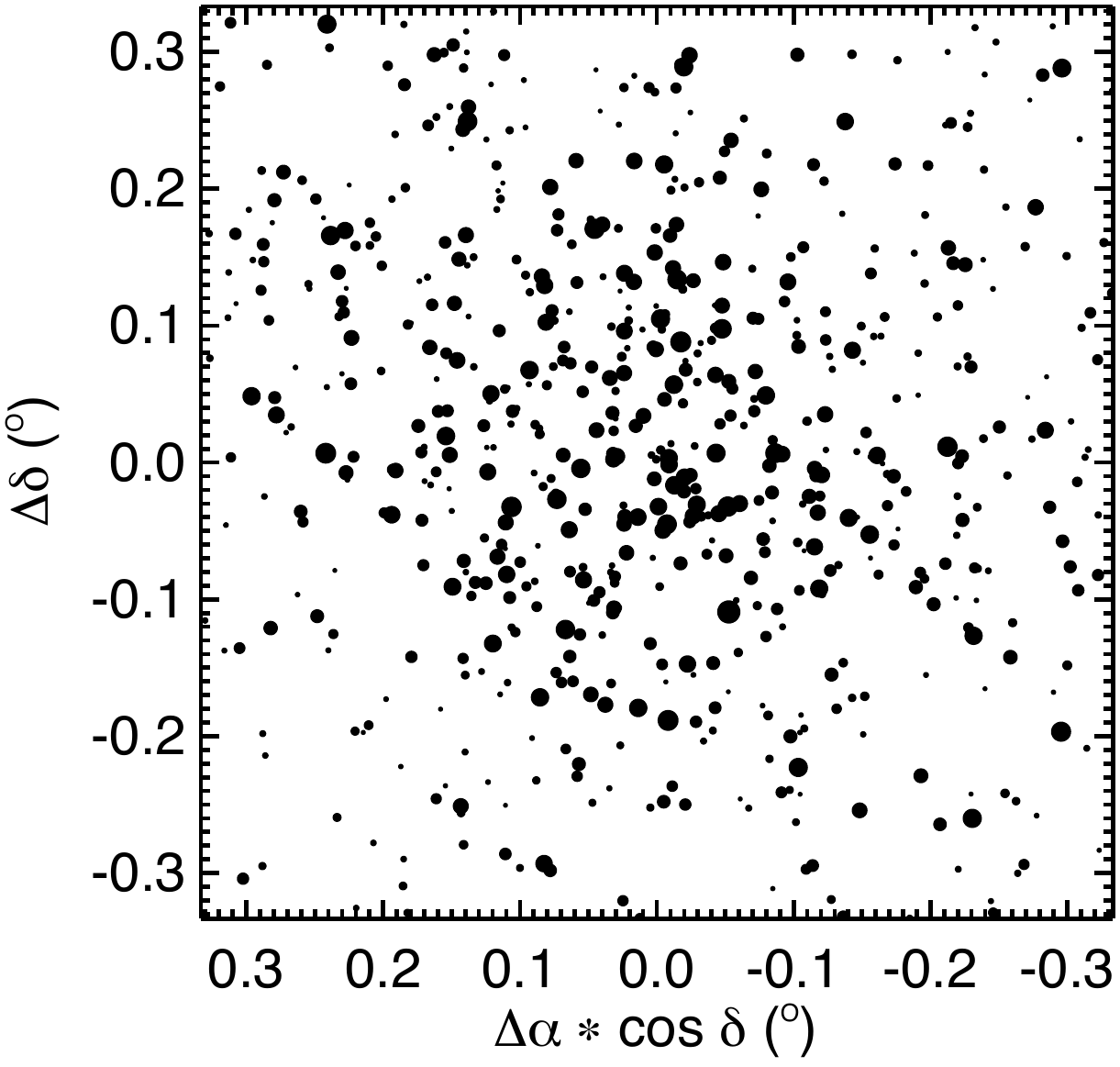}   
    \end{center}    
  }
\caption{   Left panel: Skymap for stars in a field of $40'\times40'$ centred on the OC NGC\,6940. The stars sizes are proportional to their brightness in the $G$-band. Middle panel: VPD for the sample of stars in the left-hand panel; the green box shows the proper motions filter. Right-hand panel: Reconstructed skymap for NGC\,6940, after applying the proper motions filter.   }

\label{fig:pre_analysis}
\end{center}
\end{figure*}

In this initial step of our procedure, we looked for the signature of each investigated OC in the vector-point diagram (VPD), as illustrated in Figure~\ref{fig:pre_analysis}. The left panel exhibits a skymap, centered on NGC\,6940's coordinates (Table~\ref{tab:investig_sample}), for all stars within an area of $40'\,\times\,40'$ and consistent with the quality filters outlined in Section~\ref{sec:sample_data}. It is noticeable a poor contrast between the cluster and field populations, since NGC\,6940 is located at a low Galactic latitude ($b\sim-7{\degr}\,$) and therefore projected against a dense background. The middle panel shows the VPD for this sample, where the overdensity at the centre, around ($\mu_{\alpha}\,\textrm{cos}\,\delta$ , $\mu_{\delta}$) \, $\simeq$ \, (-2.0 , -9.5) \, mas.yr$^{-1}$, is defined mostly by cluster member stars.   

In order to alleviate the contamination by field stars present in the cluster area, we reconstructed the OC skymap after considering a subsample of stars consistent with a box-shapped proper motion filter, as indicated by the green square (size equal to $\sim1.1\,$mas.yr$^{-1}$ in this case) in Figure~\ref{fig:pre_analysis}. The filtered skymap is shown in the right panel, where it is evident an improved contrast between the object and the field. An analogous procedure was employed for all investigated clusters. When a clear overdensity in the VPDs could not be obtained (due to, e.g., the centroid defined by cluster member stars being comparable to the bulk motion of the local disc field), the proper motions filter was defined after restricting the cluster skymap to squared areas of $\sim1-2$\,$\times$ the cluster radius as catalogued in  DMML21 and \cite{Cantat-Gaudin:2020}. After identifying the cluster centroid in the VPD, this spatial constraint was dismissed. 

In each case, the size of the box-shapped proper motion filter is larger than 5 times the intrinsic dispersions in $\mu_\alpha\,\textrm{cos}\,\delta$ and $\mu_\delta$, as inferred after setting memberships (see Section~\ref{sec:decontam_CMD_analysis} and Table~\ref{tab:investig_sample}). These filters are large enough to encompass the cluster member stars, but small enough to eliminate most of the contamination by the disc population.  The filtered skymaps were then employed in the subsequent steps for the structural analysis (Section~\ref{sec:struct_params}).

%%%%%%%%%%%%%%%%%%
\subsection{Structural parameters}
\label{sec:struct_params}
%%%%%%%%%%%%%%%%%%

The structural parameters (core and tidal radii, respectively $r_c$ and $r_t$) are derived from the fit of K62 law

\begin{equation}
   \sigma\,(r)\,\propto\,\left(\frac{1}{\sqrt{1+(r/r_c)^2}} - \frac{1}{\sqrt{1+(r_t/r_c)^2}} \right)^2
   \label{eq:King_profile}
\end{equation}

%Versão inicial
%\noindent 
%to each cluster RDP. Previously, the set of data was proper motion filtered, in order to improve the cluster-field contrast, and the $\alpha, \delta$ coordinates of each star were projected on the plane of the sky, according to the relations of \citeauthor{van-de-Ven:2006}\,\,(\citeyear{van-de-Ven:2006}, their section 2.3). The detailed procedure is outlined in section~2 of Paper\,I and we present here the main steps:

\noindent 
to each cluster RDP. In this step, we employed the proper motion filtered skymaps (Section~\ref{sec:pre_analysis}), for which the $\alpha, \delta$ coordinates of each star were projected on the plane of the sky, according to the relations of \citeauthor{van-de-Ven:2006}\,\,(\citeyear{van-de-Ven:2006}, their section 2.3). The detailed procedure is outlined in section~2 of Paper\,I and we present here the main steps:

\begin{itemize}

   \item construction of a grid of central coordinates around the literature centre of the cluster;  \\
   
   \item for each tentative centre, the cluster skymap is divided in concentric annuli of varying sizes; the corresponding stellar density is $\sigma(r)=N_{*}/A(r)$, where $N_{*}$ is the counted number of stars and $A(r)$ is the ring area; \\
   
   \item the background level ($\sigma_{\textrm{bg}}$) and associated dispersion are obtained from the mean density value of the more external bins, where the density values fluctuate around a nearly constant value; \\ 
      
   \item the background-subtracted RDP obtained for each $(\alpha, \delta)$ pair is fitted (by means of $\chi^2$ minimization) using the K62 profile (equation~\ref{eq:King_profile}); the adopted central coordinates are those that result in the highest central density with minimal residuals. This procedure allows to build a smooth RDP, with a significant contrast with the field. 

\end{itemize}

\begin{figure*}
\begin{center}

\parbox[c]{1.00\textwidth}
  {
   \begin{center}
       \includegraphics[width=0.430\textwidth]{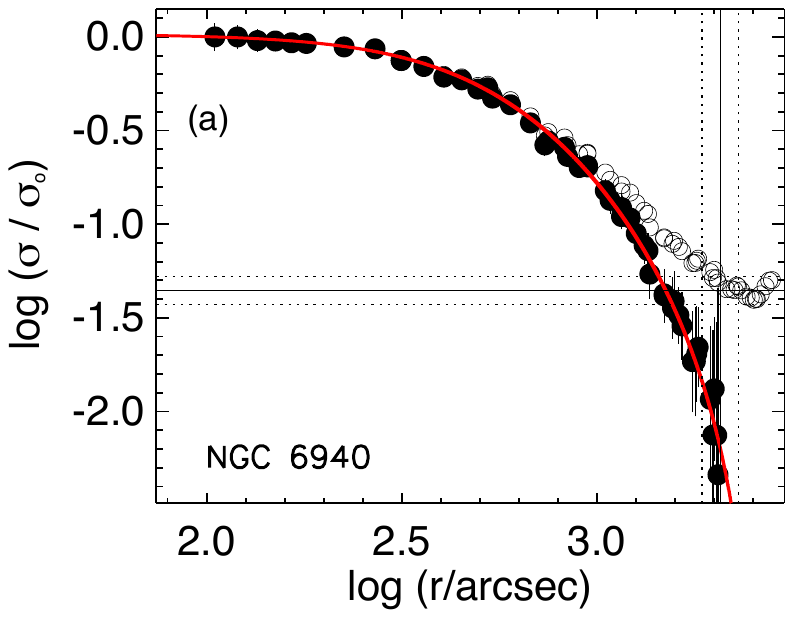}
       \includegraphics[width=0.400\textwidth]{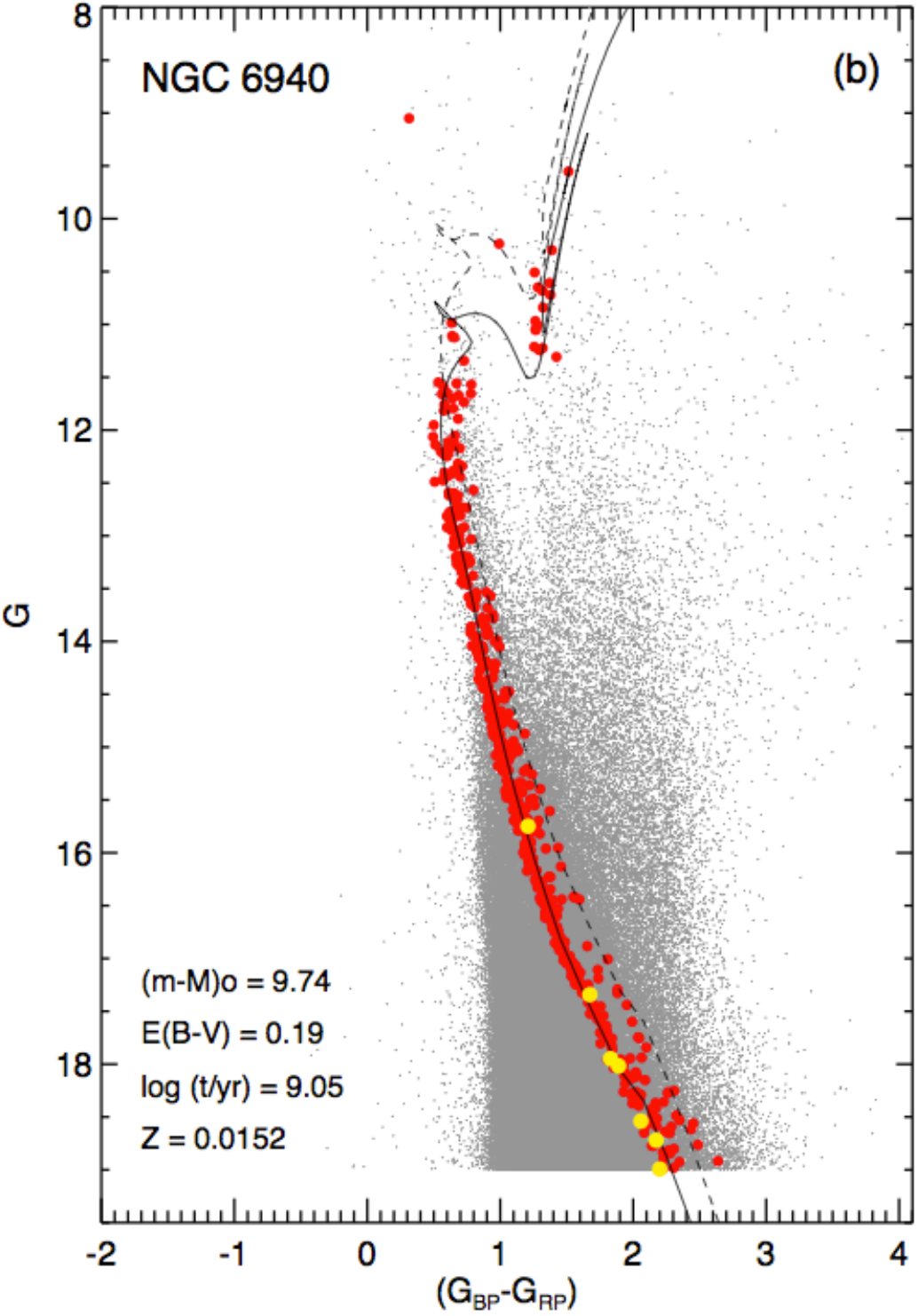}   
       \includegraphics[width=0.330\textwidth]{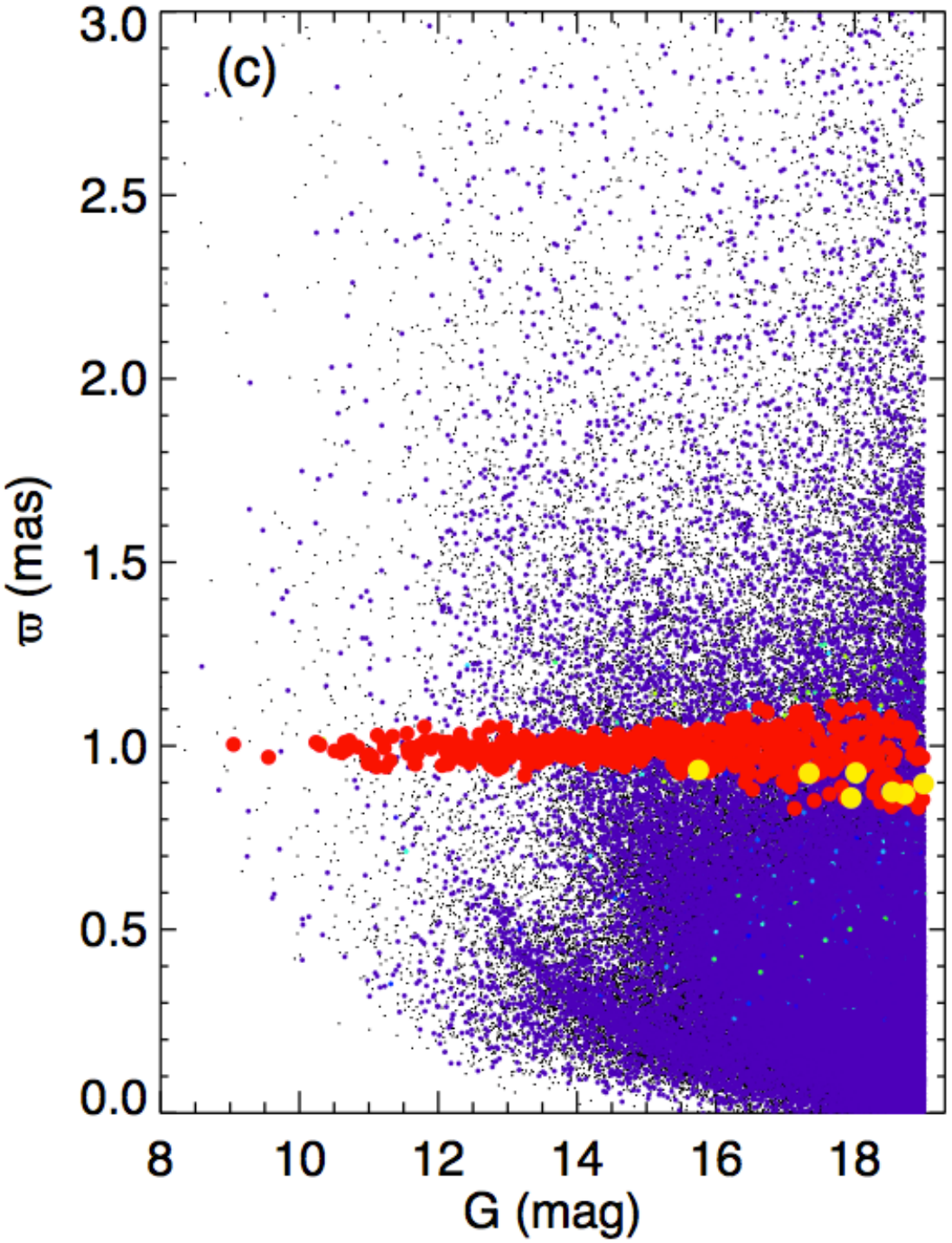}
       \includegraphics[width=0.330\textwidth]{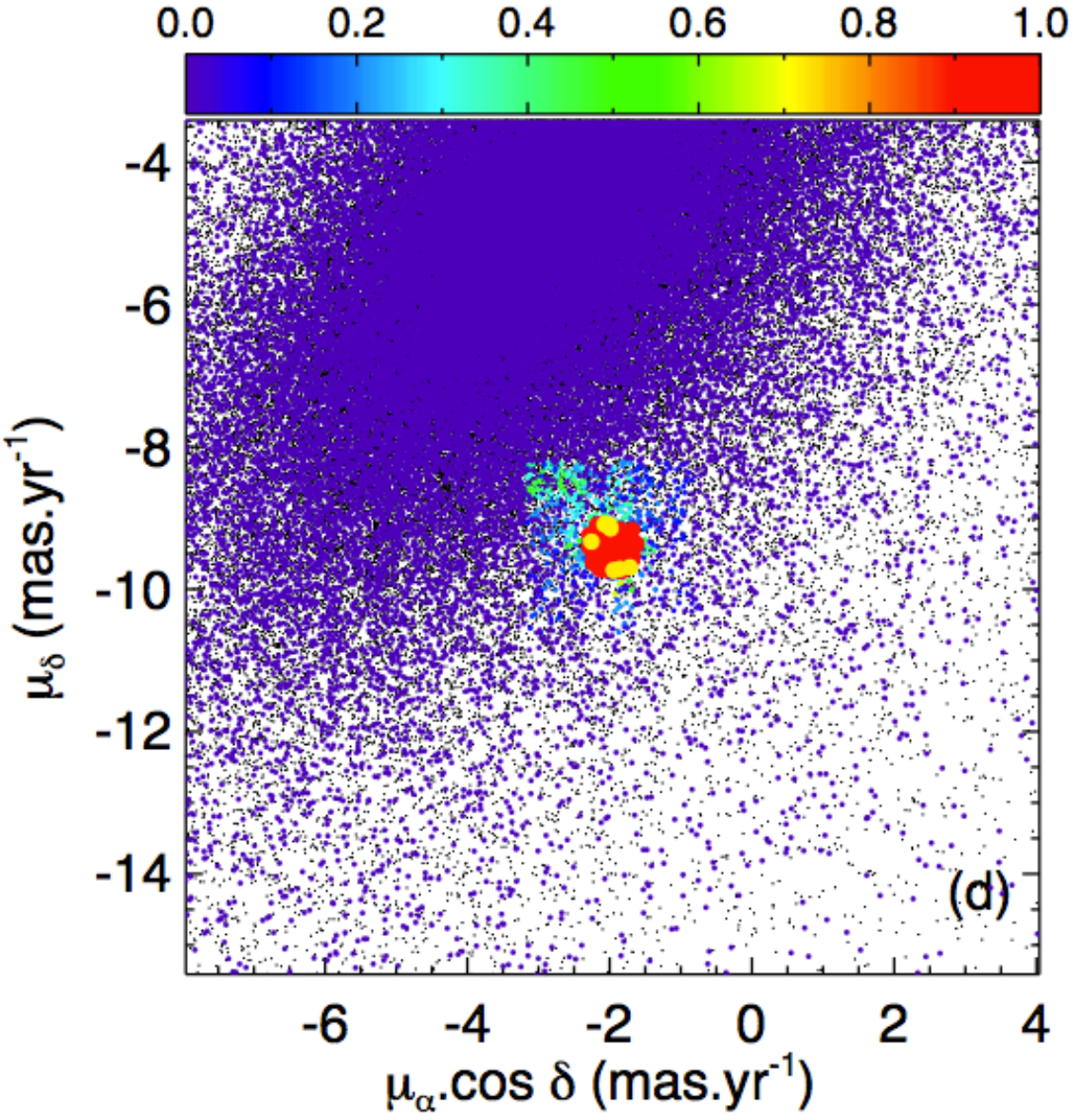}   
       \includegraphics[width=0.330\textwidth]{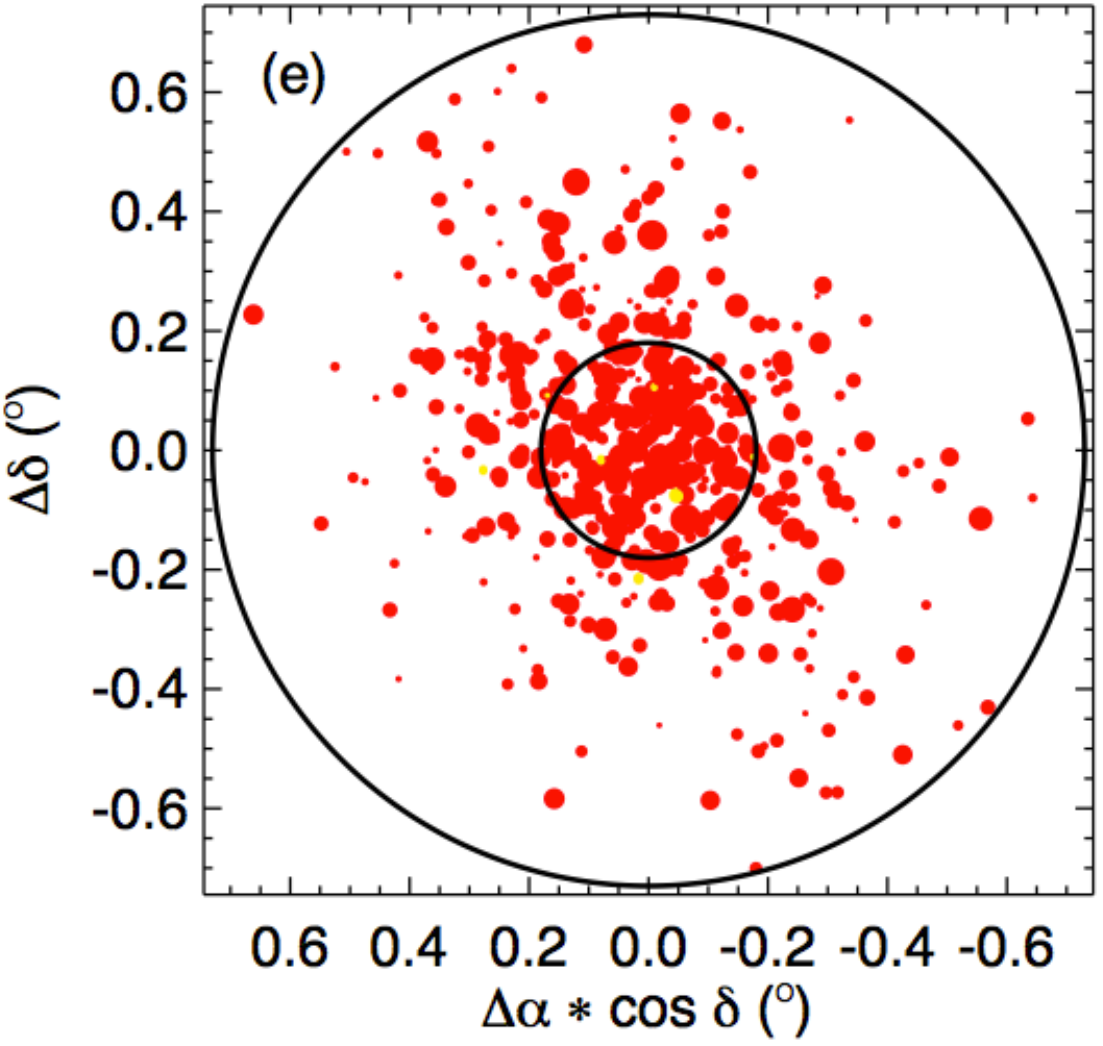}   
    \end{center}    
  }
\caption{ Results for the OC NGC\,6940. Panel (a): cluster RDP. (Open) Filled symbols represent the (non-) background subtracted profile, normalized to the central density. The mean background density ($\sigma_{\textrm{bg}}$) is indicated by the continuous horizontal line. The vertical line identifies the limiting radius ($R_{\textrm{lim}}$), defined here as the distance from the cluster centre where the observed density profile intersects the background level (uncertainties in $\sigma_{\textrm{bg}}$ and $R_{\textrm{lim}}$ are indicated by dotted lines). The red line is the fitted K62 profile. Panel (b): Decontaminated $G\times(G_{\textrm{BP}}-G_{\textrm{RP}})$ CMD. Red and yellow symbols represent member stars; small grey dots are stars in a comparison field. The continuous line is a $t\simeq1.1\,$Gyr solar metallicity PARSEC \citep{Bressan:2012} isochrone. The fundamental astrophysical parameters are indicated. The dashed line represents the locus of binaries with equal-mass components (original isochrone vertically shifted by -0.75\,mag). The brightest member ($G\,=\,9.05\,$mag) is the star \textit{Gaia} DR3 1857459766532229248 (HD\,196095; $\alpha$\,=\,20:34:13, $\delta\,=$\,28:09:51; spectral type A2, \citeauthor{Cannon:1993}\,\,\citeyear{Cannon:1993}), classified as a blue straggler by \citeauthor{Rain:2021}\,\,(\citeyear{Rain:2021}). Panels (c) and (d): $\varpi\,\times\,G\,$magnitude plot and cluster VPD, respectively. The larger red and yellow filled symbols represent member stars; symbol colours in panels (b), (c), (d) and (e) are assigned according to the membership scale, identified by the colourbar in panel (d). The tiny grey dots in panels (b), (c) and (d) are stars in a comparison field. Panel (e): Skymap for NGC\,6940. The smaller and larger circles represent, respectively, $r_c$ and $r_t$ for this cluster. Symbol size is assigned according to each star $G$ magnitude. }

\label{fig:results_clusterexample}
\end{center}
\end{figure*}

The result of this procedure is illustrated in panel (a) of Figure~\ref{fig:results_clusterexample} for the OC NGC\,6940 (see the online Supplementary material for additional RDPs). The filled circles represent the background-subracted RDP (normalized to the central density; the open circles represent the original, that is, non-background subtracted densities) and the best-fitted King profile (red line). In each radial bin, the dispersion of $\sigma_{\textrm{bg}}$ (horizontal dotted lines in the cluster RDP) has been summed in quadrature with the uncertainty derived from Poisson counting statistics. The 3D half-light radii ($r_h$; Table~\ref{tab:investig_sample}) were obtained from $r_c$ and $r_t$ using the calibrations outlined in section 6 of \cite{Santos:2020}.

The uncertainties in the core and tidal radii were estimated from their dispersion, weighted by the RMS of the residuals in the King profile fit, based on a grid of $r_t$ and $r_c$ values centred on the best fitted parameters. A bootstrap procedure is incorporated to take into account the stellar density errors in the radial profile bins.

The K62 empirical profile has been employed here instead of dynamical models (e.g., \citeauthor{King:1966}\,\,\citeyear{King:1966}, \citeauthor{Wilson:1975}\,\,\citeyear{Wilson:1975}) since the latter are mainly applied to globular clusters (GCs), for which the large number of stars allows a detailed analysis yielding a robust inference of clusters' parameters. Besides our need for a uniform analysis procedure for the whole sample and the analytical simplicity of the K62 profile, this empirical model is similar to the dynamical one for $W_0\leq7$ \citep{King:1966}, corresponding to concentration parameters log\,($r_t/r_c$)\,$\lesssim1.5$, which is the range encompassed by our investigated OCs (Table~\ref{tab:investig_sample}).

%%%%%%%%%%%%%%%%%%%%%%%%%
\subsection{Decontamination and CMD analysis}   
\label{sec:decontam_CMD_analysis} 
%%%%%%%%%%%%%%%%%%%%%%%%%

%%%%%%%%%%%%%%%%%%%%%%
\subsubsection*{Membership assignment}
%%%%%%%%%%%%%%%%%%%%%%

The disentanglement between cluster and field stars is a critical step in stellar populations analysis, indispensable to the proper recognition of evolutionary sequences on decontaminated CMDs and to optimize the determination of astrophysical parameters (e.g., \citeauthor{Maia:2010}\,\,\citeyear{Maia:2010}). Our basic strategy here to accomplish this task consists in evaluating the dispersion of the astrometric data (within a three-dimensional space, composed of parallax, $\varpi$, and proper motion components: $\mu_{\alpha}\,\textrm{cos}\,\delta$ and $\mu_{\delta}$) for stars in the cluster area compared to a representative set of \textit{observed} field stars (instead of a randomly chosen comparison sample). 

The comparison field was chosen from an annular region, concentric to the cluster, with area equal to 3 times the cluster area (that is, $A_{\rm{fld}}$ = 3\,$A_{\rm{clu}}$, where $A_{\rm{clu}}$ = $\pi\,r_t{^2}$) and with inner radius equal to 3 times the cluster $r_t$. This way, we are sampling a group of field stars located reasonably far from the object, but not too far so as to risk losing the local field-star signature in terms of proper motion and parallax distributions. 

Our method (named ANDORRA code, an acronym for ANgelo Decontamination methOd fRom astRometric dAta) is described in detail in \citeauthor{Angelo:2019a}\,\,(2019a) and has been employed in some recent papers (\citeauthor{Angelo:2019b}\,\,2019b; Angelo, Santos Jr. \& Corradi\,\,\citeyear{Angelo:2020}; \citeauthor{Angelo:2021}\,\,\citeyear{Angelo:2021}). Here we summarize the main steps of the algorithm:

\begin{itemize}

   \item we build the 3D astrometric space defined by $\varpi$, $\mu_{\alpha}\,\textrm{cos}\,\delta$, $\mu_{\delta}$ data collected for stars within the cluster empirical tidal radius\footnote[7]{Restricting the search of member stars to the $r\,\leq\,r_t$ region (where gravitationally bound cluster stars dominate) was a necessary step specially in the case of OCs presenting lower contrast with the field in the astrometric space (that is, clusters with average proper motion and parallax values not significantly different from the average values of field stars). Although external tidal structures are reported in the literature for some OCs in our sample (e.g., NGC\,1039, NGC\,1528; \citeauthor{Bhattacharya:2022}\,\,\citeyear{Bhattacharya:2022}), extending the search radius to values considerably larger than $r_t$ usually results in a number of false-positives, therefore decreasing the performance of the decontamination method.} (i.e., for $r\,\leq\,r_t$) and for a large annular external comparison field; \\
   
   \item the astrometric space is divided into small cells of varying sizes, proportional to the mean uncertainties ($\sim1$$\times$$\langle\Delta\varpi\rangle$ , $\sim$$10\times$$\langle\Delta\mu_{\alpha}^*\rangle$ , $\sim$$10\times$$\langle\Delta\mu_{\delta}\rangle$) of the overall sample employed in the decontamination procedure; \\
   
   \item within each cell, membership likelihoods are derived for stars in both cluster and field samples by means of multivariate gaussians, which incorporate the correlations among the astrometric parameters and their uncertainties;  \\ 
   
   \item entropy-like functions are then employed to identify those cells within which the ensemble of astrometric parameters for cluster stars is statistically more concentrated in comparison to the field. As long as these cells present significant overdensities comparatively to the whole grid, stars within them receive large final membership probabilities ($P$). 
   
\end{itemize}

In summary, the algorithm searches for significant overdensities, statistically distinguishable from the general Galactic field, defined by cluster stars in the astrometric space. Figure~\ref{fig:results_clusterexample} presents the results of this strategy for the OC NGC\,6940 (see the online Supplementary material for additional figures); respectively, panels (b) to (e) show the decontaminated $G\times(G_{\textrm{BP}}-G_{\textrm{RP}})$ CMD and the best-fitted isochrone (see below), the $\varpi\,\times\,G_{\textrm{mag}}$ plot, the cluster VPD and the skymap. In these panels, member stars ($P\gtrsim0.70$) are highlighted with filled symbols.  Symbol colours are assigned according to the membership scale, represented by the colourbar in panel (d); small grey dots in panels (b), (c) and (d) are stars in the comparison field. As expected, high membership stars define recognizable evolutionary sequences in the cluster CMD and prominent concentrations in the astrometric space.

In the case of clusters that are not well separated from the field population in the astrometric space (that is, severely contaminated OCs), we start our procedure employing a decontamination radius comparable to $r_c$, where there is a larger contrast with the field. When necessary to obtain cleaner results, we limited the procedure to stars in the range $G\lesssim18\,$mag or even $G\lesssim17\,$mag, depending on the contamination level. This is an important step to make sure that we are properly identifying fiducial evolutionary sequences in the CMD and real overdensities in the astrometric space defined by these more central stars. 

After finding the signature of the cluster in the CMD and astrometric diagrams, the procedure was applied again, this time with decontamination radius equal to the cluster $r_t$ and with no restrictions on magnitudes. We consider the final membership probability to be the result of this second run. 
%As mentioned before, the comparison field was taken from an external annular ring with area corresponding to 3 times the cluster area. This choice was made in order to provide a proper representativity of the field contamination in the cluster area and to avoid oversampling of field stars in the astrometric space.}

%%%%%%%%%%%%%%%%%%%%%%%%%%%%%
\subsubsection*{Fundamental astrophysical parameters}
%%%%%%%%%%%%%%%%%%%%%%%%%%%%%

\begin{figure*}
\begin{center}

   \includegraphics[width=0.90\textwidth]{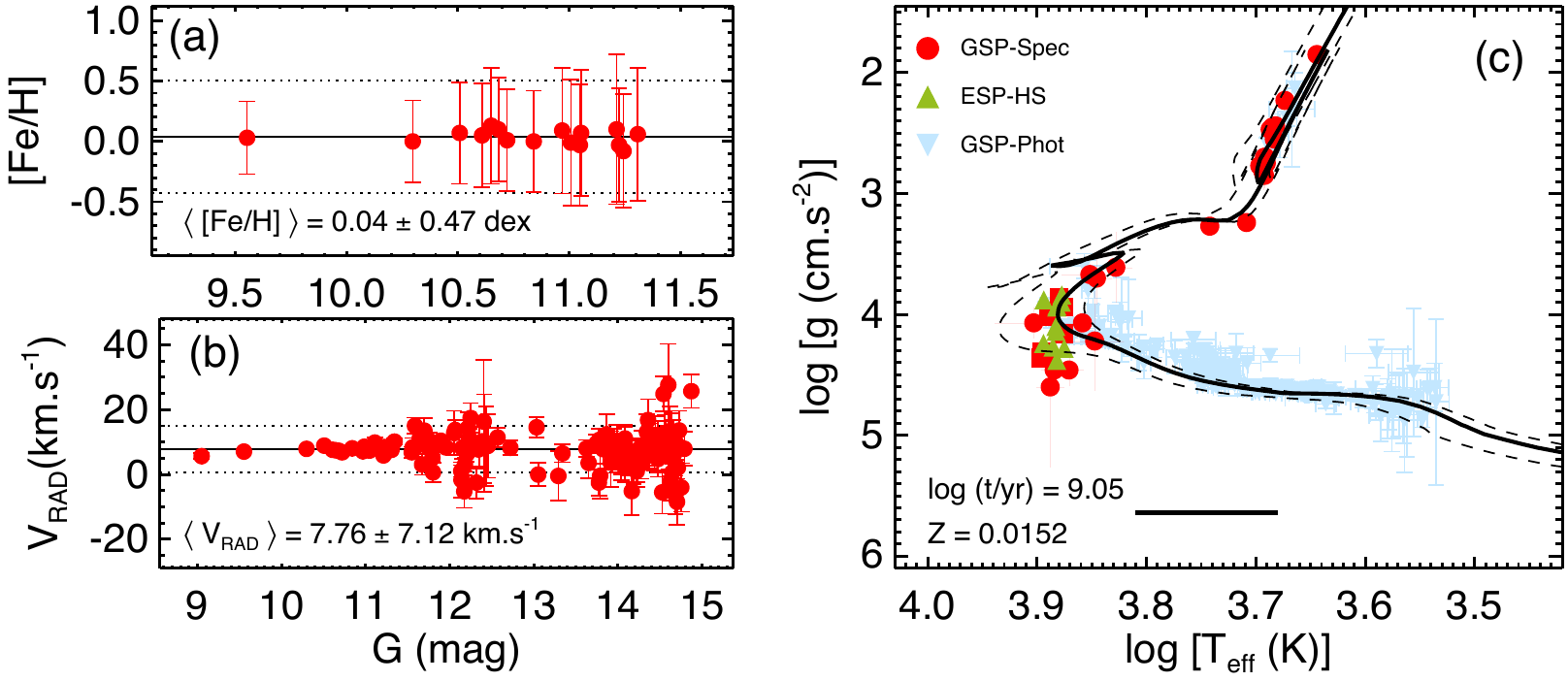}
       
\caption{ Spectroscopic plots for member stars of NGC\,6940. Panels (a) and (b) show, respectively, the metallicity $[Fe/H]$ and $V_{\textrm{rad}}$ as function of the $G\,$magnitude. The horizontal continuous lines represent the median of $[Fe/H]$ and $V_{\textrm{rad}}$ (also indicated in the legend) for the plotted samples and the dotted ones indicate the median absolute deviation, summed in quadrature with the mean uncertainty (see text for details). Panel (c): Spectroscopic \textit{Hertzsprung-Russell} diagram. Symbols and colours represent stars analysed by different algorithms within the \textit{Gaia's} inference system, namely: {\fontfamily{ptm}\selectfont GSP-Spec} (filled circles; colours assigned according to the star membership; see Figure~\ref{fig:results_clusterexample}), ESP-HS (green triangles) and {\fontfamily{ptm}\selectfont GSP-Phot} (upside down blue triangles). The continuous line is the same solar metallicity isochrone of panel (b) in Figure~\ref{fig:results_clusterexample}. For illustration purposes, the dashed lines represent two other PARSEC isochrones of log $t$\,=\,9.05, with different metallicities: $Z=0.0252$ (cooler turnoff) and $Z=0.0052$ (hotter turnoff).}

\label{fig:HRD}
\end{center}
\end{figure*}

For each investigated OC, we took those member stars presenting spectroscopic data and plotted their $[Fe/H]$, $V_{\textrm{rad}}$, log\,$g$ and $T_{\textrm{eff}}$, as shown in Figure~\ref{fig:HRD} (for the OC NGC\,6940, taken here as an example). Panel (a) in this figure allows to infer the cluster metallicity (which is connected to the metal abundance ratio $Z$ via the approximate relation $[Fe/H]$\,$\simeq$\,log$(Z/Z_{\odot})$, with $Z_{\odot}=0.0152$; \citeauthor{Bonfanti:2016}\,\,\citeyear{Bonfanti:2016}), while panel (b) shows the dispersion of the radial velocities. In both panels, the horizontal continuous line indicates the median value. The dotted ones are the absolute deviation from the median, to which we have summed in quadrature the mean uncertainty of each plotted sample, in order to properly take into account the systematic uncertainties coming from the $[Fe/H]$ and $V_{\rm{RAD}}$ recalibration procedure (Section~\ref{sec:sample_data}). In turn, panel (c) allows an initial guess for the cluster age from the intrinsic evolutionary sequences (i.e., independent of the cluster distance and interstellar reddening) shown in the spectroscopic \textit{Hertzsprung-Russell} diagram. Symbols and colours identify parameters estimated from different algorithms within the \textit{Gaia's Apsis} pipeline (see the figure legend and Section~\ref{sec:sample_data}). An initial distance estimate was obtained from simply inverting the mean parallax of the member stars.  

For the isochrone fitting procedure (Figure~\ref{fig:results_clusterexample}, panel $b$), we followed the same procedure of Paper\,I: we firstly fixed the values of the overall metallicity $Z$ and log\,$t$, obtained from the above initial estimates; then we allowed for variations in the distance modulus, $(m-M)_0$, and colour excess, $E(B-V)$, by successively shifting the isochrone in small steps of 0.05\,mag and 0.01\,mag, respectively (the extinction relations of \citeauthor{Cardelli:1989}\,\,\citeyear{Cardelli:1989} and \citeauthor{ODonnell:1994}\,\,\citeyear{ODonnell:1994} have been employed). In each step, the distance of each member star from the nearest isochrone point is determined and the residuals are registered. After determining the best solution for these both parameters (from the minimum overall residue), we allowed for variations in $Z$ and log\,$t$, within steps of, respectively, 0.002\,dex and 0.05\,dex. The procedure is iterated until a proper match of the key evolutionary sequences (the main sequence, the turnoff point, the subgiant and red giant branches and the red clump, if present) is established. The derived solutions were inspected for all OCs. Beyond these uncertainties, overall displacements and changes in the isochrone morphology would result in poor fits of the key evolutionary sequences along the CMD. The results are registered in Table~\ref{tab:investig_sample}.       
%The parameter uncertainties correspond to shifts of the isochrone around the best solution that result in poor fits of the evolutionary sequences along the cluster CMD. 

  %%%%%%%%%%
\section{Analysis}
\label{sec:analysis}
%%%%%%%%%%

%%%%%%%%%%%%%%%%%%%%%%%%%%%%%%%%%%
\subsection{Mass function}
%%%%%%%%%%%%%%%%%%%%%%%%%%%%%%%%%%

\begin{figure}
\begin{center}

\parbox[c]{0.48\textwidth}
  {
   \begin{center}
       \includegraphics[width=0.48\textwidth]{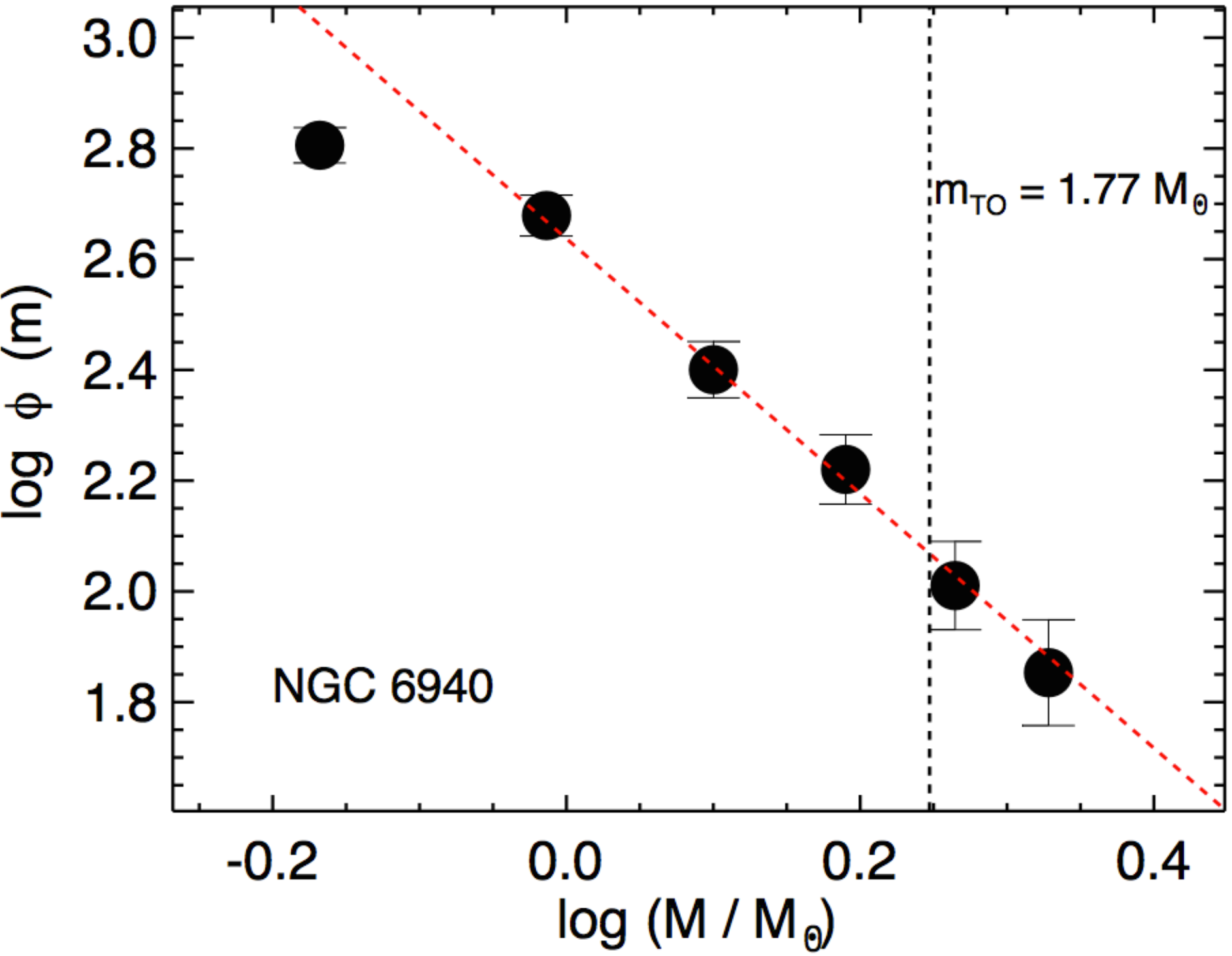}  
    \end{center}    
  }

\caption{ Observed mass function for the OC NGC\,6940. The vertical line indicates the turnoff point mass ($m_{\textrm{TO}}$; see also panel $b$ of Figure~\ref{fig:results_clusterexample}). The red line is the normalized initial mass function of \citeauthor{Kroupa:2001}\,\,(\citeyear{Kroupa:2001}; see text for details). Poisson error bars are shown. }

\label{fig:MF}
\end{center}
\end{figure}

For each cluster, the individual mass of the member stars was estimated from interpolation of their $G\,$magnitude along the fitted isochrone (see Figure~\ref{fig:results_clusterexample}, panel $b$). Then the cluster mass function (MF) was derived by counting the number of stars within different mass bins (i.e., $\phi\,(m)$=$dN/dm$; uncertainties come from Poisson counting statistics), as shown in Figure~\ref{fig:MF} for the OC NGC\,6940. In general, the slope of the observed MFs in the higher mass domain is compatible with  Kroupa's\,\,(\citeyear{Kroupa:2001}) initial mass function (IMF,  normalized according to the mass summed up within the higher mass bins of the observed MF) for most of the investigated OCs, a result that is compatible with the outcomes from \citeauthor{Hunt:2024}\,\,(\citeyear{Hunt:2024}, their section 5.3). 

Some of the investigated OCs (e.g., NGC\,6940) show signals of depletion of lower mass stars, since their observed MFs present one or more lower mass bins that deviate from Kroupa's IMF (see also Figure~\ref{fig:mbreak_versus_rhoamb} and the discussion following it). This deviation, when present, occurs at mass bins superior to the limiting mass (corresponding to $G=\,19\,$mag; see Table~\ref{tab:masses_and_other_params}) and, therefore, should not be attributed to photometric incompleteness. The cluster total mass ($M_{\textrm{clu}}$) was derived by integrating the normalized IMF until the inferior mass limit of $\simeq0.1\,M_{\odot}$, in order to consider possible member stars below the photometric completeness limit (Section~\ref{sec:sample_data}). We also performed an (over)estimation of mass possibly kept in the cluster in the form of dark stellar remnants (white dwarfs, neutron stars and black holes), following the procedure outlined in the Appendix~B of \cite{Maia:2014}. For all investigated ages, their fractional contribution to the total mass is small (see also Appendix~A of Paper\,I). Uncertainties in $M_{\textrm{clu}}$ come from error propagation.

%We did not try to fit the observed MF slope in the lower mass domain due to the scarcity of points.

%%%%%%%%%%%%%%%%%%%%%%%%%%%%%%%%%%
\subsection{Jacobi radius and Galactic potential}
%\label{sec:phig_rhoamb_Rj}
%%%%%%%%%%%%%%%%%%%%%%%%%%%%%%%%%%

Following \cite{Renaud:2011}, the Jacobi radius\footnote[8]{$R_J$ should not be confused with $r_t$, which is the truncation radius of the K62 profile (Section~\ref{sec:struct_params} and panel $a$ of Figure~\ref{fig:results_clusterexample}), that is, $r_t$ defines an empirical scale length for the total cluster size. As stated by \cite{Portegies-Zwart:2010}, there are no \textit{a priori} reasons to assume $r_t$ equal to $R_J$ (see also the discussion in section~2 of \citeauthor{Baumgardt:2010}\,\,\citeyear{Baumgardt:2010}).} ($R_J$) for each of our investigated clusters is derived from the expression 

\begin{equation}
  R_J = \left( \frac{G\,M_{\textrm{clu}}}{\lambda_{e,1}} \right)^{1/3},
  \label{eq:Rj}
\end{equation}

\noindent where $M_{\textrm{clu}}$ is the cluster mass and $\lambda_{e,1}$ is the largest eigenvalue of the tidal tensor, which comes out in the expression for the net acceleration of a member star in the cluster non-inertial reference frame. The formula for $\lambda_{e,1}$ is

\begin{equation}
   \lambda_{e,1} = -\left(\frac{\partial^2\phi_G}{\partial x'^2}\right)_{R_G} - \left(-\frac{\partial^2\phi_G}{\partial z'^2}\right)_{R_G},
   \label{eq:lambda_e1}
\end{equation}

\noindent where the derivatives of the Galactic potential (see below) are taken with respect to coordinates in a right-handed $x', y', z'$ cartesian system centred on the cluster. The $x'$-axis is oriented along the Galactic centre $-$ cluster direction, with the Galactic centre located at $x'=-R_\textrm{G}$. 
%(see figure~1 of \citeauthor{Renaud:2011}\,\,\citeyear{Renaud:2011}). 

In the present paper, the Galactic potential ($\phi_\textrm{G}$) is modeled as the sum of bulge ($\phi_B$), disk ($\phi_D$) and dark matter halo potentials ($\phi_H$), given by the following expressions (taken from, respectively, \cite{Hernquist:1990}, \cite{Miyamoto:1975} and \citeauthor{Sanderson:2017}\,\,\citeyear{Sanderson:2017}):

\begin{flalign}
    & \phi_{B} = -\frac{G\,M_B}{r+r_B}   \label{eq:phi_B}  & \\
    & \phi_{D} = -\frac{G\,M_D}{  \sqrt{x^2+y^2 + (a + \sqrt{z^2 + b^2})^2}    }   \label{eq:phi_D} &   \\
    & \phi_{H} = -\frac{G\,M_s}{(\textrm{ln}\,2 - 1/2)}\frac{\textrm{ln}(1+r/r_s)}{r}\,\,\,,   \label{eq:phi_H}   &      
\end{flalign}

\noindent where $r=\sqrt{x^2+y^2+z^2}$ is the Galactocentric distance. The following parameters have been employed: $M_B=2.5\times10^{10}\,$M$_{\odot}$, $r_B=0.5\,$kpc, $M_D=7.5\times10^{10}\,$M$_{\odot}$, $a=5.4\,$kpc and $b=0.3\,$kpc, obtained from \cite{Haghi:2015}; the scale radius $r_s=15.19\,$kpc and mass $M_s=1.87\times10^{11}\,$M$_{\odot}$ come from \cite{Sanderson:2017}.

Since equation~\ref{eq:lambda_e1} is adequate for circular orbits (equation~\ref{eq:Rj}, in turn, applies to all galactic potentials), we have employed the corrections outlined by \cite{Webb:2013} in order to properly take into account the influence of the orbital eccentricity (orbital parameters taken from \citeauthor{Tarricq:2021}\,\,\citeyear{Tarricq:2021}) on the determination of $R_J$. The uncertainty in $R_J$ corresponds to the dispersion of a set of ten thousand redrawings, where we randomly sampled values for each parameter ($\theta_i$) that enters in its determination, within the respective uncertainty (that is, within the interval $\theta_i\,\pm\,\Delta\theta_i$; section 6 of Paper\,I).

%%%%%%%%%%%%%%%%%
\subsection{Ambient density}
%%%%%%%%%%%%%%%%%

\begin{figure}
\begin{center}

\parbox[c]{0.48\textwidth}
  {
   \begin{center}
       \includegraphics[width=0.48\textwidth]{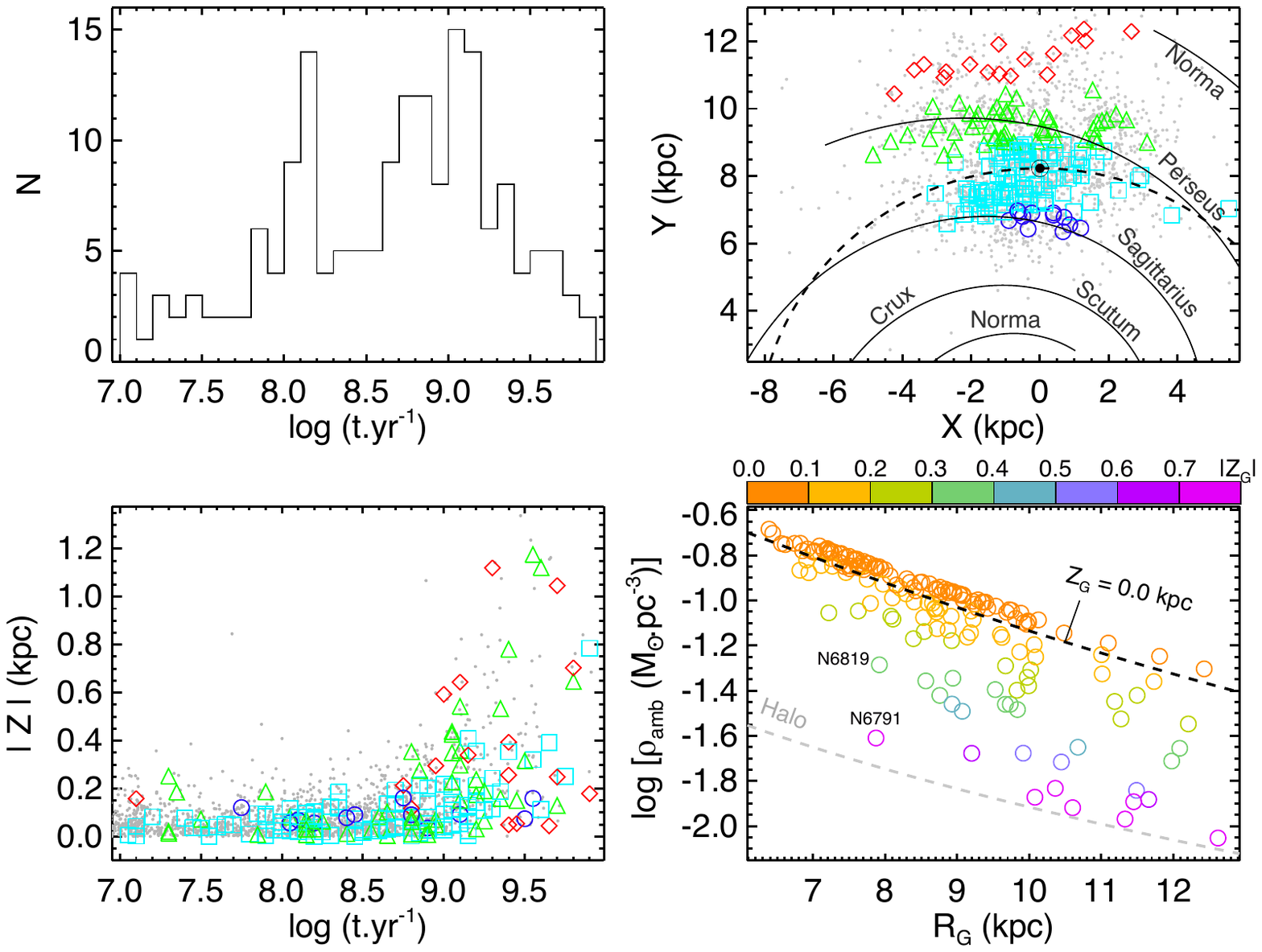}  
    \end{center}    
  }

\caption{ Ambient density ($\rho_{\textrm{amb}}$) as function of $R_\textrm{G}$ for the investigated sample, as inferred from the analytical expression for the full Galactic potential ($\phi_G$). Symbol colours were assigned according to different $\vert Z_\textrm{G}\vert$ bins, as indicated in the colourbar. The contributions of the disc ($\phi_{D}$; equation~\ref{eq:phi_D}) and halo ($\phi_H$; equation~\ref{eq:phi_H}) potentials, isolately, are indicated by dashed lines. Two of the investigated OCs (NGC\,6819 and NGC\,6791) are indicated (see Section~\ref{sec:discussion} for details). }

\label{fig:rho_amb_versus_Rgal}
\end{center}
\end{figure}

The ambient density ($\rho_{\textrm{amb}}$) to which each cluster is subject is derived from $\phi_G$ and it is employed here as a proxy for the strength of the Galactic tidal field\footnote[9]{This can be illustrated from \citeauthor{Gieles:2008}'s\,\,(\citeyear{Gieles:2008}, hereafter GB08) simulations (their section 3) of the stellar evaporation rate ($\dot N$) for a cluster evolving in a tidal field. The expression derived for $\dot N$ is proportional to the orbital frequency $\omega=V_\textrm{G}/R_\textrm{G}$ (where $V_\textrm{G}$ is the circular velocity), depending also on the number of stars ($N$) and cluster  structure. In the simple case of an external logarithmic potential (\citeauthor{Baumgardt:2003}\,\,\citeyear{Baumgardt:2003}; \citeauthor{Lamers_tdis_gal}\,\,\citeyear{Lamers_tdis_gal}) of the form $\phi(R_\textrm{G})$\,=\,$V_\textrm{G}^2$\,ln\,($R_\textrm{G}$), we have $\nabla^2\phi=\omega^2$ and thus $\rho_{\textrm{amb}}=\omega^2/(4\pi G)$. Therefore, considering GB08's approach, larger $\rho_{\textrm{amb}}$ values contribute to larger mass loss by tidal effects.}, since it is related to the divergence of the MW's local gravitational acceleration via Poisson's equation (i.e., $4\pi G\,\rho_{\textrm{amb}}$=$-\nabla\cdot\vec{g}$=$\nabla^2\phi_G$). In Figure~\ref{fig:rho_amb_versus_Rgal}, $\rho_\textrm{amb}$ is plotted as function of $R_\textrm{G}$ for all investigated OCs. Since the disc potential ($\phi_D$) is also a function of the $Z$ Galactic coordinate, the symbol colours were assigned according to the vertical distance to the plane ($\vert Z_\textrm{G}\vert$). The overplotted lines represent the dependence of $\rho_\textrm{amb}$ with $R_\textrm{G}$ taking into account solely the contributions of the disc (black dashed line) and halo (grey dashed line) potentials. The bulge contribution ($\rho_{\textrm{amb}}^{\textrm{Bulge}}$) is negligible in the range of interest for $R_\textrm{G}$ (log\,$\rho_{\textrm{amb}}^{\textrm{Bulge}}\lesssim-3.0$ for $R_\textrm{G}\gtrsim\,$6\,kpc).

%Another evidence to employ $\rho_{\textrm{amb}}$ as an indicator of the external tidal field strength is shown in Figure~\ref{fig:mbreak_versus_rhoamb}. We inspected the 174 cluster mass functions (MFs; Section~\textcolor{red}{a definir}) and selected those for which there are evidence of low-mass star depletion. In these cases, we determined the inferior observed mass bin ($M_{\textrm{break}}$) below which the observed MF departs significantly from \citeauthor{Kroupa:2001}'s\,\,(\citeyear{Kroupa:2001}) law. This quantity was plotted as function of the ambient density in Figure~\ref{fig:mbreak_versus_rhoamb}. Symbols and colours were assigned according to the scheme outlined in Section~\ref{sec:discussion}. The inset shows the Pearson  correlation coefficient $r$ (open circles connected by lines) between $M_{\textrm{break}}$ and $\rho_{\textrm{amb}}$ as we progressively restrict the sample to clusters for which $\rho_{\textrm{amb}}\,>\,\rho_{\textrm{amb}}^{\textrm{cut}}$. The number of objects ($N_{\textrm{clusters}}$) satisfying this condition is indicated at the top for each $\rho_{\textrm{amb}}^{\textrm{cut}}$, until a minimum of 10 clusters (corresponding to $\rho_{\textrm{amb}}^{\textrm{cut}}\simeq0.16\,M_{\odot}\,\textrm{pc}^{-3}$) in the correlation calculation. 

\begin{figure}
\begin{center}

\parbox[c]{0.48\textwidth}
  {
   \begin{center}
       \includegraphics[width=0.48\textwidth]{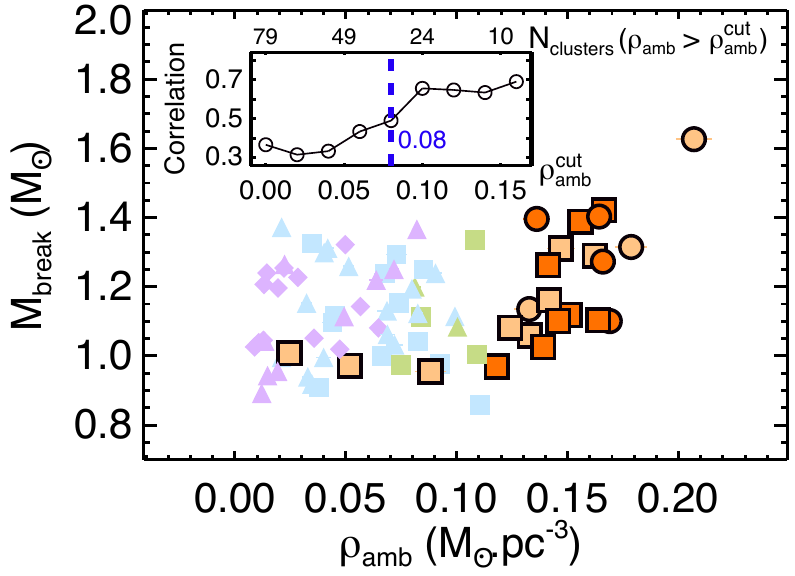}  
    \end{center}    
  }

\caption{ Inferior mass bins ($M_{\textrm{break}}$; see text for details) for the observed MFs (Figure~\ref{fig:MF}) as function of the ambient density ($\rho_{\textrm{amb}}$). Symbol colours are assigned according to the clusters $R_\textrm{G}$ and structure (see more details in Section~\ref{sec:discussion}): the orange and red contoured symbols identify, respectively, \textit{compact} and \textit{loose} clusters located at $R_\textrm{G}\,\leq\,8\,$kpc;  the blue and green symbols represent clusters, respectively, with $r_h/R_J$ ratio inferior and superior to $\sim$0.4 (see also  Figure~\ref{fig:rh_Rj_and_rho_hm_versus_Rgal})  and located in the range 8$\,<\,$\,$R_\textrm{G}\,(\textrm{kpc})$\,$\leq$ 10. The purple symbols are clusters located at $R_\textrm{G}\,>\,10\,$kpc. The inset shows the correlation $r$ (open circles) between $M_{\textrm{break}}$ and $\rho_{\textrm{amb}}$ for clusters in the range $\rho_{\textrm{amb}}\,>\,\rho_{\textrm{amb}}^{\textrm{cut}}$, for progressively larger values of $\rho_{\textrm{amb}}^{\textrm{cut}}$. The number of objects used in each case is indicated at the top ($N_{\textrm{clusters}}$). The $r$ values become larger than $\sim0.5$ for $\rho_{\textrm{amb}}$\,$\gtrsim$\,0.08\,$M_{\odot}.$pc$^{-3}$. }

\label{fig:mbreak_versus_rhoamb}
\end{center}
\end{figure}

An evidence of the impact of the external tidal field strength on the cluster dynamics is suggested in Figure~\ref{fig:mbreak_versus_rhoamb}. We inspected the 174 cluster MFs (see Figure~\ref{fig:MF}) and selected those for which there is evidence of low-mass star depletion. In these cases, we determined the inferior observed mass bin ($M_{\textrm{break}}$) below which the observed MF departs significantly from the \citeauthor{Kroupa:2001}'s\,\,(\citeyear{Kroupa:2001}) law. This quantity was plotted as function of the ambient density in Figure~\ref{fig:mbreak_versus_rhoamb}. Symbols and colours were assigned according to the scheme outlined in Section~\ref{sec:discussion} (see Table~\ref{tab:symbols_convention} below). The inset shows the Pearson  correlation coefficient $r$ (open circles connected by lines) between $M_{\textrm{break}}$ and $\rho_{\textrm{amb}}$ as we progressively restrict the sample to clusters for which $\rho_{\textrm{amb}}\,>\,\rho_{\textrm{amb}}^{\textrm{cut}}$. The number of objects ($N_{\textrm{clusters}}$) satisfying this condition is indicated at the top for each $\rho_{\textrm{amb}}^{\textrm{cut}}$, until a minimum of 10 clusters (corresponding to $\rho_{\textrm{amb}}^{\textrm{cut}}\simeq0.16\,M_{\odot}\,\textrm{pc}^{-3}$) in the correlation calculation.

The $r$ values become greater than $\sim$0.5 for ambient densities larger than $\sim$0.08\,$M_{\odot}\,\textrm{pc}^{-3}$ (log\,$\rho_{\textrm{amb}}\gtrsim-1.1$, blue dashed line in Figure~\ref{fig:mbreak_versus_rhoamb}). At this domain, the positive correlation is an indication that those OCs subject to stronger tidal stresses tend to have their lower-mass stellar content more efficiently depleted. The increase of $M_{\textrm{break}}$ with $\rho_{\textrm{amb}}$ is particularly evident for OCs located at $R_G\leq8\,$kpc (contoured symbols, for which $r\simeq0.70$). At the weaker external tidal field domain ($\rho_{\textrm{amb}}\lesssim0.08\,M_{\odot}\,\textrm{pc}^{-3}$), the plotted quantities present no clear correlations. 

%We also did not find significant correlations with $R_{\textrm{G}}$ or $Z_{\textrm{G}}$, thus suggesting that $M_{\textrm{break}}$ for these OCs seems determined mostly by their internal evolution, as set by the formation conditions followed by two-body relaxation.    

\cite{De-Marchi:2010} obtained a relation between a tapered mass function characteristic mass (their equation 1 and figure 1) and the dynamical age for 30 objects, from embedded clusters to GCs. Taking $M_{\textrm{break}}$ as a proxy for the characteristic mass, the present study, focused on a much larger sample but restricted to open clusters, does not show the same relation as that derived by \cite{De-Marchi:2010}. At least for the age range investigated here ($\log\ (t.{\rm yr}^{-1})\sim 7-10$), the environment density (as inferred from the local local gravitational potential) is the main factor determining the stellar mass where depletion starts to be significant.

%%%%%%%%%%%%%%%%%%%%%
\subsection{Half-light relaxation time}
\label{sec:trh}
%%%%%%%%%%%%%%%%%%%%%

Furthermore, we also derived the half-light relaxation time ($t_{rh}$) for the investigated OCs. This internal timescale, which can be interpreted as the time interval for stars within a gravitationally bound system to reach dynamical equilibrium, was estimated from the expression \citep{Spitzer:1971}:

\begin{equation}
   t_{rh}=(8.9\times10^5\,\textrm{yr})\,\frac{M_{\textrm{clu}}^{1/2}\,r_h^{3/2}}{\langle m\rangle\,\textrm{log}_{10}(0.4M_{\textrm{clu}}/\langle m\rangle)},
   \label{eq:trh}
\end{equation}

\noindent 
where $M_{\textrm{clu}}$ and $N_{\textrm{clu}}$ are, respectively, the cluster mass and number of stars (Table~\ref{tab:masses_and_other_params}) and $\langle m\rangle=M_{\textrm{clu}}/N_{\textrm{clu}}$.

%%%%%%%%%%%%%%%%%%%%%%%%%%%%%%
\subsection{Initial mass and dissolution time estimates}
%%%%%%%%%%%%%%%%%%%%%%%%%%%%%%

Initial mass estimates ($M_{\textrm{ini}}$) for the investigated OCs were obtained from equation 7 of \citeauthor{Lamers_tdis_anal}\,\,(\citeyear{Lamers_tdis_anal}, and references therein), which employs the cluster age and present-day mass (Table~\ref{tab:investig_sample}). We also estimated each cluster dissolution time ($t_{\textrm{95}}$), assumed here as the time interval after which the cluster has lost $\sim95\%$ of their initial mass content by tidal effects, combined with stellar evolution. From equation 6 of \cite{Lamers_tdis_anal}, we can numerically solve the expression:

\begin{equation}
    0.05 = \left[ \mu_{\textrm{ev}}^\gamma (t_{95}) - \frac{\gamma}{t_0}\,t_{95}\,M_{\textrm{ini}}^{-\gamma} \right]^{1/\gamma}  
    \label{eq:t95}
\end{equation} 

\noindent
where $\gamma=0.62$, $t_0=810\,\textrm{Myr}\,(1-\epsilon)\,10^{-4\gamma}\,\rho_{\textrm{amb}}^{-1/2}$ (Lamers, Gieles \& Portegies Zwart\,\,\citeyear{Lamers_tdis_gal}), $\epsilon$ being the cluster orbital eccentricity and $\rho_{\textrm{amb}}$ is the ambient density at the apogalactic radius. In turn, $\mu_{\textrm{ev}}\,(t) = 1-q_{\textrm{ev}}(t)$. The function $q_{\textrm{ev}} (t)$ (see equation 2 of \citeauthor{Lamers_tdis_anal}\,\,\citeyear{Lamers_tdis_anal}) is the fraction of the initial cluster mass lost by stellar evolution only (derived from the GALEV evolutionary models; \citeauthor{Schulz:2002}\citeyear{Schulz:2002}; \citeauthor{Anders:2003}\,\,\citeyear{Anders:2003}). 

The errors in both $M_{\textrm{ini}}$ and $t_{95}$ are obtained from a set of random resamplings of all parameters employed in their determination, as described in section 6 of Paper\,I.

  %%%%%%%%%%
\section{Discussion}
\label{sec:discussion}
%%%%%%%%%%

%%%%%%%%%%%%%%%%%%%%%%%%%
\subsection{Spatial distribution in the Galaxy}
%%%%%%%%%%%%%%%%%%%%%%%%%

\begin{figure*}
\begin{center}

\parbox[c]{1.00\textwidth}
  {
   \begin{center}
    \includegraphics[width=0.49\textwidth]{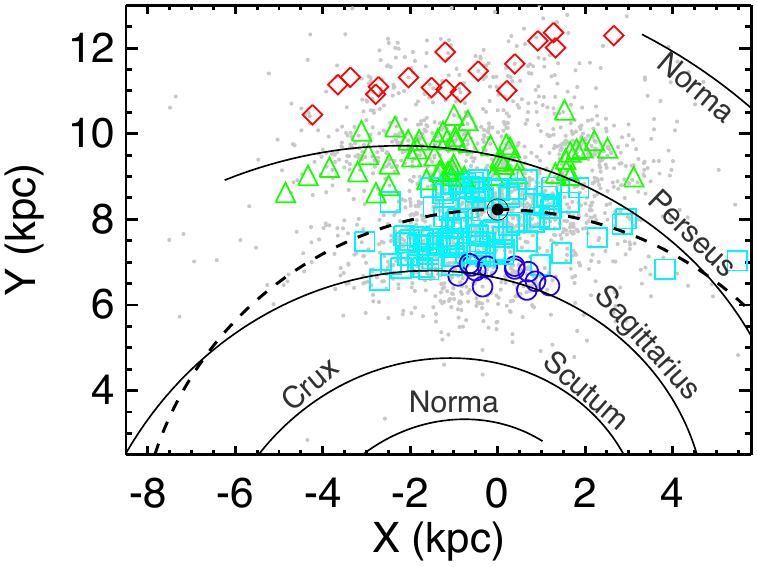} 
    \includegraphics[width=0.495\textwidth]{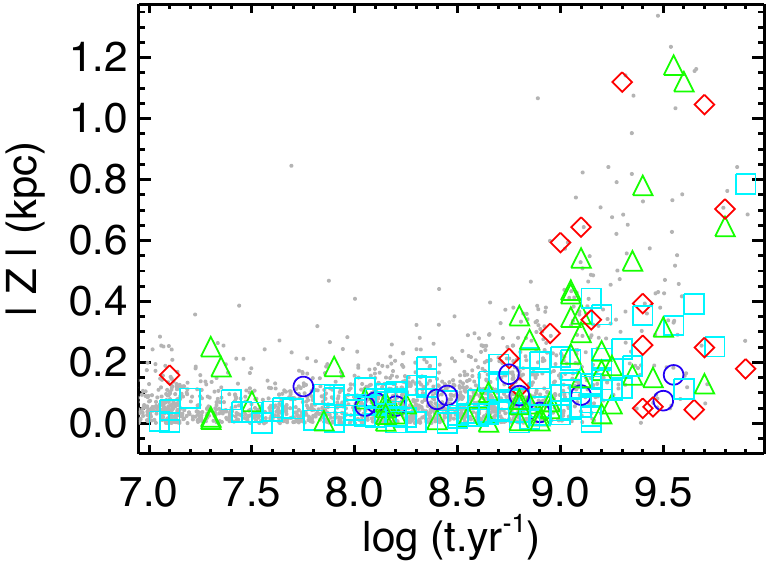}   
    \end{center}    
  }
\caption{ Left panel: distribution of the investigated OCs in the Galactic plane. The solar symbol ($X=0.0\,$; $Y\simeq8.0\,$kpc; \citeauthor{Reid:1993a}\,\,\citeyear{Reid:1993a}), the solar circle (dashed line) and the schematic location of the spiral arms (taken from \citeauthor{Vallee:2008}\,\,\citeyear{Vallee:2008}) are indicated. Symbols and colours were assigned according to $R_\textrm{G}$ (see text for details). Right panel: distance to the Galactic plane as a function of log\,$t$. In both panels, the small grey circles represent OCs from the DMML21 catalogue. }

\label{fig:Gal_plane_Z_versus_logt}
\end{center}
\end{figure*}

Figure~\ref{fig:Gal_plane_Z_versus_logt}, left panel, shows the distribution of the complete sample of investigated OCs (174 objects) in the Galactic plane. For better visualization, different colours and symbols have been assigned according to the following $R_\textrm{G}$ bins: $R_\textrm{G}$\,$\le\,$7\,kpc (blue circles), 7\,$<$\,$R_\textrm{G}\,$(kpc)$\,\le$\,9 (turquoise squares), 9\,$<$$\,R_\textrm{G}$\,(kpc)\,$\le$\,11 (green triangles) and $R_\textrm{G}$\,$>\,$11\,kpc (red diamonds). 

The same scheme has been employed in the right panel, which shows the vertical distance ($\vert Z\vert$) to the Galactic plane as a function of log\,$t$. As expected, there is an overall trend (with similar dispersion in comparison to the literature OCs, taken from DMML21; small grey dots) in which the oldest investigated OCs tend to be found at higher $\vert Z\vert$. In what follows, we discuss evolutionary connections between structural and time-related parameters, besides enlightening some possible relations with the Galactic tidal field.

%%%%%%%%%%%%%%%%%%%
\subsection{Core and half-light radii}
%%%%%%%%%%%%%%%%%%%
    
\begin{figure}
\begin{center}

\parbox[c]{0.48\textwidth}
  {
   \begin{center}
       \includegraphics[width=0.48\textwidth]{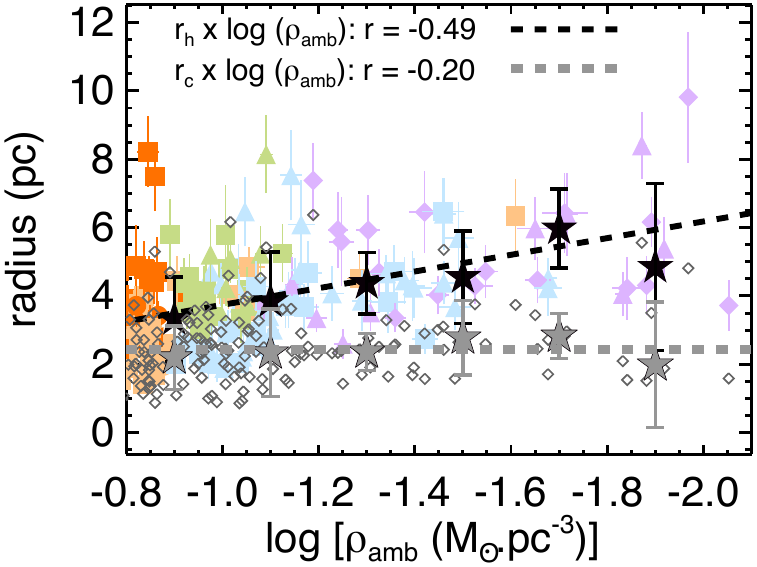}  
    \end{center}    
  }

\caption{ Half-light (coloured symbols) and core (open diamonds) radii as function of the ambient density. In both cases, the log\,$\rho_{\textrm{amb}}$ values were binned in intervals of 0.2\,dex, within which the median values of $r_h$ and $r_c$ were determined (filled stars); the thick error bars represent the median absolute deviations. The dashed black line is a linear fit to the $r_h$ versus log\,$(\rho_{\rm{amb}})$ relation (see text for details). The dashed grey line represents a constant value of $\langle r_c\rangle\sim2.5\,$pc.  }

\label{fig:rh_and_rc_versus_rhoamb}
\end{center}
\end{figure}

In Figure~\ref{fig:rh_and_rc_versus_rhoamb}, the half-light radius is plotted as function of the ambient density (the convention for the coloured symbols is outlined below). We have binned the log\,$\rho_{\textrm{amb}}$ values in intervals of 0.2\,dex and determined the median of $r_h$ within each bin, as indicated by the black filled stars (each one plotted at the centre of the respective bin). There is an overall trend (the legend indicates the Pearson correlation coefficient for the whole sample) in which the $r_h$ values tend to increase as $\rho_{\textrm{amb}}$ diminishes. This result suggests that as clusters are subject to weaker external tidal fields, they can extend their stellar content over greater distances without being tidally disrupted. The overall correlation between $r_h$ and log\,$(\rho_{\rm{amb}})$ can be verified by means of a linear fit between both quantities (black dashed line in Figure~\ref{fig:rh_and_rc_versus_rhoamb}), according to the expression

\begin{equation}
     r_h (\textrm{pc})\,=\,1.30(\pm0.37)-2.44(\pm0.33)\,\rm{log}\,(\rho_{\rm{amb}})\,.
\end{equation}

\begin{figure}
\begin{center}

\parbox[c]{0.48\textwidth}
  {
   \begin{center}
       \includegraphics[width=0.48\textwidth]{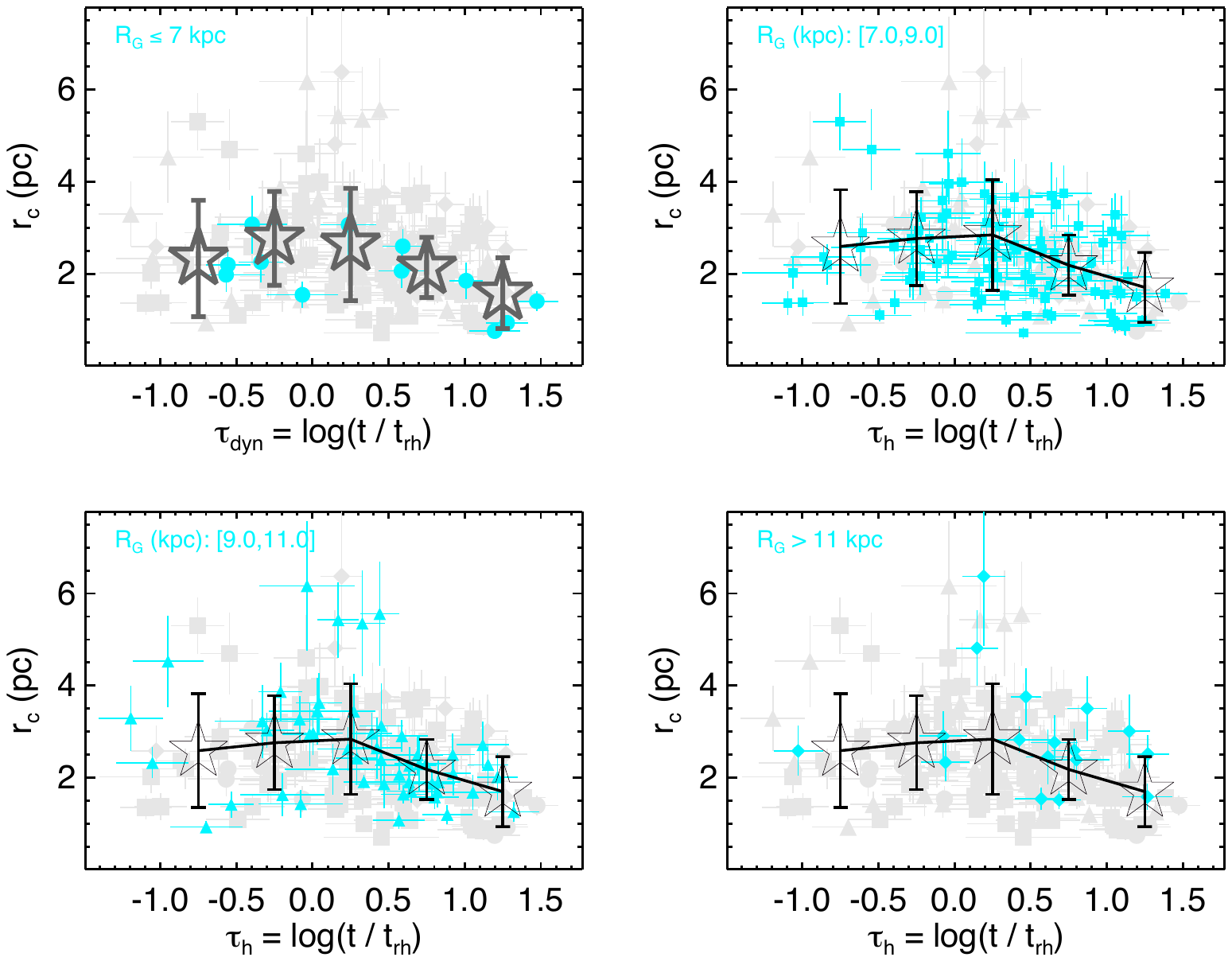}  
    \end{center}    
  }

\caption{ Core radius as function of the dynamical age, in log scale ($\tau_{\textrm{dyn}}$). The $\tau_{\textrm{dyn}}$ domain was binned in intervals of 0.5\,dex. In each case, the median $r_c$ values (open grey stars, together with the associated dispersion) were determined. The turquoise symbols identify those OCs located at $R_\textrm{G}\leq7\,$kpc. }

\label{fig:rc_versus_taudyn}
\end{center}
\end{figure}

A weaker correlation is found in the case of the core radii, which tend to fluctuate around a mean value of $\sim2.5\,$pc (grey dashed line). This result suggests that the innermost regions are less sensitive to external changes in the MW gravitational field. This is an expected result considering, for instance, the outcomes from \cite{Miholics:2014}, who performed $N$-body simulations to investigate how the clusters' size (half-mass and core radii) is modified as they are submitted to changes in the galactic potential (including accretion from a nearby dwarf galaxy). They found that the core radius is insensitive to such changes. In turn, the tidal radii derived in the present work (Table~\ref{tab:investig_sample}) show an overall dispersion with $\textrm{log}\,\rho_{\textrm{amb}}$ (moderate correlation of $r\simeq-0.56$) analogous to $r_h$ and have not been represented in Figure~\ref{fig:rh_and_rc_versus_rhoamb} for visualization purposes. This outcome means that the clusters' outer structure is affected by the external potential (e.g., \citeauthor{Nilakshi:2002}\,\,\citeyear{Nilakshi:2002}).         

Figure~\ref{fig:rc_versus_taudyn} exhibits a slight decreasing trend of the core radius with the cluster dynamical age (defined as the logarithm of the $t/t_{rh}$ ratio, that is $\tau_{\textrm{dyn}}$ $=$ log\,$t/t_{rh}$) in the $\tau_{\textrm{dyn}}\gtrsim0$ interval. This result is more evident by looking at the median $r_c$ values (open stars) obtained for the complete sample, binned in intervals of 0.5\,dex in the $\tau_{\textrm{dyn}}$ domain. The error bars represent the associated 1$\sigma$ dispersion. This behaviour of the $r_c$ values is reminiscent of results from simulations (e.g., \citeauthor{Makino:1996}\,\,\citeyear{Makino:1996}; \citeauthor{Spitzer:1969}\,\,\citeyear{Spitzer:1969}; \citeauthor{Lynden-Bell:1980}\,\,\citeyear{Lynden-Bell:1980}), which indicate that, as the internal relaxation process takes place, energy is transferred from the central parts to the outer halo, causing mass segregation and core contraction. 

For dynamically unevolved systems ($\tau_{\textrm{dyn}}\lesssim0$, that is, $t\,\lesssim\,t_{rh}$), no clear trends are evident. Moreover, Figure~\ref{fig:rc_versus_taudyn} shows that almost no clusters with extended cores ($r_c\gtrsim3\,$pc) are found in our sample at $R_\textrm{G}\lesssim7\,$kpc, regardless of the dynamical age. This way, inner orbits within the Galaxy seem to favour the presence of centrally more compact structures.

%%%%%%%%%%%%%%%%%%%%%    
\subsection{Tidal filling ratio ($r_h/R_J$)}   
%%%%%%%%%%%%%%%%%%%%%    

%%%%%%%%%%%%%%%%%%%%%%%%%%%%%    
\subsubsection*{Dependence on the external tidal field}   
%%%%%%%%%%%%%%%%%%%%%%%%%%%%%

The half-light to Jacobi radius ratio (i.e., $r_h/R_J$) is an indicator of how the cluster's main body fills the allowed tidal volume (e.g., \citeauthor{Alexander:2014}\,\,\citeyear{Alexander:2014}), therefore being a useful parameter to evaluate if a stellar system is more or less susceptible to tidal effects (e.g., \citeauthor{Ernst:2013}\,\,\citeyear{Ernst:2013}; \citeauthor{Ernst:2015}\,\,\citeyear{Ernst:2015}). The tidal volume filling ratio is related to the fraction ($\xi$) of evaporated stars at each $t_{\textrm{rh}}$ according to ln\,$\xi\propto r_h/R_J$, as found by \cite{Lee:2002} and GB08 from estimates of the fraction of stars above the system escape velocity for tidally limited clusters with different $r_h/R_J$ ratios. This scaling is valid for $r_h/R_J$\,$\gtrsim$\,0.05 (the ``tidal regime"), which is the range encompassed by all OCs in our sample.

\begin{table}
 \small
% \hskip-2.0cm
\begin{minipage}{85mm}
  \caption{ Symbol convention and colours used in Section~\ref{sec:discussion}. For $R_\textrm{G}\leq 8\,$kpc: the \textit{compact} (\textit{loose}) group refers to OCs below (above) the division line in Figure~\ref{fig:rh_Rj_and_rho_hm_versus_Rgal}, panel (a). }
  \label{tab:symbols_convention}
 \begin{tabular}{ccccc}

  \hline
  \multicolumn{5}{c}{$R_{\textrm{G}}$ intervals (in kpc)}                                                                  \\ 
  \hline
  \hline	
  6.0$-$7.0         & 7.0$-$9.0               &  9.0$-$11.0               & 11.0$-$12.5              &                     \\ 
  \hline
  \Huge{$\bullet$}  & \Large{$\blacksquare$}  & \Large{$\blacktriangle$}  & \Large{$\blacklozenge$}  &                     \\                                                                                                                                                
  \hline
  \hline
  \multicolumn{5}{c}{ Colours (see Figure~\ref{fig:rh_Rj_and_rho_hm_versus_Rgal}, panel \textit{a}) }                      \\ 
  \hline
  \textcolor{orange}{orange (group 1)}   &  \multicolumn{4}{c}{\textcolor{orange}{$R_{\textrm{G}}\leq8.0\,$kpc (compact)}} \\  
                                         &  \multicolumn{4}{c}{   }                                                        \\

  \textcolor{red}{red (group 2)}         &  \multicolumn{4}{c}{\textcolor{red}{$R_{\textrm{G}}\leq8.0\,$kpc (loose)}}      \\  
                                         &  \multicolumn{4}{c}{   }                                                        \\

	\textcolor{blue}{blue (group 3)}       &  \multicolumn{4}{c}{\textcolor{blue}{$8<R_{\textrm{G}}\,$(kpc)$\leq10$}}        \\ 
                                         &  \multicolumn{4}{c}{\textcolor{blue}{$r_h/R_J\leq0.40$}}                        \\ 
                                         &  \multicolumn{4}{c}{   }                                                        \\

	\textcolor{green}{green (group 4)}     &  \multicolumn{4}{c}{\textcolor{green}{$8<R_{\textrm{G}}\,$(kpc)$\leq10$}}       \\  
                                         &  \multicolumn{4}{c}{\textcolor{green}{$r_h/R_J>0.40$}}                          \\
                                         &  \multicolumn{4}{c}{   }                                                        \\

	\textcolor{amethyst}{purple (group 5)}   &  \multicolumn{4}{c}{\textcolor{amethyst}{$R_{\textrm{G}}>10\,$kpc}}           \\

\hline

\end{tabular}

\end{minipage}
\end{table}

\begin{figure*}
\begin{center}

\parbox[c]{1.00\textwidth}
  {
   \begin{center}
    \includegraphics[width=0.49\textwidth]{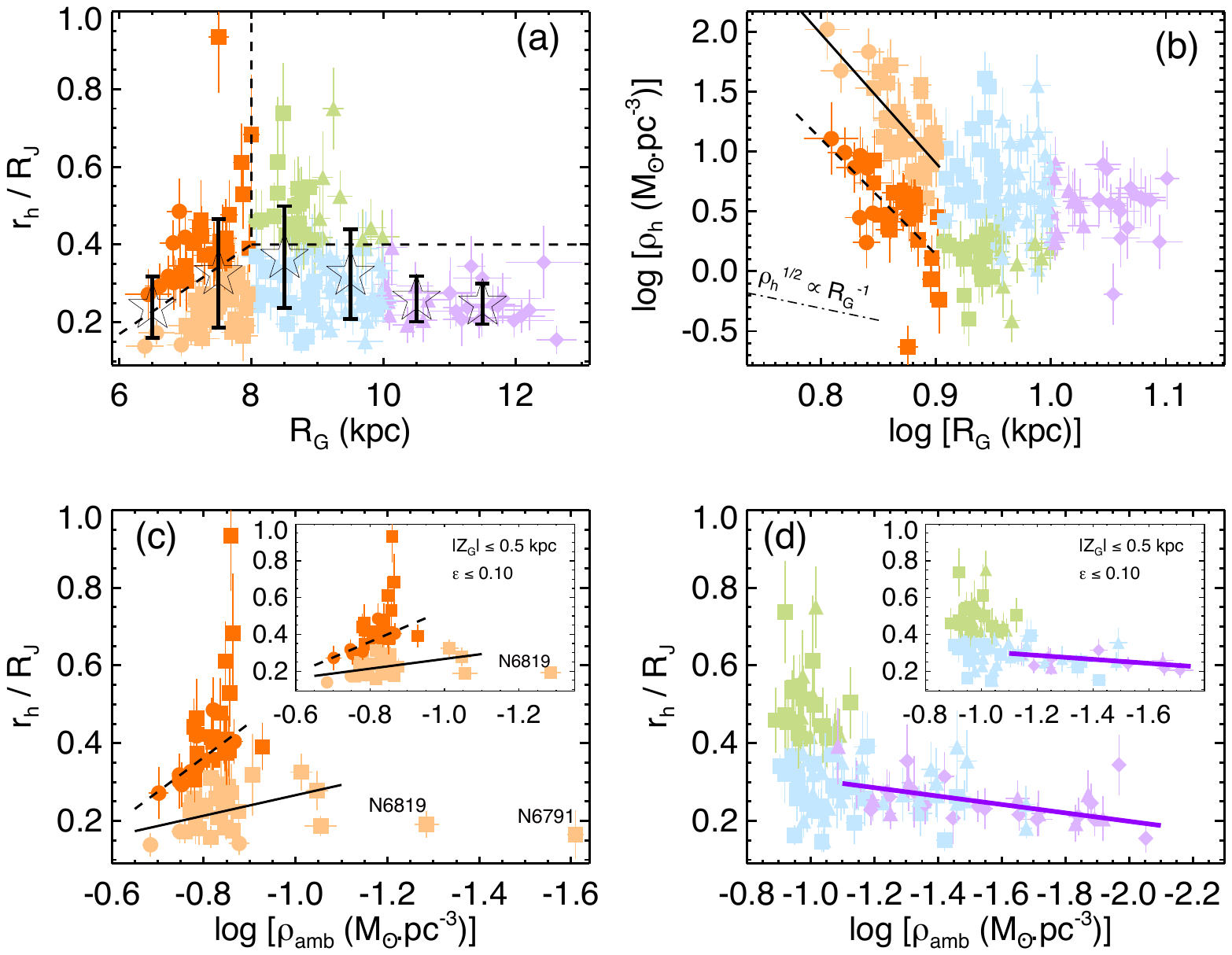} 
    \includegraphics[width=0.49\textwidth]{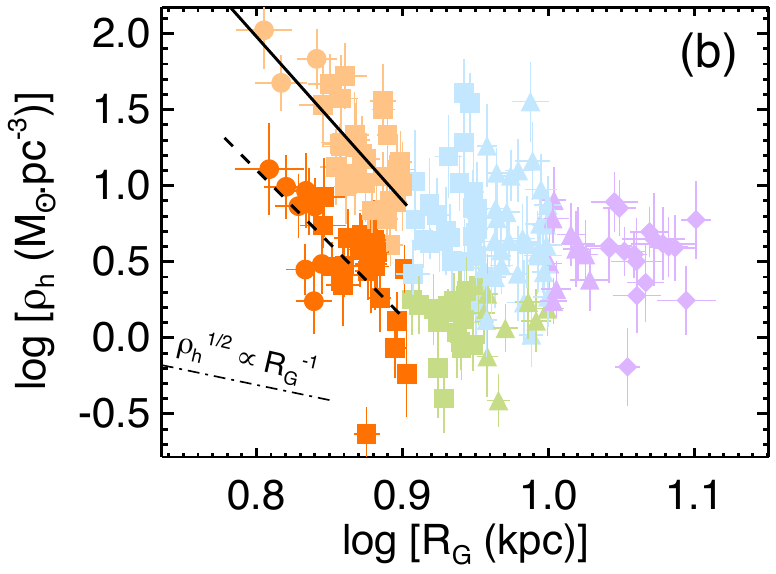}   
    \includegraphics[width=0.505\textwidth]{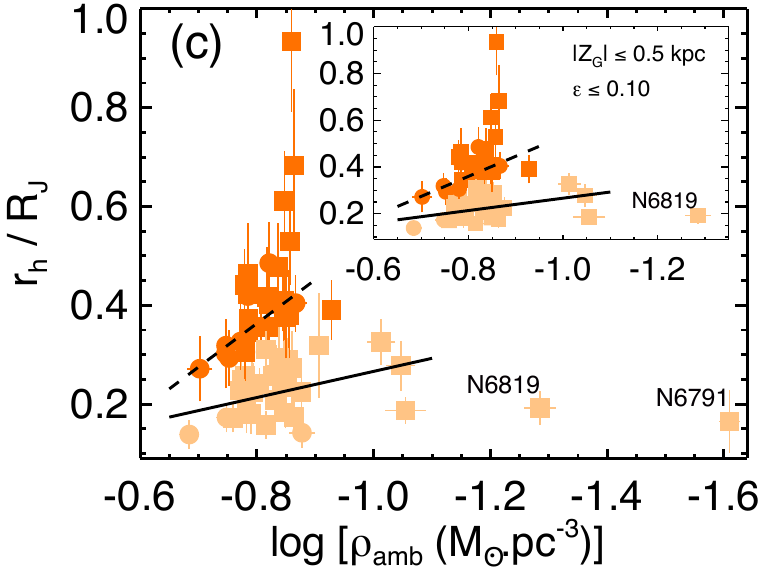}   
    \includegraphics[width=0.490\textwidth]{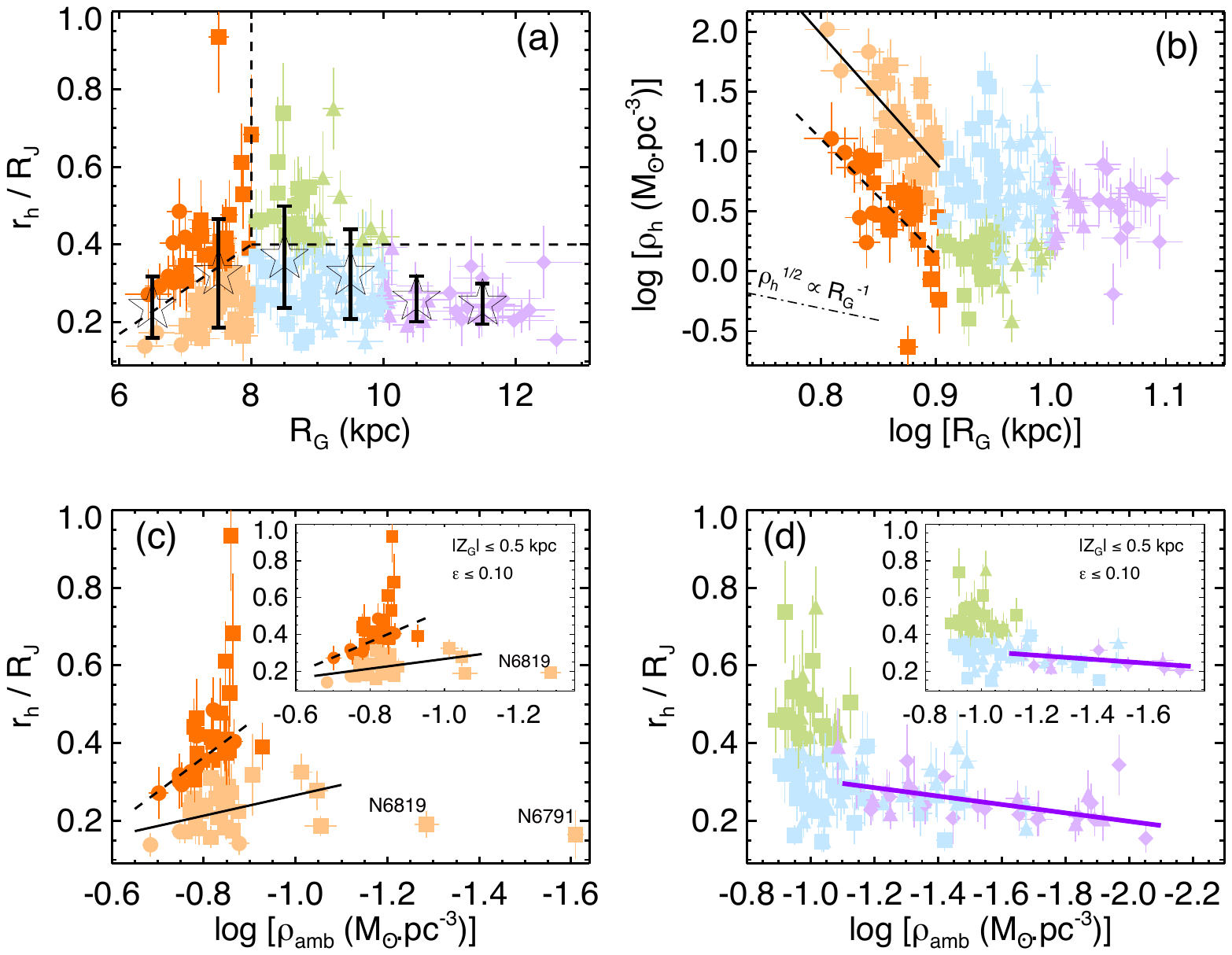}   
    
    \end{center}    
  }
\caption{   Panel (a): Tidal filling ratio as function of the Galactocentric distance ($R_\textrm{G}$). Symbols and colours were assigned according to the different $r_h/R_J$ ratios and $R_{\textrm{G}}$ values, following the convention of Table~\ref{tab:symbols_convention}. The dashed lines have been plotted to clarify the sample separation. The open stars represent the mean values (and associated dispersion, as indicated by the error bars) of $r_h/R_J$ obtained within six $R_{\textrm{G}}$ bins: $R_{\textrm{G}}\leq7\,$kpc, $7 < R_{\textrm{G}} \leq 8\,$kpc, $8 < R_{\textrm{G}} \leq 9\,$kpc, $9 < R_{\textrm{G}} \leq 10\,$kpc, $10 < R_{\textrm{G}} \leq 11\,$kpc and $R_{\textrm{G}} > 11\,$kpc. Panel (b): half-light density versus $R_{\textrm{G}}$. The trends followed by groups 1 and 2 are overplotted (respectively, continuous and dashed lines) to guide the eye. The dot-dashed line shows the trend expected for clusters in homologous evolution (see text for details). Panels (c) and (d): respectively, $r_h/R_J$ versus log\,$\rho_{\textrm{amb}}$ for clusters located at $R_{\textrm{G}}\leq8\,$kpc (groups 1 and 2) and $R_{\textrm{G}} > 8\,$kpc (groups 3, 4 and 5). The lines indicate general trends followed by part of our sample. The insets highlight clusters following almost circular orbits in the Galactic plane.   }

\label{fig:rh_Rj_and_rho_hm_versus_Rgal}
\end{center}
\end{figure*}    

%Panel (d): The trend line is a linear fit to all objects of groups 3 and 5 for which log\,$\rho_{\textrm{amb}}\lesssim-1.1$.    

%Submitted manuscript
%In panel (a) of Figure~\ref{fig:rh_Rj_and_rho_hm_versus_Rgal}, the tidal volume filling ratio is plotted as function of $R_\textrm{G}$. The complete sample has been split into five groups (see the colours and symbol scheme in Table~\ref{tab:symbols_convention}), which will be employed in the subsequent analyses. For $R_\textrm{G}$\,$\leq$\,8\,kpc, the orange (more compact) and red (less compact) groups are statistically distinct from each other in terms of the $r_h/R_J$ distribution, presenting resemblance probability smaller than $\sim$0.1\% according to the Kolmogorov-Smirnov (K-S) two-sample test. This result points to two subsamples of OCs submitted to intrinsically distinct evaporation regimes. Both groups present an overall positive correlation with $R_\textrm{G}$, as indicated by the black open stars (representing the average of $r_h/R_J$ for clusters within two intervals: $R_\textrm{G}$\,$\leq\,$7\,kpc and 7\,$<$\,$R_\textrm{G}\,$(kpc)$\le$\,8; the dashed line connecting the mean values separates groups 1 and 2). 

In panel (a) of Figure~\ref{fig:rh_Rj_and_rho_hm_versus_Rgal}, the tidal volume filling ratio is plotted as function of $R_\textrm{G}$. The complete sample has been split into five groups (see the colours and symbol scheme in Table~\ref{tab:symbols_convention}), which will be employed in the subsequent analyses. The red and orange symbols identify, respectively, more and less tidally influenced groups (GB08) in the range $R_\textrm{G}$\,$\leq$\,8\,kpc. Both groups present an overall positive correlation with $R_\textrm{G}$, as indicated by the black open stars (representing the average of $r_h/R_J$ for clusters within two intervals: $R_\textrm{G}$\,$\leq\,$7\,kpc and 7\,$<$\,$R_\textrm{G}\,$(kpc)$\le$\,8; the dashed line connecting the mean values separates groups 1 and 2). 

Groups 3 (blue; more compact) and 4 (green; less compact) are located in the range 8\,$<$\,$R_\textrm{G}\,$(kpc)$\le$\,10 and the division line between them ($r_h/R_J\simeq0.4$) corresponds to nearly the maximum value of $r_h/R_J$ reached by the more external clusters (purple symbols, group 5; $R_\textrm{G}$\,$>$\,10\,kpc). The range of $R_\textrm{G}$ between $\sim8-9\,$kpc corresponds nearly to the location of the MW corotation radius\footnote[10]{Galactocentric distance at which the rotational speed of the stars matches the rotational speed of the spiral arms. At this distance, the arms change from leading to trailing, since the interstellar matter penetrates the arms in opposite directions \citep{Dias:2019}.} ($R_\textrm{C}=8.51\pm0.64\,$kpc; \citeauthor{Dias:2019}\,\,\citeyear{Dias:2019}), considering its uncertainty. This $R_\textrm{G}$ interval seems a transient region, where it is noticeable a large dispersion of the $r_h/R_J$ ratios; beyond $R_\textrm{G}\sim9\,$kpc, the ensemble of values tend to be less scattered and with no significant evidence of correlation between the plotted quantities.      
%Corotation is about midway between the inner and outer Lindblad resonances \citep{Canzian:1998}.

%Mass loss by spiral arm shocks will occur just at the moment the cluster crosses the spiral arm. Assuming that spiral arms move with a constant pattern speed (Ωp) and that the matter in the disk has a constant circular velocity (Vdisk), the relative velocity between the two (Vdrift) depends on the location in the galaxy (R). Density waves that pass with a low velocity have a large effect on the star clusters (e.g. Ostriker et al. 1972). Therefore, the disruptive effect of spiral arm shocks is most im- portant close to the corotation radius (RCR), i.e. the point where the disk and the spiral arms have the same rotational velocity.  Lamers & Gieles sobre disruption na vizinhança solar.    

%\begin{figure*}
%\begin{center}

%\parbox[c]{1.00\textwidth}
%  {
%   \begin{center}
%      \includegraphics[width=0.49\textwidth]{histogram_masses_part1.pdf} 
%      \includegraphics[width=0.49\textwidth]{histogram_masses_part2.pdf}        
%    \end{center}    
%  }
%\caption{ \textbf{Left panel}: afsdfaf \textbf{Right panel}: afsfd }

%\label{fig:histogram_masses}
%\end{center}
%\end{figure*}

Groups 1 and 2 exhibit an overall anticorrelation between the half-light density, defined as $\rho_{h}$\,=\,$3\,M_{\textrm{clu}}/(8\pi\,r_{h}^{3}$), and $R_\textrm{G}$ (absolute value of the Pearson correlation coefficient is greater than 0.65 in both cases), as indicated, respectively, by the continuous and dashed lines in panel (b) of Figure~\ref{fig:rh_Rj_and_rho_hm_versus_Rgal}, plotted in log-scale. Both lines have been superimposed on the data to guide the eye. Based on these trends, the stronger external tidal forces at smaller $R_\textrm{G}$ (Figure~\ref{fig:rho_amb_versus_Rgal}) seem to be more effective in constraining the OCs mass distribution within the allowed tidal volume, in contrast with groups 3, 4 and 5, for which no significant trends between $\rho_h$ and $R_\textrm{G}$ are verified. It is noticeable the absence of objects with $\rho_h$ smaller than $\sim1\,M_{\odot}/$pc$^3$ at $R_\textrm{G}\lesssim7\,$kpc. At this same $R_\textrm{G}$ range, clusters with $r_h/R_J\gtrsim0.5$ are absent. Such clusters would be subject to strong tidal stresses and have consequently small dissolution times \citep{Baumgardt:2010}. 

For comparison purposes, the dot-dashed line at the bottom of panel (b) represents the $\rho_h$\,$\propto\,$$R_\textrm{G}^{-2}$ relation. This scaling (figure 6 of \citeauthor{Gieles:2011a}\,\,\citeyear{Gieles:2011a}) corresponds to GCs in homologous evolution (where $r_h\propto R_J$), subject to an external potential modeled by an isothermal sphere (their appendix B), being in an evaporation-dominated phase. In comparison to these GCs, the investigated OCs located at $R_\textrm{G}\leq8\,$kpc present much steeper variations of $\rho_h$ with $R_\textrm{G}$ (and no evident trends are noted for larger $R_\textrm{G}$), thus indicating that the evaporation + internal relaxation processes do not result in homologous evolution of their structure. Moreover, according to \citeauthor{Henon:1961}'s\,\,(\citeyear{Henon:1961}) models, clusters evolving homologously present $r_h/R_J\sim0.15$; as shown in Figure~\ref{fig:rh_Rj_and_rho_hm_versus_Rgal}, almost all of our OCs present tidal volume filling ratios higher than this limit, therefore being less compact and more tidally influenced. This is not an unexpected result, since the $r_h/R_J$ ratio for OCs is $\sim3-5$ times larger than the typical values for GCs \citep{Ernst:2013}. 

%Moreover, according to \citeauthor{Henon:1961}'s\,\,(\citeyear{Henon:1961}) models, clusters with small mass spectrum (representative of GCs) evolving homologously present $r_h/R_J\sim0.15$; as shown in Figure~\ref{fig:rh_Rj_and_rho_hm_versus_Rgal}, almost all of our OCs present tidal volume filling ratios higher than this limit, therefore being less compact and more tidally influenced. This is not an unexpected result, since the $r_h/R_J$ ratio for OCs is $\sim3-5$ times larger than the typical values for globulars \citep{Ernst:2013}\textcolor{red}{...at\'e aqui]}. 

In panels (c) and (d) of Figure~\ref{fig:rh_Rj_and_rho_hm_versus_Rgal}, the tidal filling ratio is plotted as function of the ambient density ($\rho_{\textrm{amb}}$). In panel (c), tentative linear fits (continuous and dashed lines superimposed on the data) have been performed in the case of groups 1 and 2 for clusters presenting $r_h/R_J\lesssim0.6$ and log\,$\rho_{\textrm{amb}}$\,$\gtrsim$\,$-1.1$. The steeper variation of $r_h/R_J$ in the case of group 2 (less compact) suggests different (and more intense) evaporation regimes in comparison to their more compact counterparts of group 1. Following the trends evidenced by both groups, larger fractions of the tidal volume can be progressively reached as the OCs are exposed to weaker tidal forces.

Interestingly, almost all clusters for which log\,$\rho_{\textrm{amb}}$\,$\lesssim$\,$-1.1$ present ($r_h/R_J)_{\textrm{max}}$\,$\sim$\,0.4 (panels $c$ and $d$). Particularly, all clusters of group 5 (purple symbols) are located within this $\rho_{\textrm{amb}}$ domain (according to Figure~\ref{fig:rho_amb_versus_Rgal}, log\,$\rho_{\textrm{amb}}$\,$\lesssim$\,$-1.1$ for all clusters with $R_\textrm{G}\gtrsim\,$10\,kpc). In fact, clusters with a given mass subject to less intense external tidal fields will present more extended Roche lobes and, consequently, tend to occupy smaller fractions of the allowed tidal volume. In comparison to $r_h$, at this smaller ambient density domain, the Jacobi radius seems to have a larger increment as $\rho_{\textrm{amb}}$ diminishes, causing the $r_h/R_J$ ratio to decrease (trend line in panel $d$ of Figure~\ref{fig:rh_Rj_and_rho_hm_versus_Rgal}). This change in the behaviour of the $r_h/R_J$ ratios at log\,$\rho_{\textrm{amb}}$\,$\sim$\,$-1.1$ is consistent with the outcomes of Figure~\ref{fig:mbreak_versus_rhoamb} (vertical line in this figure's inset). Based on the results of GB08, those OCs in the log\,$\rho_{\textrm{amb}}$\,$\lesssim$\,$-1.1$ interval tend to present smaller evaporation rates and their evolution may be more importantly ruled by the internal relaxation process (\citeauthor{Madrid:2012}\,\,\citeyear{Madrid:2012}; \citeauthor{Piatti:2019a}\,\,\citeyear{Piatti:2019a}) compared to their counterparts subject to higher $\rho_{\textrm{amb}}$ (see also Figure~\ref{fig:rc_rh_versus_taudyn} and the discussion following it).

Comments analogous to those of the previous paragraph can be stated for NGC\,6819 and NGC\,6791 (panel $c$ of Figure~\ref{fig:rh_Rj_and_rho_hm_versus_Rgal}): they are the only OCs in our sample located at $R_\textrm{G}\leq8\,$kpc for which log\,$\rho_{\textrm{amb}}$\,$\lesssim$\,$-1.1$ (see Figure~\ref{fig:rho_amb_versus_Rgal}) and they present small $r_h/R_J$ ratios ($<$\,0.2), thus the fractional loss of their stellar content at each $t_{rh}$ is expected to be smaller than their counterparts at compatible $R_\textrm{G}$. Besides, both are among the most massive objects in our sample (log\,$M_{\textrm{N6819}}\simeq4.0$ and log\,$M_{\textrm{N6791}}\simeq4.4$). The insets in panels (c) and (d) of Figure~\ref{fig:rh_Rj_and_rho_hm_versus_Rgal} highlight those clusters located close to the Galactic plane ($\vert Z_\textrm{G}\vert\lesssim0.5\,$kpc) and describing nearly circular orbits (eccentricity $\epsilon\lesssim0.1$). The previous statements do not change if we take this subsample of clusters subject to almost static external potentials. 

\subsubsection*{Dependence on the cluster mass}
%%%%%%%%%%%%%%%%%%%%%%%%%%

\begin{figure*}
\begin{center}

\parbox[c]{1.00\textwidth}
  {
   \begin{center}
      \includegraphics[width=0.49\textwidth]{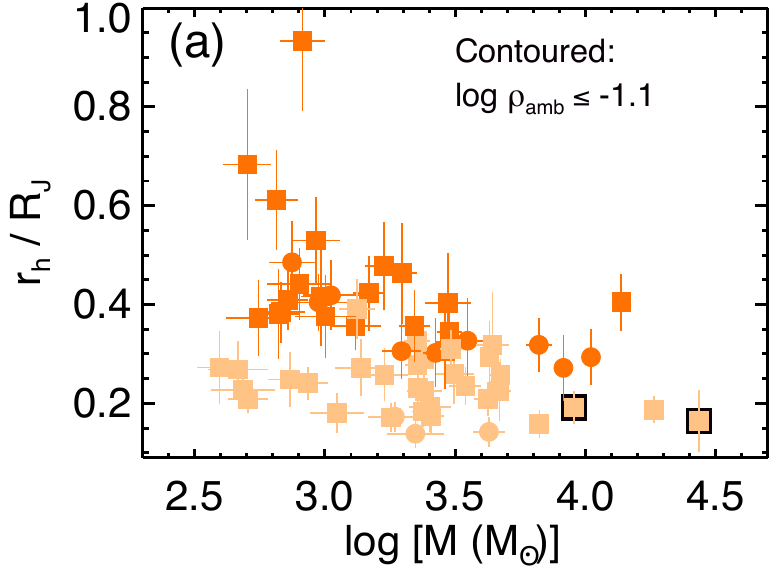} 
      \includegraphics[width=0.49\textwidth]{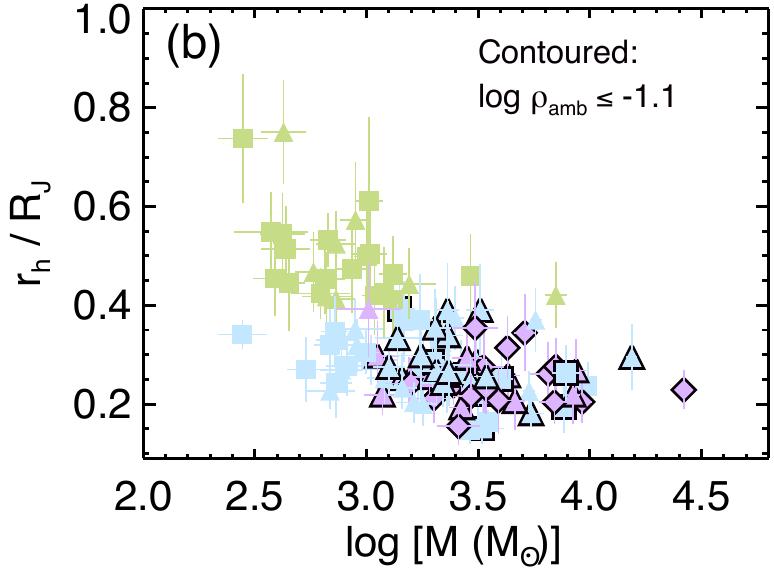}        
    \end{center}    
  }
\caption{ $r_h/R_J$ ratio as function of mass for clusters located at $R_{\textrm{G}}\leq8\,$kpc (panel a) and $R_{\textrm{G}} > 8\,$kpc (panel b). The contoured symbols represent OCs for which log\,$\rho_{\textrm{amb}} \lesssim -1.1$. }

\label{fig:rhm_rJ_logmass}
\end{center}
\end{figure*}

Figure~\ref{fig:rhm_rJ_logmass} allows to verify possible connections between the $r_h/R_J$ ratio and the cluster mass. For clarity, clusters of groups 1 and 2 and those of groups 3, 4 and 5 (Table~\ref{tab:symbols_convention}) were plotted in panels $a$ and $b$, respectively. A qualitatively similar data dispersion can be seen in both cases: it is noticeable an overall anticorrelation between the plotted quantities for each group.

In the left panel, the decreasing trend is more evident in the case of the less compact group 2, due to the larger range encompassed by the $r_h/R_J$ ratios. The contrast with group 1 (more compact) is larger in the domain of smaller masses, where more tidally influenced clusters are found. In the right panel, the ensemble of $r_h/R_J$ ratios presents the largest anticorrelation with mass in the domain log\,$M$\,$\lesssim3.5$. Beyond this limit there is an apparent plateau, where the $r_h/R_J$ values tend to fluctuate around a nearly constant value ($\simeq0.25$). 

We have highlighted (contoured symbols) those clusters for which log\,$\rho_{\textrm{amb}}\lesssim-1.1$; these less tidally affected clusters are moderately massive ($M\gtrsim10^3\,M_{\odot}$) and most of them are dynamically evolved ($t/t_{rh}\gtrsim1$; see also Figure~\ref{fig:rc_rh_versus_taudyn}). In both panels, it is evident the absence of clusters in the upper right region of the plots, that is, in the higher mass and higher tidal filling ratio intervals.

A possible interpretation for the outcomes of Figure~\ref{fig:rhm_rJ_logmass} is that, for higher masses, the cluster potential well becomes progressively deeper and prevents its mass content from extending out to considerably large fractions of the allowed tidal volume, therefore making its overall mass distribution to become denser (\citeauthor{Fukushige:2000}\,\,\citeyear{Fukushige:2000}; \citeauthor{Tarricq:2022}\,\,\citeyear{Tarricq:2022}).

\subsubsection*{Dependence with the evolutionary stage}
%%%%%%%%%%%%%%%%%%%%%%%%%%%%%

\begin{figure*}
\begin{center}

\parbox[c]{1.00\textwidth}
  {
   \begin{center}
      \includegraphics[width=0.49\textwidth]{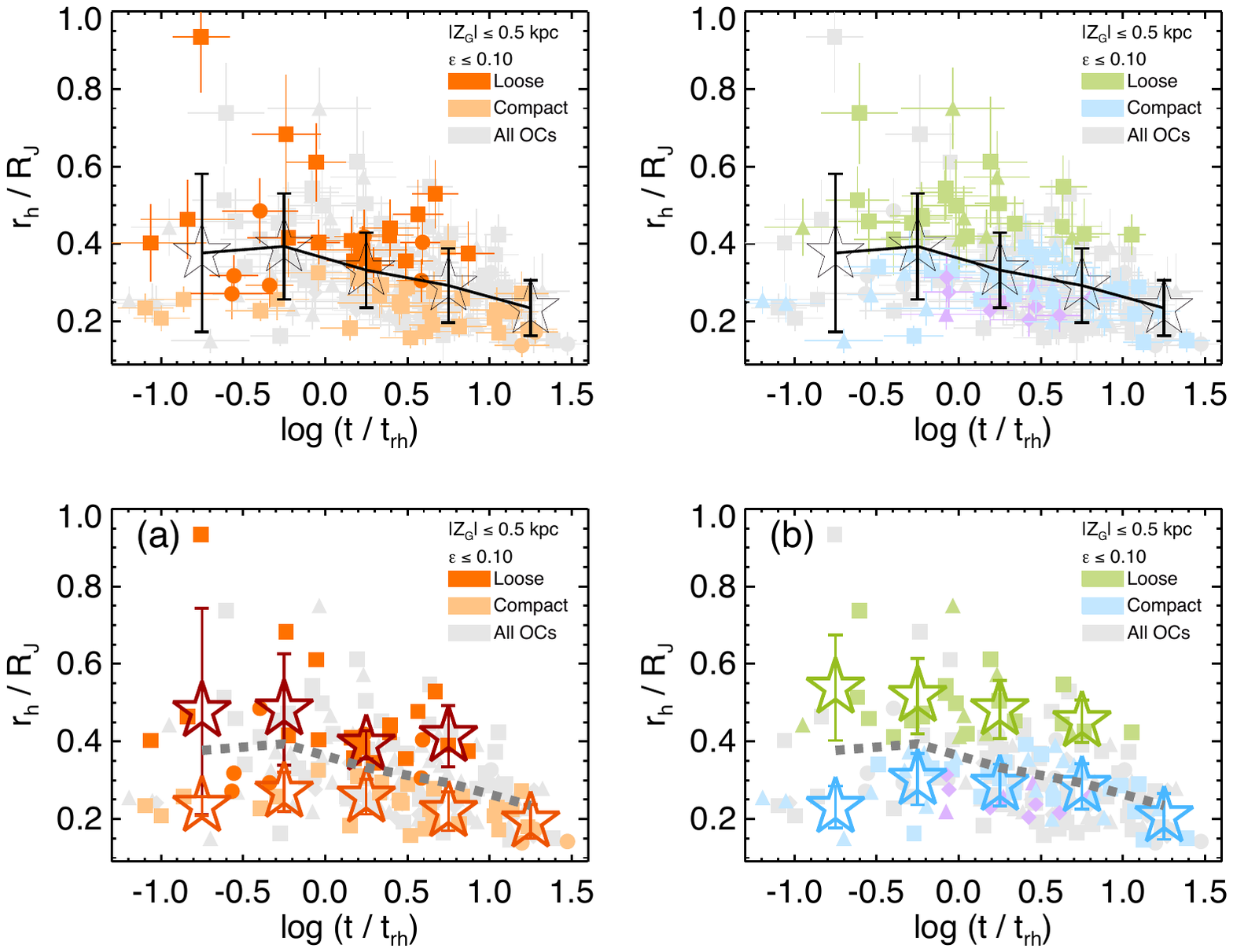} 
      \includegraphics[width=0.49\textwidth]{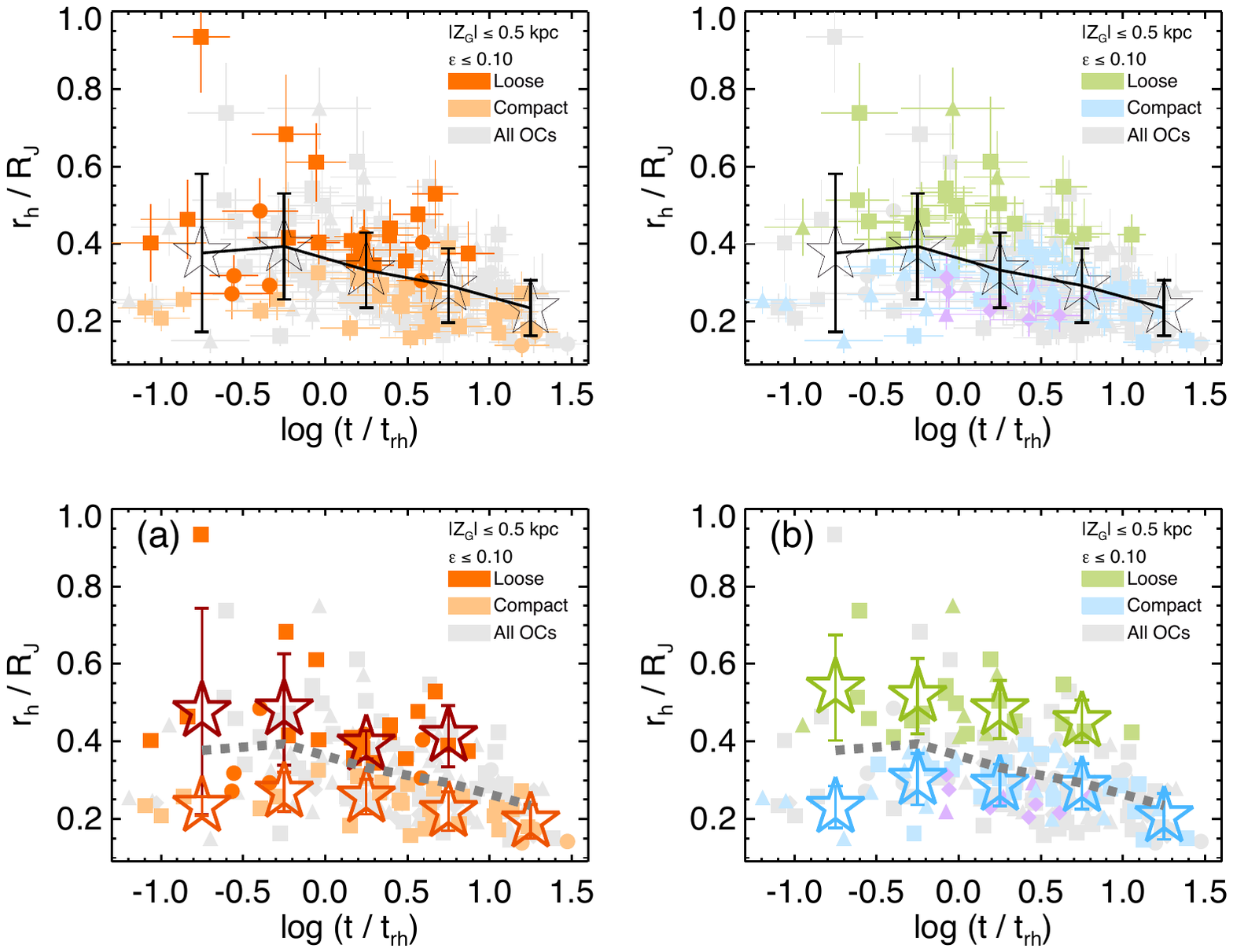}        
    \end{center}    
  }
\caption{  $r_h/R_J$ ratio as function of the dynamical age ($\tau_{\textrm{dyn}}$ = log\,$(t/t_{rh})$) for clusters in nearly circular orbits (as indicated in the legend) located at $R_{\textrm{G}}\,\leq\,8\,$kpc (left panel) and $R_{\textrm{G}}\,>\,8\,$kpc (right panel). In both panels, the $\tau_{\textrm{dyn}}$ axis was split into 5 bins of 0.5\,dex each ($\tau_{\textrm{dyn}}$\,$<$\,-0.5, -0.5\,$\leq$\,$\tau_{\textrm{dyn}}$\,$<$\,0.0, 0.0\,$\leq$\,$\tau_{\textrm{dyn}}$\,$<$\,0.5, 0.5\,$\leq$\,$\tau_{\textrm{dyn}}$\,$<$\,1.0, \,$\tau_{\textrm{dyn}}$\,$\ge\,$1.0) and the $\langle r_h/r_J\rangle$ value (together with the associated dispersion) was determined for each group of OCs (as indicated by the open stars and associated error bars). The grey dashed line connects the $\langle r_h/r_J\rangle$ values determined within the same $\tau_{\textrm{dyn}}$ bins, but considering the whole OCs sample (grey filled symbols). }

\label{fig:rh_Rj_versus_taudyn}
\end{center}
\end{figure*}

Figure~\ref{fig:rh_Rj_versus_taudyn} allows to evaluate how the tidal filling ratio is related to the clusters' internal dynamical evolutionary stage. In both panels, the grey symbols represent the complete OCs sample and the thick dashed lines represent the mean values of $r_h/R_J$ (i.e., $\langle r_h/R_J\rangle$) determined within five $\tau_{\textrm{dyn}}=$log\,$(t/t_{rh})$ bins (indicated in the figure caption). The coloured symbols identify OCs in groups 1 to 5 describing nearly circular orbits along the Galactic disc, with the $\langle r_h/R_J\rangle$ values (together with the associated dispersion) indicated by the open stars (and error bars). In general, for $\tau_{\textrm{dyn}}\gtrsim-0.5$ we can verify a slight anticorrelation between the plotted quantities (less noticeable in the case of group 2, due to larger scatter in $r_h/R_J$); since these systems are not subject to strong variations in the external tidal conditions, the trends indicate that internal interactions tend to shrink the OCs main bodies as they evolve dynamically. 

The slight decrease of $r_h/R_J$ with $\tau_{\textrm{dyn}}$ is reminiscent of the results from the simulations of \citeauthor{Heggie:2003}\,\,(\citeyear{Heggie:2003}; their figure 33.2). They found that, in the case of initially tidally filled clusters, the $r_h/r_J$ ratio\footnote[11]{Actually, in \cite{Heggie:2003} the term \textit{tidal radius} defines the limit of the last closed equipotential surface, i.e., the distance from the cluster centre, subject to a galactic potential, to the first Lagrangian point. In the present paper, this definition is equivalent to the \textit{Jacobi radius} ($R_J$; Section~\ref{sec:analysis}). In our case, $r_t$ is the empirical tidal radius  (Section~\ref{sec:struct_params}).} is typically larger than $\sim$0.20 and this ratio does not vary significantly during their evolution (see also section 5.1 of \citeauthor{Glatt:2011}\,\,\citeyear{Glatt:2011}). Nearly $\sim90\%$ of our OCs present $r_h/R_J\gtrsim0.20$, so they are consistent with this scenario. In contrast, during the initial evolutionary stages, the half-mass radius (or equivalently the half-light radius, as employed here under the assumption that light traces mass) undergoes an expansion triggered by stellar evolution and feedback mechanisms (\citeauthor{Madrid:2012}\,\,\citeyear{Madrid:2012}; \citeauthor{Miholics:2016}\,\,\citeyear{Miholics:2016}). This could explain, at least partially, the much larger scatter of the $r_h/R_J$ ratios for the dynamically young clusters in the range $\tau_{\textrm{dyn}}\lesssim-0.5$ (that is, $t\,\lesssim\,0.3\,t_{rh}$), where 17 clusters ($\sim10\%$ of the complete sample) are found.

%Heggie & Hut (definição de rt): This is analogous to the existence of an escape energy for stars in an isolated cluster. The lowest value of C at which the surface opens out to infinity is C = CL, and so it is usual to adopt the “radius” of this last closed surface, i.e. rL, as the limiting radius of a cluster in a galactic potential. It is usually called the tidal radius, denoted by rt.

The different groups of OCs showed in each panel of Figure~\ref{fig:rh_Rj_versus_taudyn} encompass similar dynamical ages (except for the last bins, i.e., $\tau_{\textrm{dyn}}\gtrsim1.0$). For a given $\tau_{\textrm{dyn}}$, we can find OCs located at similar positions within the Galaxy and presenting significantly different tidal filling ratios, therefore being more or less tidally influenced. We speculate that their intrinsically different tidal evaporation regimes may be due to differing initial formation conditions. In both panels, OCs of groups 1 and 3 within the more dynamically evolved bins ($\tau_{\textrm{dyn}}\gtrsim1.0$) present relatively compact main bodies ($r_h/R_J\lesssim0.25$). This can be interpreted as a consequence of their denser structures resulting in shorter internal timescales, which accelerates their dynamical evolution (e.g., \citeauthor{Spitzer:1971}\,\,\citeyear{Spitzer:1971}; \citeauthor{Portegies-Zwart:2010}\,\,\citeyear{Portegies-Zwart:2010}). Considering the group of compact OCs (group 1) shown in the left panel of Figure~\ref{fig:rh_Rj_versus_taudyn}, nine of them present $\tau_{\textrm{dyn}}\gtrsim1.0$, of which 8 (namely, Pismis\,18, NGC\,5715, NGC\,6134, NGC\,6208, NGC\,6253, IC\,4651, Dias\,6 and Berkeley\,81) describe inner orbits ($R_\textrm{G}\lesssim7.4\,$kpc) within the Galaxy. Interestingly, the most dynamically evolved OC (NGC\,6253; log\,($t/t_{rh}$)$\,\cong$\,1.5) in our sample is also located at $R_\textrm{G}\lesssim7\,$kpc. These results may imply that the OCs dynamical evolution is differentially affected by the MW tidal field, which can be understood from the outcomes of Figure~\ref{fig:rh_and_rc_versus_rhoamb} (i.e., shrinkage of $r_h$ for larger $\rho_{\textrm{amb}}$) combined with the dependence of $t_{rh}$ with $r_h$ ($t_{rh}\propto r_{h}^{3/2}$, for a given cluster mass and number of stars; Equation~\ref{eq:trh}).

\begin{figure*}
\begin{center}

\parbox[c]{1.00\textwidth}
  {
   \begin{center}
      \includegraphics[width=0.49\textwidth]{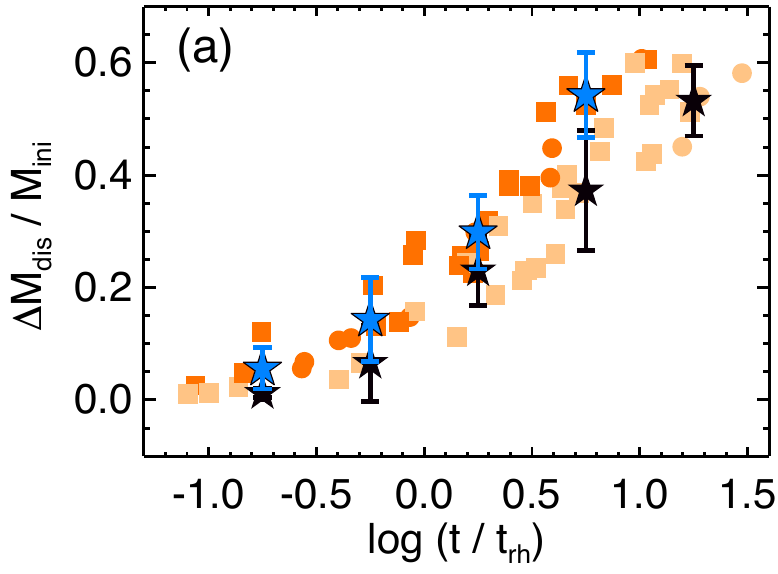} 
      \includegraphics[width=0.49\textwidth]{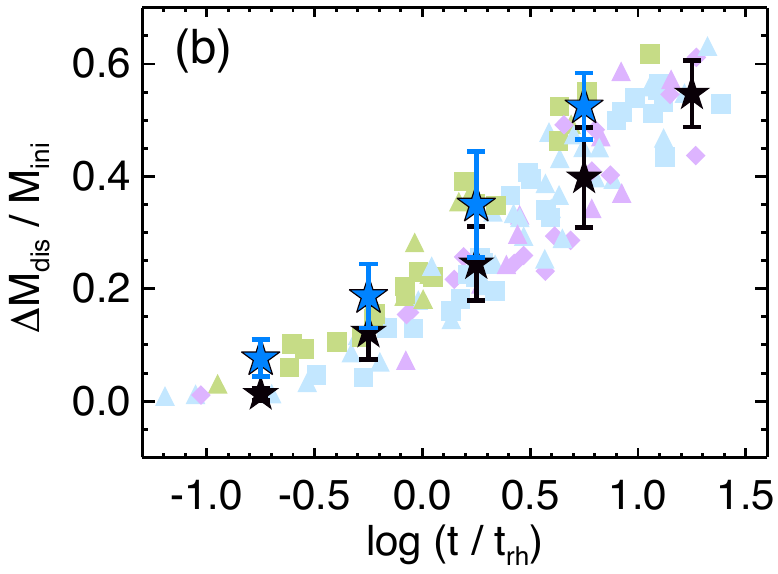}        
    \end{center}    
  }
\caption{ Panel (a): Fraction of mass loss by disruption as function of the dynamical age for OCs located at $R_\textrm{G}\,\leq\,8\,$kpc. The blue (\textit{black}) stars represent the median of the $\Delta\,M_{\textrm{din}}/M_{\textrm{ini}}$ values within the same $\tau_{\textrm{dyn}}$ bins of Figure~\ref{fig:rh_Rj_versus_taudyn} for group 2 (\textit{group 1}). Panel (b): Same as for panel (a), but representing groups 3 and 5 (black stars) and group 4 (dark blue stars). Error bars represent the median absolute deviation within each $\tau_{\textrm{dyn}}$ bin. }

\label{fig:Mdis_Mini_taudyn}
\end{center}
\end{figure*}

%%%%%%%%%%%%%%%%%%%%%
\subsubsection*{Mass-loss by disruption}
%%%%%%%%%%%%%%%%%%%%%

Since dynamical evolution implies a given fraction of stars evaporated at each $t_{rh}$, we expect some trend between  mass lost due to relaxation and tidal heating (that is, due to disruptive effects; $\Delta M_{\textrm{dis}}$), relative to the cluster initial mass ($M_{\textrm{ini}}$; Section~\ref{sec:analysis}), and the $t/t_{rh}$ ratio. This is confirmed in Figure~\ref{fig:Mdis_Mini_taudyn}, which shows a clear positive correlation between $\Delta M_{\textrm{dis}}/M_{\textrm{ini}}$ and the dynamical age. 

The values of $\Delta M_{\textrm{dis}}$ were estimated from the expression:

\begin{equation}
   M_{\textrm{ini}} = M(t) + (\Delta M)_{\textrm{dis}} + (\Delta M)_{\textrm{ev}},
   \label{eq:Mini}
\end{equation}

\noindent where $(\Delta M)_{\textrm{ev}}$, the mass lost by stellar evolution, was obtained from equation 2 of \cite{Lamers_tdis_anal}. $M(t)$ is the cluster total mass ($M_{\textrm{clu}}$; Table~\ref{tab:masses_and_other_params}). The symbol scheme is the same of the previous figures. The filled stars represent the median of the $\Delta M_{\textrm{dis}}/M_{\textrm{ini}}$ values within five $\tau_{\textrm{dyn}}$ bins for each group (analogously to Figure~\ref{fig:rh_Rj_versus_taudyn}). In panel (b), the black stars identify the median values for groups 3 and 5 combined. 

In each panel, the slight vertical separation between the blue and black filled stars indicates that, for a given evolutionary stage, those clusters presenting more inflated main structures tend do be more affected by tidal heating, resulting in preferentially larger mass-loss fractions. This result is somewhat contrasting with those presented by \cite{Piatti:2020} in the case of GCs, for which clusters that have lost relatively more mass by disruption do not seem to have preferentially larger $r_h/R_J$ (their section 3). Some degeneracy with mass is expected along the sequences in Figure~\ref{fig:Mdis_Mini_taudyn}, since evaporation rates also depend on the number of stars (e.g., \citeauthor{Spitzer:1987}\,\,\citeyear{Spitzer:1987}; GB08).

%%%%%%%%%%%%%%%%%%
\subsubsection*{Dissolution time}
%%%%%%%%%%%%%%%%%%

\begin{figure*}
\begin{center}

\parbox[c]{1.00\textwidth}
  {
   \begin{center}
      \includegraphics[width=0.49\textwidth]{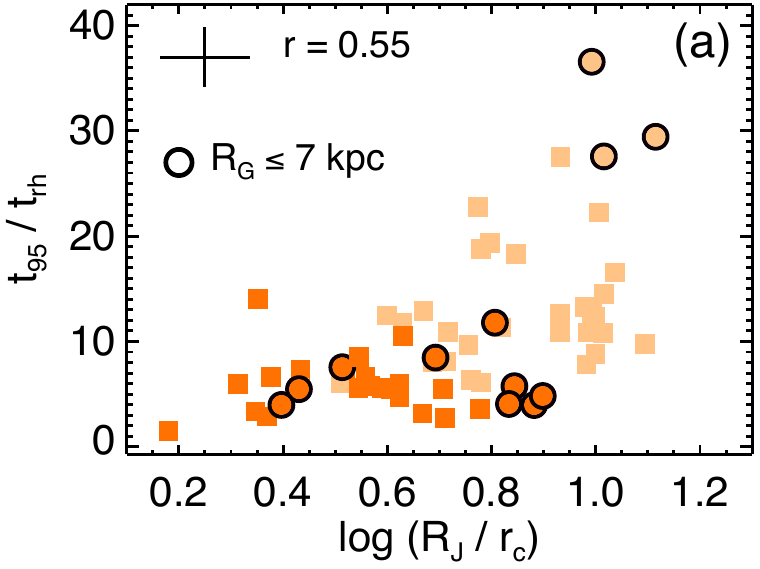} 
      \includegraphics[width=0.49\textwidth]{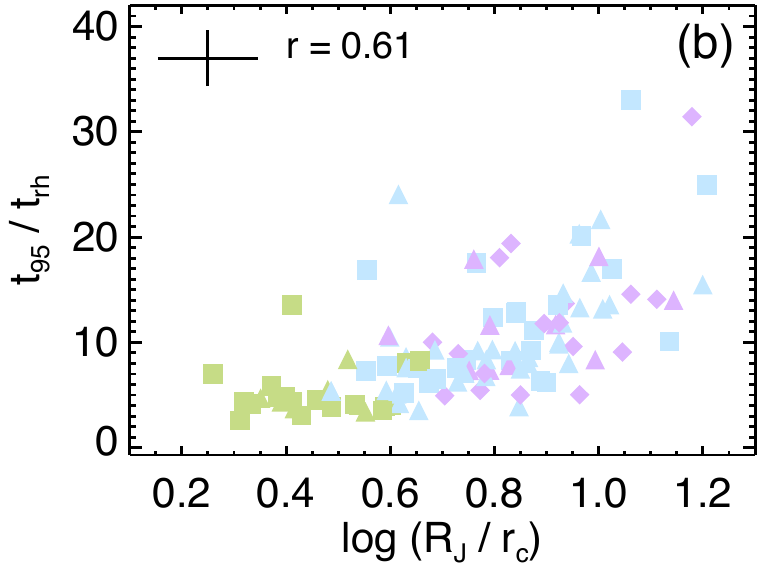}        
    \end{center}    
  }
\caption{ Dissolution time ($t_{95}$), in units of $t_{rh}$, versus cluster concentration for OCs located at $R_\textrm{G}\leq8\,$kpc (panel a) and $R_\textrm{G}\,>\,8\,$kpc (panel b). The Pearson correlation coefficient and the mean error bars are indicated in each panel. The circled symbols highlight those clusters in the range $R_\textrm{G}\leq7\,$kpc. }

\label{fig:t95_trh_cparam}
\end{center}
\end{figure*}

In Figure~\ref{fig:t95_trh_cparam}, we represent the dissolution time ($t_{95}$; equation~\ref{eq:t95}), in units of the half-light relaxation time, versus the cluster concentration, defined here as the logarithm of the $R_J/r_c$ ratio (see also section 3 of \citeauthor{Gnedin:1997}\,\,\citeyear{Gnedin:1997}). Both panels show that the more centrally concentrated clusters tend to take a larger number of relaxation times to be disrupted. This result is particularly true for clusters of group 1 located at the upper right part of panel $a$ (namely, NGC\,6253, Dias\,6 and Berkeley\,81): the more centrally concentrated ones (log\,($R_J/r_c$)$\,\gtrsim$\,0.95) with larger $t_{95}/t_{rh}$ ratios ($\gtrsim$\,25) are located at inner Galactic orbits ($R_\textrm{G}\leq7\,$kpc; circled symbols). Their compact structures seem to make them stable against tidal disruption for many relaxation times. 

The correlation value, considering all OCs plotted in each panel, is moderately high (larger than 0.50, as indicated in the legend). In turn, the more loosely bound clusters (groups 2 and 4 in panels $a$ and $b$, respectively) tend to be more easily disrupted, since $t_{95}/t_{rh}$ is smaller than $\sim7$ for most of them. These results presented in Figure~\ref{fig:t95_trh_cparam} remain almost unaltered (very similar dispersion and $r$ values) if we consider exclusively the cluster dissolution by dynamical effects, that is, disregarding the mass loss due to stellar evolution (by setting $\Delta M_{\textrm{ev}}=0$ in equation~\ref{eq:Mini} and $\mu_{\textrm{ev}}(t)\equiv1.0$ in equation~\ref{eq:t95}).

%%%%%%%%%%%%%%%%%
\subsection{The $r_c/r_h$ ratio}
%%%%%%%%%%%%%%%%%

\begin{figure*}
\begin{center}

\parbox[c]{1.00\textwidth}
  {
   \begin{center}
      \includegraphics[width=0.49\textwidth]{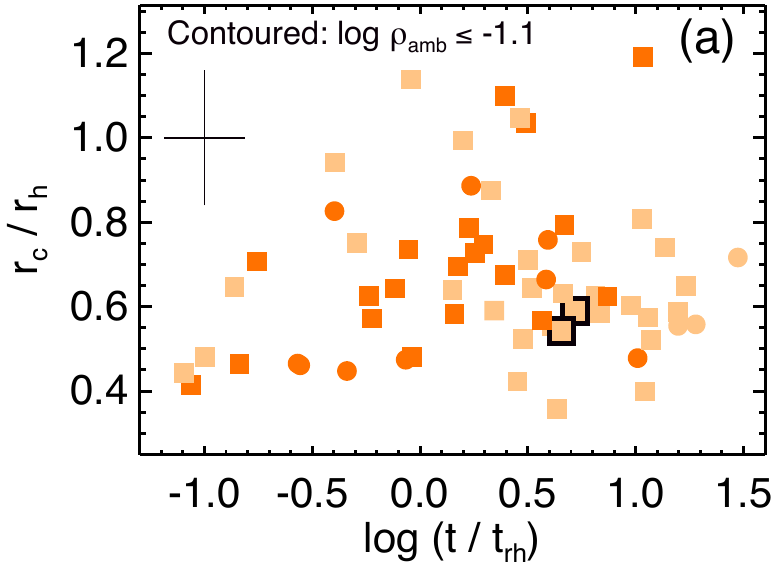} 
      \includegraphics[width=0.49\textwidth]{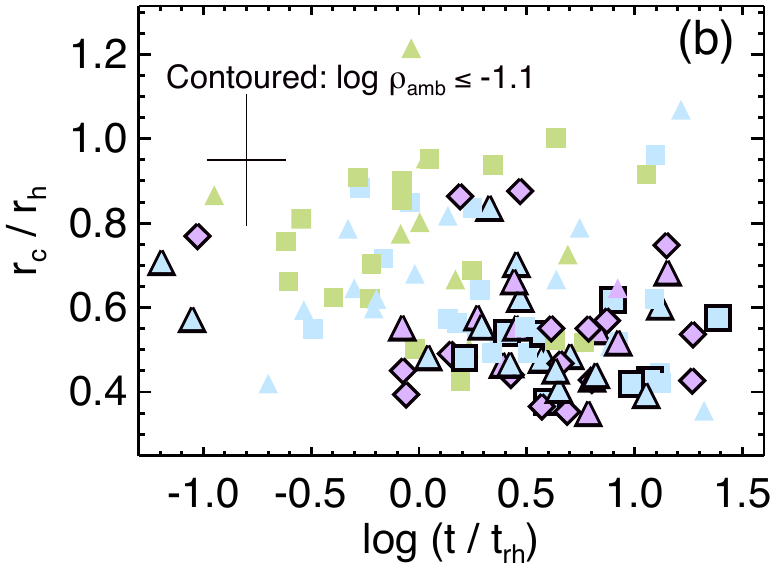}        
    \end{center}    
  }
\caption{ $r_c/r_h$ ratios as function of the dynamical age for OCs located at $R_\textrm{G}\,\leq\,8\,$kpc (panel a) and $R_\textrm{G}\,>\,8\,$kpc (panel b). The contoured symbols represent OCs located in the $\rho_{\textrm{amb}}$ indicated in the legend. Mean error bars are also shown. }

\label{fig:rc_rh_versus_taudyn}
\end{center}
\end{figure*}

The $r_c/r_h$ ratio is also a useful parameter to evaluate the degree of internal dynamical evolution of star clusters. As described by \cite{Heggie:2003}, $r_c/r_h$ is expected to decrease as consequence of re-virialization through violent relaxation at early times (e.g., \citeauthor{Darma:2021}\,\,\citeyear{Darma:2021} and references therein), followed by two-body relaxation, mass segregation and, possibly, core-collapse \citep{Baumgardt:2003}. In this sense, in Figure~\ref{fig:rc_rh_versus_taudyn} the $r_c/r_h$ ratio is plotted as a function of $\tau_{\textrm{dyn}}$. Error bars have been supressed, for better visualization (instead, the mean uncertainties have been indicated). 

At first sight, no clear correlations can be verified. The non-trivial connections between internal evolution, different evaporation rates and initial conditions (thus resulting in different evolutionary paths) seem to erase any distinct trends among the plotted quantities. Despite this, a closer inspection reveals that those OCs subject to less intense external tidal fields (contoured symbols, according to the $\rho_{\textrm{amb}}$ interval indicated in the legend) and presenting signals of dynamical evolution ($\tau_{\textrm{dyn}}\gtrsim0.0$, that is, older than their respective $t_{rh}$) tend to have systematically smaller $r_c/r_h$ ratios (typically $\lesssim0.6$). In addition, all highlighted clusters at these $\tau_{\textrm{dyn}}$ and $r_c/r_h$ ranges are relatively old (log\,$t\,\gtrsim8.8$). As stated previously, the dynamical evolution of these systems seems more importantly determined by internal interactions (which tend to compact their central structures; Figure~\ref{fig:rc_versus_taudyn}) comparatively to other OCs subject to higher $\rho_{\textrm{amb}}$. Besides, both panels of Figure~\ref{fig:rc_rh_versus_taudyn} show that no core-collapsed clusters ($r_c/r_h$\,$\lesssim\,$0.2; e.g., \citeauthor{Piatti:2019a}\,\,\citeyear{Piatti:2019a}) are expected to be found within our sample.

%All encircled clusters in Figure~\ref{fig:rc_rh_versus_taudyn} present tidal filling ratios smaller than 0.4 (see also panels (c) and (d) of Figure~\ref{fig:rh_Rj_and_rho_hm_versus_Rgal}) and, as stated previously, their dynamical evolution may be more importantly determined by internal interactions (which tend to compact their central structures, as seen in Figure~\textcolor{red}{a definir}) comparatively to other OCs subject to higher $\rho_{\textrm{amb}}$.        

%%%%%%%%%%%%%%%%
\subsection{Mass segregation}
\label{sec:mass_segregation}
%%%%%%%%%%%%%%%%

\begin{figure*}
\begin{center}

\parbox[c]{1.00\textwidth}
  {
   \begin{center}
       \includegraphics[width=0.49\textwidth]{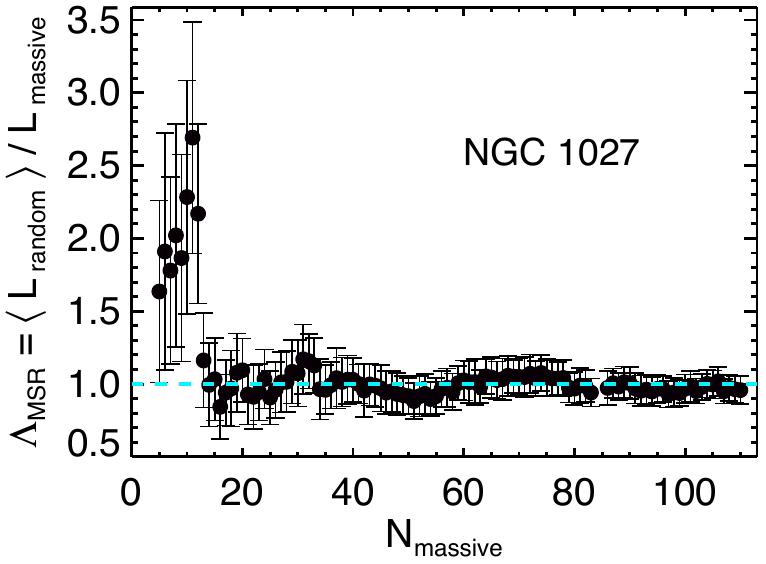} 
      \includegraphics[width=0.49\textwidth]{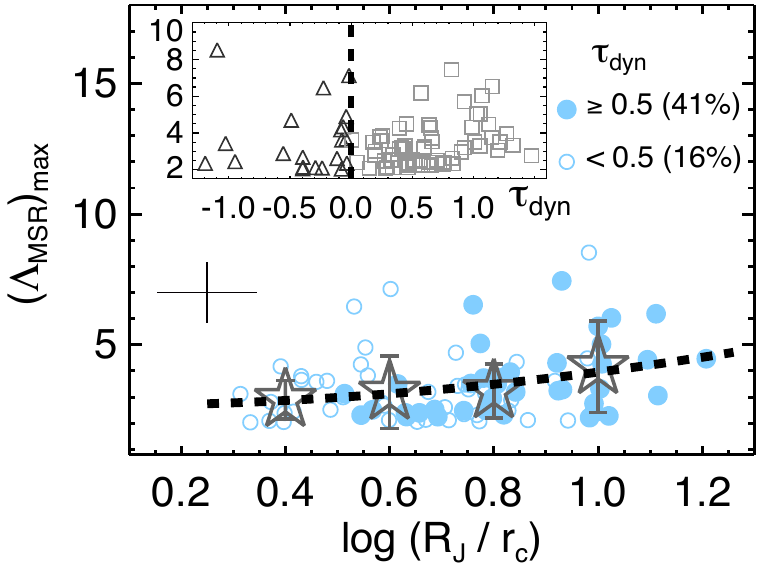}        
    \end{center}    
  }
\caption{  Left panel: mass segregation ratio ($\Lambda_{\textrm{MSR}}$) as function of the $N$ most massive member stars ($N_{\textrm{massive}}$) of NGC\,1027. The $\Lambda_{\textrm{MSR}}=1.0$ value (dashed line) indicates no mass segregated stars. Right panel: maximum value of $\Lambda_{\textrm{MSR}}$ as function of the cluster concentration for 85 clusters presenting $\Lambda_{\textrm{MSR}}^{\textrm{max}}\ge2.0$. The open stars (and associated error bars) represent the mean values (and dispersion) of $\Lambda_{\textrm{MSR}}$ within the following bins: log($R_J/r_c$) $\le$ 0.5, 0.5 $<$ log($R_J/r_c$) $\le$ 0.7, 0.7 $<$ log($R_J/r_c$) $\le$ 0.9 and log($R_J/r_c$) $>$ 0.9. The filled and open blue circles represent clusters with $\tau_{\textrm{dyn}}\gtrsim0.5$ and  $\tau_{\textrm{dyn}}\lesssim0.5$, respectively; the numbers in parentheses indicate the correlation between $\Lambda_{\textrm{MSR}}^{\textrm{max}}$ and log($R_J/r_c$) for each subsample. Mean error bars are shown, for better visualization. The thick dashed line is a polynomial fit to the plotted sample (see text for details). The inset shows the $\Lambda_{\textrm{MSR}}^{\textrm{max}}$ as function of $\tau_{\textrm{dyn}}$: triangles identify OCs younger than their respective $t_{rh}$ (therefore, $\tau_{\textrm{dyn}}\lesssim0.0$) and squares represent those with $t/t_{rh}\,\gtrsim\,1$ ($\tau_{\textrm{dyn}}\gtrsim0.0$).  }

\label{fig:mass_segreg}
\end{center}
\end{figure*}

In order to evaluate if there is statistically significant evidence of mass segregation within each investigated OC, we firstly took a subset of members ($N_{\textrm{massive}}$) more massive than a given mass threshold ($M_{\textrm{cut}}$). Within this sample, we computed the nearest neighbour graph (e.g., \citeauthor{Eppstein:1997}\,\,\citeyear{Eppstein:1997}) and then evaluated the total summed length $L_{\textrm{massive}}=\sum\limits_{i=1}^{N_{\textrm{massive}}}\ell_{i}$, where $\ell_i$ is the distance of each star to its closest neighbour with no repetitions.

In a second step, we randomly selected a set of $N_{\textrm{random}}$ stars (where $N_{\textrm{random}}=N_{\textrm{massive}}$), among the complete list of cluster members, and evaluated their corresponding summed length ($L_{\textrm{random}}$). This same procedure was repeated 100 times\footnote[12]{This number of runs was established to save computational time. We have also performed a set of runs employing a  thousand of redrawings, with results very similar to those presented here.} and the mean value, $\langle L_{\textrm{random}}\rangle$, was computed together with the associated dispersion, $\sigma_ {L_{random}}$. We then computed the mass segregation ratio\footnote[13]{Our prodedure is analogous to that of \cite{Tarricq:2022} and \citeauthor{Allison:2009b}\,\,(2009b), who employed the length of the minimum spanning tree to obtain $L_{\textrm{massive}}$ and $L_{\textrm{random}}$. In our alternative method, we do not impose the constraint that the different ``branches" of the spanning tree be connected, since we simply utilized a quantity that indicates how close the more masive stars are from each other.} ($\Lambda_{\textrm{MSR}}$) and its uncertainty, defined as (see \citeauthor{Allison:2009b}\,\,2009b):

\begin{equation}
    \Lambda_{\textrm{MSR}} = \frac{\langle L_{\textrm{random}}\rangle \pm \sigma_{L_{\textrm{random}}}}{L_{\textrm{massive}}} 
\end{equation}

\noindent
With this definition, a given group of $N_{\textrm{massive}}$ stars for which $\Lambda_{\textrm{MSR}}$ exceeds 1.0 (value that corresponds to the absence of mass segregation), considering uncertainties, is compatible with being mass segregated. 

Then, progressively smaller $M_{\textrm{cut}}$ values are taken and the corresponding $\Lambda_{\textrm{MSR}}$ is obtained. The results of this procedure are illustrated in Figure~\ref{fig:mass_segreg}, left panel, for the OC NGC\,1027, one of the investigated clusters with the largest peak value of $\Lambda_{\textrm{MSR}}$. A minimum of 5 massive stars is considered in the procedure to avoid small number statistics. For larger $N$, the ensemble of $\Lambda_{\textrm{MSR}}$ converges to values close to unity. Larger error bars are present for smaller $N_{\textrm{massive}}$, due to stochasticity in the sampling procedure, and become smaller for larger samples. For NGC\,1027, the eleven most massive stars (in this case, $m\gtrsim3\,M_{\odot}$) present the highest level of mass segregation, since their $L_{\textrm{massive}}$ value is, on average, $\sim2.7$ times smaller compared to a random sample of 11 stars taken from the members list.

%We then selected those OCs containing groups of mass segregated stars for which $\Lambda_{\textrm{MSR}}\ge2$ (85 clusters in our case) and took the largest value of $\Lambda_{\textrm{MSR}}$ for each one (for example, in the case of NGC\,1027, $\Lambda_{\textrm{MSR}}^{\textrm{max}}\simeq2.7$). This quantity was plotted as function of log\,($R_J/r_c$) in the right panel of Figure~\ref{fig:mass_segreg}. The open stars represent the mean values obtained within four log\,($R_J/r_c$) bins, as indicated in the figure caption. Filled blue circles represent the more dynamically evolved clusters ($\tau_{\textrm{dyn}}\gtrsim0.5$), for which a moderate correlation ($\simeq$40\%) between the plotted quantities was found. In the case of the less dynamically evolved ones ($\tau_{\textrm{dyn}}\lesssim0.5$; open blue circles in the main panel), the correlation decreases to less than 20\%. 

We then selected those OCs containing groups of stars with significant evidence of mass segregation (in our case, $\Lambda_{\textrm{MSR}}\ge2$, resulting in 85 selected clusters) and took the largest value of $\Lambda_{\textrm{MSR}}$ for each one (for example, in the case of NGC\,1027, $\Lambda_{\textrm{MSR}}^{\textrm{max}}\simeq2.7$). In the right panel of Figure~\ref{fig:mass_segreg}, this quantity was plotted as function of the cluster concentration, defined here as the logarithm of the Jacobi to core radius ratio (that is, log\,$R_J/r_c$). The open stars represent the mean values obtained within four log\,($R_J/r_c$) bins, as indicated in the figure caption. Filled blue circles represent clusters in a relatively high degree of dynamical evolution ($\tau_{\textrm{dyn}}\gtrsim0.5$, that is, $t/t_{rh}\,\gtrsim\,3$), for which a moderate correlation ($\simeq$40\%) between the plotted quantities was found. For OCs with $\tau_{\textrm{dyn}}\lesssim0.5$ (open blue circles in the main panel), the correlation decreases to less than 20\%.

The mild increase in the mean $\Lambda_{\textrm{MSR}}^{\textrm{max}}$ values for more centrally concentrated clusters (relatively to the first bin, the last one presents and increase of $\sim43\%$) is somewhat consistent with the overall mass segregation scenario (i.e., compactness of the central structure due to the presence of an increasingly larger fraction of massive stars; e.g., \citeauthor{Mackey:2008}\,\,\citeyear{Mackey:2008}; \citeauthor{Portegies-Zwart:2002}\,\,\citeyear{Portegies-Zwart:2002}; \citeauthor{Gurkan:2004}\,\,\citeyear{Gurkan:2004}). In turn, the less concentrated OCs tend to present slightly smaller and less dispersed $\Lambda_{\textrm{MSR}}^{\textrm{max}}$ values. To reinforce this trend, we have performed a second order polynomial fit, by means of $\chi^2$ minimization, which resulted in the relation 

\begin{equation}
      \Lambda_{\textrm{MSR}}^{\textrm{max}} = 2.66(\pm0.23) + 1.29(\pm0.34)\,[\textrm{log}(R_J/r_c)]^2
\end{equation}

\noindent
It is also noticeable that the bin of more centrally concentrated OCs (log\,($R_J/r_c$) $\gtrsim$ 0.9) is preferentially occupied by clusters in more advanced stages of dynamical evolution (filled blue circles), while the opposed is verified at the other extreme (log\,($R_J/r_c$) $\lesssim$ 0.5 interval, defined mainly by open blue circles).     

%It is also noticeable that the bin of less centrally concentrated OCs (log\,($R_J/r_c$) $\lesssim$ 0.5) is preferentially occupied by less evolved clusters (open blue circles), while the opposed is verified at the other extreme (log\,($R_J/r_c$) $\gtrsim$ 1.0 interval, defined mainly by blue filled circles).     

%a given group of $N_{\textrm{massive}}$ stars is considered mass segregated if the associated $\Lambda_{\textrm{MSR}}$ is larger than 1.0 (value that corresponds to the absence of mass segregation), considering its uncertainty. Progressively smaller $M_{\textrm{cut}}$ values are taken, and the corresponding $\Lambda_{\textrm{MSR}}$ is obtained. The results of this procedure are illustrated in Figure ... for the OC... A minimum of 5 massive stars is considered in the procedure, to avoid small number statistics. Typically, for $N_{\textrm{massive}}>100$, the ensemble of $\Lambda_{\textrm{MSR}}$ converges to unity. 

Although it is expected an increase of $\Lambda_{\textrm{MSR}}$ as a cluster evolves (e.g., \citeauthor{Sanchez:2009}\,\,\citeyear{Sanchez:2009}), the inset in Figure~\ref{fig:mass_segreg}, right panel, reveals no particular behaviour of $\Lambda_{\textrm{MSR}}^{\textrm{max}}$ as function of the dynamical age. This plot shows that even dynamically unevolved systems ($\tau_{\textrm{dyn}}\lesssim0.0$) can present very different levels of mass segregation (see also \citeauthor{Dib:2018}\,\,\citeyear{Dib:2018}); the dispersion of the ensemble of $\Lambda_{\textrm{MSR}}$ values for the dynamically unevolved systems ($\tau_{\textrm{dyn}}\lesssim0.0$) is comparable to what is observed for the evolved ones ($\tau_{\textrm{dyn}}\gtrsim0.0$). The presence of mass segregation at young ages can impact significantly the further evolution of a cluster (e.g., \citeauthor{Pang:2021}\,\,\citeyear{Pang:2021}; \citeauthor{Portegies-Zwart:2010}\,\,\citeyear{Portegies-Zwart:2010} and references therein) and could be the result of the merging of smaller substructures during the cluster formation process: multiple clumps can mass segregate locally in short timescales and their merging can originate a cluster that inherited the substructure's segregation (\citeauthor{Allison:2009a}\,\,2009a; \citeauthor{McMillan:2007}\,\,\citeyear{McMillan:2007}). Besides, mass segregation occurs in timescales that differ from cluster to cluster, since the time taken by the more massive stars to slow down and sink towards the cluster centre is $t_s\sim t_{rh}\,\langle m\rangle/m$ \citep{Spitzer:1969}, where $\langle m\rangle$ is the cluster mean stellar mass. The combination of these effects apparently causes each cluster to follow a distinct individual evolution, thus erasing possible general correlations between $\Lambda_{\textrm{MSR}}$ and $\tau_{\textrm{dyn}}$ (or age) for all clusters.          

%Obviously, $t_s$ varies from cluster to cluster, even considering those in compatible evolutionary stages.

  %%%%%%%%%%%%%%%%%%%%%%
\section{Concluding remarks}
\label{sec:conclusions}
%%%%%%%%%%%%%%%%%%%%%%

The present paper was devoted to a thorough investigation of a relatively large sample of 174 Galactic OCs (114 of them analysed here, combined with 60 other objects from a previous investigation). After a proper analysis of the RPDs and astrometrically decontaminated CMDs, we performed a joint exploration of structural and evolutionary-related parameters. Since the imprints of the evolutionary processes on the clusters' morphology present a multifactorial dependence, we segregated the complete sample in terms of the OCs physical properties and location within the Galaxy.

Our main conclusions can be summarized as follows:

\begin{itemize}
   
   \item The tidal filling ratio ($r_h/R_J$) is a useful parameter to evaluate the cluster dynamical state, since it is determined by both the internal evolution and the external tidal field conditions. It tends to decrease with the cluster dynamical age; we suggest that this result may be a consequence of the internal relaxation process, which causes the clusters' main body to be progressively more compact. Besides, larger masses tend to produce smaller tidal filling ratios;
   
   \item Regarding the dependence of $r_h/R_J$ with the external conditions, we identified different evaporation regimes: for $R_\textrm{G}\,\leq\,8\,$kpc, those OCs with more internal orbits are also denser, which favours their survival against more intense mass loss due to tidal stripping. Besides, smaller $r_h/R_J$ implies smaller fraction of evaporated stars at each $t_{rh}$; 
   
   \item The range of $R_\textrm{G}$ between $\sim8-9$\,kpc is apparently a transition region (compatible with the location of the MW corotation radius), where a large spread in $r_h/r_J$ is verified; for $R_\textrm{G}\gtrsim9\,$kpc, the ensemble of $r_h/r_J$ seems less dispersed. There is a slight decreasing trend with $\rho_{\textrm{amb}}$ in the domain log\,$\rho_{\textrm{amb}}\lesssim-1.1$ (followed particularly by those OCs at $R_\textrm{G}$\,$>$\,10\,kpc), such that they become progressively less tidally influenced as they are exposed to a weaker external gravitational field;

\item At a given dynamical stage, as inferred from $\tau_{\textrm{dyn}}$, clusters with larger $r_h/r_J$ present, expectedly, larger fraction of mass loss by disruption;  

 \item The core radius tends to decrease with $\tau_{\textrm{dyn}}$, presumibly as a consequence of mass segregation. In this sense, OCs with higher degree of mass segregation (as inferred from the mass segregation ratio $\Lambda_{\textrm{MSR}}$) tend to present slightly larger cluster central concentration. The $r_c$ values present smaller dependence with the external conditions compared to $r_h$ and $r_t$, indicating that the clusters' central structure is more sensible to the internal evolution; 

\item In the log\,$(\rho_{\textrm{amb}})\lesssim-1.1$ domain, the clusters dynamical evolution seems more regulated by internal relaxation, which causes a decrease in their $r_c/r_h$;

\item There are clusters within our sample located at compatible $R_\textrm{G}$, subject to compatible $\rho_{\textrm{amb}}$ and presenting similar $\tau_{\textrm{dyn}}$, but in different evaporation regimes, thus indicating that the initial formation conditions play a role in their dynamical evolution;

 \item Clusters with larger degree of central concentration are more stable against disruptive effects (tidal heating combined with two-body interactions), since they tend to live for larger number of relaxation times compared to the less concentrated ones.

 \item The mass segregation ratio ($\Lambda_{\textrm{MSR}}$) does not present well-defined trends with the dynamical age; both dynamically evolved ($t/t_{rh}\gtrsim1$) and unevolved ($t/t_{rh}\lesssim1$) clusters present comparable dispersion of the ensemble of the $\Lambda_{\textrm{MSR}}$ values. This result is a possible empirical evidence that mass segregation can be produced not only dynamically, but may also be a consequence of the progenitor cloud fragmentation process. There is, however, a slight positive correlation with the cluster concentration (log\,$R_J/r_c$).

\end{itemize}

Despite the useful discussions presented here, our sample still lacks clusters at the very beginning of their evolution, that is, still embedded in their progenitor clouds. The analysis of such objects requires a proper treatment of differential reddening by means of the construction of extinction maps. The accomplishment of this task demands the characterization of the interstellar medium from, e.g., ground-based infrared and narrow-band photometry, polarimetry and spectroscopy. This is a necessary step towards establishing a more complete overview of the OCs formation and dissolution processes based on observational parameters, as well as providing progressively better constraints for theoretical investigations.

%Together with the increasingly larger precision of \textit{Gaia's} astrometry, the combination of different kind of data for the young objects will provide        

%%%%%%%%%%%%%%%%%%%%%%%%%

\section{Acknowledgments}

The authors thank the anonymous referee for useful suggestions, which helped improving the clarity of the paper. The authors acknowledge financial support from Conselho Nacional de Desenvolvimento Cient\'ifico e Tecnol\'ogico\,$-$\,CNPq (proc. 404482/2021-0). F.F.S.Maia acknowledges financial support from FAPERJ (proc. E-26/201.386/2022 and E-26/211.475/2021). W. Corradi acknowledges the support from CNPq - BRICS 440142/2022-9, FAPEMIG APQ 02493-22 and FNDCT/FINEP/REF 0180/22. This research has made use of the VizieR catalogue access tool, CDS, Strasbourg, France. This research has made use of the SIMBAD database, operated at CDS, Strasbourg, France. This work has made use of data from the European Space Agency (ESA) mission \textit{Gaia} (https://www.cosmos.esa.int/gaia), processed by the \textit{Gaia} Data Processing and Analysis Consortium (DPAC, https://www.cosmos.esa.int/web/gaia/dpac/consortium). Funding for the DPAC has been provided by national institutions, in particular the institutions participating in the \textit{Gaia} Multilateral Agreement. This research has made use of \textit{Aladin sky atlas} developed at CDS, Strasbourg Observatory, France. 

%The authors thank the Brazilian financial agencies CNPq and FAPEMIG. This study was financed in part by the Coordena\c{c}\~ao de Aperfei\c{c}oamento de Pessoal de N\'ivel Superior $-$ Brazil (CAPES) $-$ Finance Code 001.

\subsection*{Data availability}
\textit{The data underlying this article are available in the article and in its online supplementary material.}

%{\footnotesize
\bibliographystyle{mn2e}
\bibliography{referencias}
%}

\newpage
\newpage
\newpage

\appendix

%%%%%%%%%%%%%%%%%%%%
\section{Astrophysical parameters}
\label{sec:appendix_astroph_params}
%%%%%%%%%%%%%%%%%%%%

Table~\ref{tab:investig_sample} of this appendix show the fundamental parameters for the investigated OCs. Table~\ref{tab:masses_and_other_params} shows additional parameters.

\begin{table*}
\centering
\tiny
\rotcaption{ Central coordinates (Section~\ref{sec:struct_params}), Galactocentric distance ($R_\textrm{G}$, Section~\ref{sec:decontam_CMD_analysis}), structural (core, tidal and half-light radii, Section~\ref{sec:struct_params}) and fundamental parameters (distance modulus, colour excess, age and metallicity, Section~\ref{sec:decontam_CMD_analysis}), mean proper motion components (Section~\ref{sec:decontam_CMD_analysis}) and half-light relaxation time ($t_{rh}$, Section~\ref{sec:trh}) for the studied sample. }
\begin{sideways}
\begin{minipage}{225mm}

\begin{tabular}{lcrrccrrccrrrr}

 Cluster               & $RA$                            & $DEC$                             & R$_\textrm{G}^{(*)}$   & $r_c$              & $r_{h}$             & $r_t$                & $(m-M)_0$             & $E(B-V)$              & log\,$t$             & $[Fe/H]^{(**)}$         & $\langle\mu_{\alpha}\,\textrm{cos}\,\delta\rangle^{\dag}$    & $\langle\mu_{\delta}\rangle^{\dag}$     & $t_{rh}$             \\    
                       &($h$:$m$:$s$)                    & ($^{\circ}$:$'$:$''$)             &  (kpc)                 &  (pc)              &     (pc)            &  (pc)                & (mag)                 & (mag)                 & (dex)                & (dex)                   & (mas yr$^{-1}$)                                              & (mas yr$^{-1}$)                         & (Myr)                \\

\hline                                                                                                                                                                                                                                                                                                                                                                                                                              
 NGC\,103              & 00:25:16                        &  61:19:27                         & 10.0\,$\pm$\,0.2       & 2.19\,$\pm$\,0.55  & 2.68\,$\pm$\,0.39   &  7.25\,$\pm$\,1.15   & 12.31\,$\pm$\,0.25    & 0.57\,$\pm$\,0.10     & 8.15\,$\pm$\,0.30    & -0.13\,$\pm$\,0.23      &  -2.82\,$\pm$\,0.05                                           & -1.06\,$\pm$\,0.04                     &  104\,$\pm$\, 24     \\ 
 NGC\,436              & 01:15:59                        &  58:48:59                         & 10.1\,$\pm$\,0.2       & 1.44\,$\pm$\,0.30  & 2.60\,$\pm$\,0.43   &  9.65\,$\pm$\,2.29   & 12.19\,$\pm$\,0.20    & 0.57\,$\pm$\,0.06     & 7.90\,$\pm$\,0.25    & -0.13\,$\pm$\,0.23      &  -0.99\,$\pm$\,0.06                                           & -0.65\,$\pm$\,0.05                     &   94\,$\pm$\, 24     \\ 
 NGC\,457              & 01:19:40                        &  58:17:17                         &  9.9\,$\pm$\,0.2       & 2.33\,$\pm$\,0.34  & 4.07\,$\pm$\,0.80   & 14.68\,$\pm$\,4.84   & 11.92\,$\pm$\,0.30    & 0.57\,$\pm$\,0.08     & 7.35\,$\pm$\,0.20    & -0.06\,$\pm$\,0.26      &  -1.59\,$\pm$\,0.07                                           & -0.73\,$\pm$\,0.08                     &  248\,$\pm$\, 74     \\ 
 NGC\,581              & 01:33:23                        &  60:40:01                         &  9.8\,$\pm$\,0.2       & 4.53\,$\pm$\,0.99  & 5.23\,$\pm$\,0.82   & 13.53\,$\pm$\,2.87   & 11.80\,$\pm$\,0.30    & 0.48\,$\pm$\,0.09     & 7.50\,$\pm$\,0.20    &  0.00\,$\pm$\,0.23      &  -1.40\,$\pm$\,0.12                                           & -0.61\,$\pm$\,0.10                     &  285\,$\pm$\, 73     \\ 
 Trumpler\,2           & 02:36:41                        &  55:55:48                         &  8.7\,$\pm$\,0.1       & 3.21\,$\pm$\,0.76  & 3.76\,$\pm$\,0.54   &  9.84\,$\pm$\,1.61   &  9.00\,$\pm$\,0.30    & 0.39\,$\pm$\,0.08     & 8.00\,$\pm$\,0.15    &  0.00\,$\pm$\,0.23      &   1.53\,$\pm$\,0.12                                           & -5.41\,$\pm$\,0.11                     &  120\,$\pm$\, 27     \\ 
 NGC\,1039             & 02:41:60                        &  42:43:60                         &  8.6\,$\pm$\,0.1       & 1.92\,$\pm$\,0.34  & 2.99\,$\pm$\,0.60   &  9.82\,$\pm$\,3.25   &  8.32\,$\pm$\,0.35    & 0.09\,$\pm$\,0.07     & 8.30\,$\pm$\,0.35    &  0.10\,$\pm$\,0.23      &   0.66\,$\pm$\,0.18                                           & -5.79\,$\pm$\,0.22                     &  104\,$\pm$\, 32     \\ 
 NGC\,1193             & 03:05:55                        &  44:22:59                         & 12.6\,$\pm$\,0.3       & 1.59\,$\pm$\,0.29  & 3.72\,$\pm$\,0.73   & 17.33\,$\pm$\,5.62   & 13.48\,$\pm$\,0.15    & 0.22\,$\pm$\,0.04     & 9.70\,$\pm$\,0.10    & -0.22\,$\pm$\,0.19      &  -0.21\,$\pm$\,0.11                                           & -0.42\,$\pm$\,0.12                     &  271\,$\pm$\, 83     \\ 
 Trumpler\,3           & 03:12:10                        &  63:14:05                         &  8.7\,$\pm$\,0.1       & 2.76\,$\pm$\,0.56  & 3.04\,$\pm$\,0.45   &  7.59\,$\pm$\,1.50   &  9.05\,$\pm$\,0.30    & 0.40\,$\pm$\,0.10     & 7.65\,$\pm$\,0.20    & -0.06\,$\pm$\,0.26      &  -3.39\,$\pm$\,0.10                                           & -0.15\,$\pm$\,0.09                     &   86\,$\pm$\, 20     \\ 
 NGC\,1245             & 03:14:42                        &  47:13:12                         & 10.7\,$\pm$\,0.3       & 3.44\,$\pm$\,0.82  & 5.97\,$\pm$\,0.93   & 21.40\,$\pm$\,4.04   & 12.25\,$\pm$\,0.30    & 0.26\,$\pm$\,0.06     & 9.05\,$\pm$\,0.10    &  0.00\,$\pm$\,0.17      &   0.48\,$\pm$\,0.04                                           & -1.66\,$\pm$\,0.05                     &  601\,$\pm$\,142     \\ 
 NGC\,1528             & 04:15:31                        &  51:13:50                         &  9.1\,$\pm$\,0.2       & 2.63\,$\pm$\,0.47  & 5.21\,$\pm$\,1.05   & 20.91\,$\pm$\,6.99   &  9.89\,$\pm$\,0.25    & 0.31\,$\pm$\,0.06     & 8.65\,$\pm$\,0.15    &  0.10\,$\pm$\,0.14      &   2.15\,$\pm$\,0.07                                           & -2.29\,$\pm$\,0.09                     &  261\,$\pm$\, 81     \\ 
 NGC\,1664             & 04:50:60                        &  43:40:34                         &  9.4\,$\pm$\,0.2       & 2.31\,$\pm$\,0.53  & 2.92\,$\pm$\,0.39   &  8.12\,$\pm$\,1.11   & 10.43\,$\pm$\,0.25    & 0.29\,$\pm$\,0.07     & 8.80\,$\pm$\,0.10    & -0.06\,$\pm$\,0.20      &   1.60\,$\pm$\,0.11                                           & -5.80\,$\pm$\,0.07                     &  113\,$\pm$\, 23     \\ 
 Berkeley\,14          & 04:59:45                        &  43:28:58                         & 12.4\,$\pm$\,0.6       & 2.77\,$\pm$\,0.59  & 5.92\,$\pm$\,1.02   & 25.41\,$\pm$\,6.33   & 13.18\,$\pm$\,0.30    & 0.54\,$\pm$\,0.05     & 9.40\,$\pm$\,0.10    & -0.13\,$\pm$\,0.23      &   1.40\,$\pm$\,0.13                                           &  0.38\,$\pm$\,0.11                     &  554\,$\pm$\,150     \\ 
 NGC\,1778             & 05:08:07                        &  36:59:43                         &  9.7\,$\pm$\,0.2       & 2.97\,$\pm$\,0.48  & 3.70\,$\pm$\,0.68   & 10.16\,$\pm$\,3.01   & 10.85\,$\pm$\,0.25    & 0.45\,$\pm$\,0.08     & 8.15\,$\pm$\,0.15    & -0.06\,$\pm$\,0.20      &   0.46\,$\pm$\,0.06                                           & -3.20\,$\pm$\,0.08                     &  140\,$\pm$\, 40     \\ 
 NGC\,1798             & 05:11:39                        &  47:41:31                         & 12.1\,$\pm$\,0.4       & 2.46\,$\pm$\,0.60  & 4.46\,$\pm$\,0.67   & 16.60\,$\pm$\,2.81   & 13.02\,$\pm$\,0.20    & 0.58\,$\pm$\,0.06     & 9.15\,$\pm$\,0.10    & -0.32\,$\pm$\,0.24      &   0.79\,$\pm$\,0.04                                           & -0.37\,$\pm$\,0.06                     &  344\,$\pm$\, 81     \\ 
 Berkeley\,17          & 05:20:30                        &  30:34:50                         & 11.0\,$\pm$\,0.3       & 2.52\,$\pm$\,0.41  & 4.69\,$\pm$\,0.68   & 17.87\,$\pm$\,3.88   & 12.23\,$\pm$\,0.25    & 0.60\,$\pm$\,0.05     & 9.90\,$\pm$\,0.10    & -0.06\,$\pm$\,0.20      &   2.54\,$\pm$\,0.11                                           & -0.33\,$\pm$\,0.11                     &  426\,$\pm$\, 98     \\ 
 Berkeley\,70          & 05:25:49                        &  41:56:52                         & 12.2\,$\pm$\,0.4       & 2.60\,$\pm$\,0.47  & 4.72\,$\pm$\,0.79   & 17.58\,$\pm$\,4.49   & 13.04\,$\pm$\,0.20    & 0.55\,$\pm$\,0.06     & 9.40\,$\pm$\,0.10    & -0.22\,$\pm$\,0.19      &   0.81\,$\pm$\,0.15                                           & -1.87\,$\pm$\,0.08                     &  411\,$\pm$\,107     \\ 
 NGC\,1907             & 05:28:08                        &  35:19:37                         &  9.8\,$\pm$\,0.3       & 1.21\,$\pm$\,0.20  & 2.37\,$\pm$\,0.46   &  9.46\,$\pm$\,3.07   & 10.93\,$\pm$\,0.35    & 0.46\,$\pm$\,0.06     & 8.90\,$\pm$\,0.10    & -0.06\,$\pm$\,0.20      &  -0.13\,$\pm$\,0.03                                           & -3.40\,$\pm$\,0.08                     &  103\,$\pm$\, 31     \\ 
 NGC\,1912             & 05:28:38                        &  35:49:39                         &  9.3\,$\pm$\,0.2       & 2.95\,$\pm$\,0.68  & 4.33\,$\pm$\,0.62   & 13.56\,$\pm$\,2.20   & 10.14\,$\pm$\,0.25    & 0.32\,$\pm$\,0.08     & 8.40\,$\pm$\,0.15    &  0.00\,$\pm$\,0.23      &   1.55\,$\pm$\,0.09                                           & -4.43\,$\pm$\,0.11                     &  262\,$\pm$\, 57     \\ 
 NGC\,1960             & 05:36:15                        &  34:08:47                         &  9.4\,$\pm$\,0.2       & 1.42\,$\pm$\,0.28  & 2.40\,$\pm$\,0.35   &  8.37\,$\pm$\,1.69   & 10.28\,$\pm$\,0.30    & 0.30\,$\pm$\,0.08     & 7.30\,$\pm$\,0.10    &  0.00\,$\pm$\,0.23      &  -0.22\,$\pm$\,0.01                                           & -3.41\,$\pm$\,0.08                     &   68\,$\pm$\, 16     \\ 
 NGC\,2194             & 06:13:42                        &  12:48:51                         & 11.0\,$\pm$\,0.4       & 2.34\,$\pm$\,0.42  & 5.93\,$\pm$\,1.11   & 29.73\,$\pm$\,9.07   & 12.29\,$\pm$\,0.30    & 0.53\,$\pm$\,0.08     & 8.80\,$\pm$\,0.10    & -0.13\,$\pm$\,0.23      &   0.47\,$\pm$\,0.07                                           & -1.42\,$\pm$\,0.08                     &  724\,$\pm$\,205     \\ 
 NGC\,2192             & 06:15:17                        &  39:50:43                         & 11.7\,$\pm$\,0.3       & 1.52\,$\pm$\,0.20  & 4.31\,$\pm$\,0.86   & 23.92\,$\pm$\,8.38   & 12.71\,$\pm$\,0.20    & 0.20\,$\pm$\,0.06     & 9.10\,$\pm$\,0.05    & -0.22\,$\pm$\,0.19      &   0.18\,$\pm$\,0.09                                           & -1.95\,$\pm$\,0.04                     &  259\,$\pm$\, 80     \\ 
 NGC\,2236             & 06:29:40                        &  06:50:38                         & 10.5\,$\pm$\,0.3       & 2.39\,$\pm$\,0.58  & 4.34\,$\pm$\,0.81   & 16.13\,$\pm$\,4.22   & 11.92\,$\pm$\,0.25    & 0.52\,$\pm$\,0.06     & 8.95\,$\pm$\,0.10    & -0.06\,$\pm$\,0.20      &  -0.78\,$\pm$\,0.10                                           &  0.01\,$\pm$\,0.10                     &  315\,$\pm$\, 90     \\ 
 Trumpler\,5           & 06:36:28                        &  09:28:38                         & 11.1\,$\pm$\,0.4       & 6.37\,$\pm$\,1.50  & 7.38\,$\pm$\,1.11   & 19.11\,$\pm$\,3.54   & 12.42\,$\pm$\,0.30    & 0.64\,$\pm$\,0.08     & 9.45\,$\pm$\,0.10    & -0.32\,$\pm$\,0.24      &  -0.62\,$\pm$\,0.06                                           &  0.27\,$\pm$\,0.08                     & 1815\,$\pm$\,412     \\ 
 NGC\,2266             & 06:43:20                        &  26:58:55                         & 11.5\,$\pm$\,0.5       & 1.54\,$\pm$\,0.24  & 4.22\,$\pm$\,0.92   & 22.69\,$\pm$\,8.62   & 12.61\,$\pm$\,0.30    & 0.12\,$\pm$\,0.06     & 9.00\,$\pm$\,0.10    & -0.13\,$\pm$\,0.16      &  -0.49\,$\pm$\,0.11                                           & -1.25\,$\pm$\,0.02                     &  269\,$\pm$\, 89     \\ 
 Collinder\,115        & 06:46:37                        &  01:46:40                         &  9.9\,$\pm$\,0.3       & 1.63\,$\pm$\,0.46  & 2.63\,$\pm$\,0.47   &  8.87\,$\pm$\,1.91   & 11.36\,$\pm$\,0.30    & 0.43\,$\pm$\,0.08     & 7.85\,$\pm$\,0.25    & -0.06\,$\pm$\,0.26      &  -1.27\,$\pm$\,0.04                                           & -0.56\,$\pm$\,0.02                     &  112\,$\pm$\, 31     \\ 
 NGC\,2286             & 06:47:41                        & -03:09:31                         &  9.9\,$\pm$\,0.2       & 2.27\,$\pm$\,0.41  & 3.40\,$\pm$\,0.43   & 10.81\,$\pm$\,1.73   & 11.50\,$\pm$\,0.20    & 0.36\,$\pm$\,0.07     & 8.80\,$\pm$\,0.15    & -0.13\,$\pm$\,0.23      &   0.22\,$\pm$\,0.04                                           & -1.46\,$\pm$\,0.11                     &  146\,$\pm$\, 31     \\ 
 NGC\,2281             & 06:48:25                        &  41:01:30                         &  8.7\,$\pm$\,0.1       & 1.74\,$\pm$\,0.18  & 3.32\,$\pm$\,0.69   & 12.90\,$\pm$\,4.73   &  8.38\,$\pm$\,0.25    & 0.11\,$\pm$\,0.07     & 8.65\,$\pm$\,0.20    &  0.00\,$\pm$\,0.17      &  -2.97\,$\pm$\,0.15                                           & -8.28\,$\pm$\,0.16                     &  105\,$\pm$\, 33     \\ 
 LP\,930               & 06:51:15                        & -01:48:41                         & 10.1\,$\pm$\,0.2       & 2.50\,$\pm$\,0.45  & 3.88\,$\pm$\,0.54   & 12.65\,$\pm$\,2.44   & 11.72\,$\pm$\,0.25    & 0.35\,$\pm$\,0.05     & 9.20\,$\pm$\,0.10    & -0.06\,$\pm$\,0.20      &  -0.92\,$\pm$\,0.08                                           &  3.00\,$\pm$\,0.07                     &  190\,$\pm$\, 45     \\ 
 NGC\,2301             & 06:51:49                        &  00:29:54                         &  8.9\,$\pm$\,0.1       & 2.37\,$\pm$\,0.43  & 2.83\,$\pm$\,0.40   &  7.53\,$\pm$\,1.48   &  9.60\,$\pm$\,0.25    & 0.07\,$\pm$\,0.06     & 8.30\,$\pm$\,0.20    &  0.05\,$\pm$\,0.15      &  -1.36\,$\pm$\,0.12                                           & -2.18\,$\pm$\,0.11                     &  113\,$\pm$\, 24     \\ 
 NGC\,2309             & 06:56:04                        & -07:10:43                         & 10.0\,$\pm$\,0.2       & 1.87\,$\pm$\,0.53  & 3.01\,$\pm$\,0.54   & 10.14\,$\pm$\,2.18   & 11.65\,$\pm$\,0.25    & 0.43\,$\pm$\,0.06     & 8.60\,$\pm$\,0.15    & -0.13\,$\pm$\,0.23      &  -0.53\,$\pm$\,0.02                                           & -0.63\,$\pm$\,0.04                     &  135\,$\pm$\, 37     \\ 
 Berkeley\,32          & 06:58:06                        &  06:25:34                         & 11.2\,$\pm$\,0.3       & 3.01\,$\pm$\,0.80  & 4.03\,$\pm$\,0.56   & 11.67\,$\pm$\,1.39   & 12.55\,$\pm$\,0.20    & 0.18\,$\pm$\,0.04     & 9.70\,$\pm$\,0.10    & -0.32\,$\pm$\,0.24      &  -0.36\,$\pm$\,0.11                                           & -1.58\,$\pm$\,0.11                     &  356\,$\pm$\, 78     \\ 
 Tombaugh\,1           & 07:00:26                        & -20:34:44                         &  9.8\,$\pm$\,0.2       & 1.91\,$\pm$\,0.34  & 4.24\,$\pm$\,0.88   & 18.82\,$\pm$\,6.58   & 11.85\,$\pm$\,0.30    & 0.26\,$\pm$\,0.07     & 9.10\,$\pm$\,0.10    & -0.13\,$\pm$\,0.23      &  -2.56\,$\pm$\,0.06                                           &  3.84\,$\pm$\,0.06                     &  293\,$\pm$\, 93     \\ 
 NGC\,2324             & 07:04:06                        &  01:03:18                         & 11.5\,$\pm$\,0.4       & 2.91\,$\pm$\,0.54  & 6.46\,$\pm$\,1.21   & 28.64\,$\pm$\,8.61   & 12.84\,$\pm$\,0.30    & 0.25\,$\pm$\,0.06     & 8.75\,$\pm$\,0.10    & -0.22\,$\pm$\,0.28      &  -0.34\,$\pm$\,0.06                                           & -0.07\,$\pm$\,0.08                     &  667\,$\pm$\,189     \\ 
 NGC\,2335             & 07:06:46                        & -10:00:29                         &  9.3\,$\pm$\,0.2       & 3.27\,$\pm$\,0.49  & 4.23\,$\pm$\,0.51   & 11.91\,$\pm$\,2.00   & 10.82\,$\pm$\,0.30    & 0.52\,$\pm$\,0.10     & 8.15\,$\pm$\,0.25    & -0.06\,$\pm$\,0.26      &  -0.79\,$\pm$\,0.06                                           & -0.62\,$\pm$\,0.07                     &  171\,$\pm$\, 33     \\ 
 NGC\,2354             & 07:13:57                        & -25:43:43                         &  8.9\,$\pm$\,0.1       & 2.83\,$\pm$\,0.86  & 2.94\,$\pm$\,0.48   &  7.03\,$\pm$\,1.14   & 10.40\,$\pm$\,0.30    & 0.17\,$\pm$\,0.08     & 9.20\,$\pm$\,0.15    & -0.13\,$\pm$\,0.23      &  -2.87\,$\pm$\,0.07                                           &  1.86\,$\pm$\,0.09                     &  127\,$\pm$\, 33     \\ 
 NGC\,2355             & 07:17:02                        &  13:45:46                         &  9.8\,$\pm$\,0.2       & 1.58\,$\pm$\,0.26  & 3.67\,$\pm$\,0.68   & 16.94\,$\pm$\,5.16   & 11.22\,$\pm$\,0.25    & 0.11\,$\pm$\,0.07     & 9.10\,$\pm$\,0.10    &  0.00\,$\pm$\,0.17      &  -3.84\,$\pm$\,0.04                                           & -1.06\,$\pm$\,0.06                     &  198\,$\pm$\, 56     \\ 
 Haffner\,5            & 07:17:58                        & -22:38:58                         &  9.2\,$\pm$\,0.1       & 1.26\,$\pm$\,0.21  & 3.55\,$\pm$\,0.69   & 19.59\,$\pm$\,6.43   & 11.03\,$\pm$\,0.20    & 0.24\,$\pm$\,0.04     & 9.70\,$\pm$\,0.10    &  0.00\,$\pm$\,0.17      &   1.00\,$\pm$\,0.07                                           & -4.04\,$\pm$\,0.09                     &  239\,$\pm$\, 73     \\ 
 Melotte\,66           & 07:26:22                        & -47:41:28                         & 10.1\,$\pm$\,0.2       & 5.56\,$\pm$\,1.12  & 8.40\,$\pm$\,0.97   & 26.87\,$\pm$\,2.88   & 13.29\,$\pm$\,0.20    & 0.17\,$\pm$\,0.06     & 9.60\,$\pm$\,0.10    & -0.22\,$\pm$\,0.19      &  -1.44\,$\pm$\,0.10                                           &  2.77\,$\pm$\,0.10                     & 1440\,$\pm$\,255     \\ 
 NGC\,2396             & 07:27:55                        & -11:45:24                         &  9.2\,$\pm$\,0.2       & 6.17\,$\pm$\,1.40  & 5.08\,$\pm$\,0.68   & 10.24\,$\pm$\,1.59   & 10.75\,$\pm$\,0.25    & 0.17\,$\pm$\,0.07     & 8.25\,$\pm$\,0.30    & -0.06\,$\pm$\,0.20      &  -1.47\,$\pm$\,0.04                                           & -0.22\,$\pm$\,0.04                     &  193\,$\pm$\, 42     \\ 
 NGC\,2414             & 07:33:09                        & -15:26:38                         & 11.7\,$\pm$\,0.4       & 2.58\,$\pm$\,0.54  & 3.36\,$\pm$\,0.42   &  9.52\,$\pm$\,1.31   & 13.35\,$\pm$\,0.25    & 0.68\,$\pm$\,0.07     & 7.10\,$\pm$\,0.15    & -0.13\,$\pm$\,0.23      &  -1.41\,$\pm$\,0.07                                           &  1.44\,$\pm$\,0.01                     &  138\,$\pm$\, 29     \\ 
 NGC\,2423             & 07:37:15                        & -13:53:35                         &  8.8\,$\pm$\,0.1       & 2.13\,$\pm$\,0.49  & 4.13\,$\pm$\,0.74   & 16.21\,$\pm$\,4.17   &  9.63\,$\pm$\,0.20    & 0.09\,$\pm$\,0.04     & 9.10\,$\pm$\,0.10    &  0.00\,$\pm$\,0.11      &  -0.75\,$\pm$\,0.10                                           & -3.59\,$\pm$\,0.11                     &  217\,$\pm$\, 60     \\ 
 NGC\,2425             & 07:38:20                        & -14:53:13                         & 10.1\,$\pm$\,0.3       & 2.54\,$\pm$\,0.51  & 3.72\,$\pm$\,0.45   & 11.60\,$\pm$\,1.50   & 12.32\,$\pm$\,0.30    & 0.28\,$\pm$\,0.05     & 9.50\,$\pm$\,0.10    &  0.05\,$\pm$\,0.15      &  -3.57\,$\pm$\,0.03                                           &  2.01\,$\pm$\,0.04                     &  240\,$\pm$\, 47     \\ 
 Melotte\,72           & 07:38:27                        & -10:41:46                         & 10.0\,$\pm$\,0.2       & 1.91\,$\pm$\,0.46  & 3.51\,$\pm$\,0.54   & 13.23\,$\pm$\,2.40   & 11.93\,$\pm$\,0.20    & 0.17\,$\pm$\,0.05     & 9.05\,$\pm$\,0.10    & -0.13\,$\pm$\,0.16      &  -4.16\,$\pm$\,0.02                                           &  3.69\,$\pm$\,0.05                     &  167\,$\pm$\, 41     \\ 
 NGC\,2420             & 07:38:27                        &  21:35:06                         & 10.4\,$\pm$\,0.2       & 2.10\,$\pm$\,0.47  & 4.05\,$\pm$\,0.65   & 15.95\,$\pm$\,3.37   & 11.83\,$\pm$\,0.20    & 0.03\,$\pm$\,0.04     & 9.40\,$\pm$\,0.05    & -0.13\,$\pm$\,0.23      &  -1.23\,$\pm$\,0.10                                           & -2.04\,$\pm$\,0.02                     &  299\,$\pm$\, 73     \\ 
 NGC\,2428             & 07:39:23                        & -16:32:24                         &  9.1\,$\pm$\,0.1       & 2.49\,$\pm$\,0.42  & 3.44\,$\pm$\,0.68   & 10.20\,$\pm$\,3.28   & 10.60\,$\pm$\,0.25    & 0.09\,$\pm$\,0.05     & 8.80\,$\pm$\,0.10    &  0.14\,$\pm$\,0.12      &  -3.30\,$\pm$\,0.08                                           &  2.49\,$\pm$\,0.05                     &  129\,$\pm$\, 40     \\ 
 NGC\,2439             & 07:40:49                        & -31:42:05                         & 10.0\,$\pm$\,0.2       & 3.29\,$\pm$\,0.71  & 4.66\,$\pm$\,0.55   & 14.12\,$\pm$\,1.50   & 12.55\,$\pm$\,0.30    & 0.46\,$\pm$\,0.07     & 7.30\,$\pm$\,0.20    & -0.13\,$\pm$\,0.23      &  -2.25\,$\pm$\,0.06                                           &  3.18\,$\pm$\,0.05                     &  312\,$\pm$\, 57     \\ 
 Haffner\,13           & 07:41:01                        & -30:03:47                         &  8.5\,$\pm$\,0.1       & 2.89\,$\pm$\,0.42  & 4.37\,$\pm$\,0.75   & 13.97\,$\pm$\,3.87   &  8.74\,$\pm$\,0.20    & 0.06\,$\pm$\,0.05     & 7.50\,$\pm$\,0.20    &  0.10\,$\pm$\,0.18      &  -6.19\,$\pm$\,0.19                                           &  5.90\,$\pm$\,0.13                     &  129\,$\pm$\, 34     \\ 
 NGC\,2437             & 07:41:49                        & -14:50:08                         &  9.2\,$\pm$\,0.2       & 3.87\,$\pm$\,0.63  & 6.47\,$\pm$\,1.02   & 22.50\,$\pm$\,5.47   & 10.80\,$\pm$\,0.25    & 0.19\,$\pm$\,0.06     & 8.65\,$\pm$\,0.10    & -0.13\,$\pm$\,0.31      &  -3.86\,$\pm$\,0.12                                           &  0.40\,$\pm$\,0.12                     &  725\,$\pm$\,172     \\ 
 NGC\,2447             & 07:44:27                        & -23:52:41                         &  8.8\,$\pm$\,0.1       & 1.88\,$\pm$\,0.40  & 3.81\,$\pm$\,0.81   & 15.61\,$\pm$\,5.33   &  9.95\,$\pm$\,0.25    & 0.05\,$\pm$\,0.05     & 8.80\,$\pm$\,0.10    & -0.06\,$\pm$\,0.20      &  -3.57\,$\pm$\,0.12                                           &  5.09\,$\pm$\,0.12                     &  198\,$\pm$\, 64     \\ 
 Berkeley\,39          & 07:46:48                        & -04:40:20                         & 11.4\,$\pm$\,0.2       & 3.51\,$\pm$\,0.70  & 6.16\,$\pm$\,0.73   & 22.32\,$\pm$\,2.63   & 13.02\,$\pm$\,0.15    & 0.16\,$\pm$\,0.04     & 9.80\,$\pm$\,0.10    & -0.13\,$\pm$\,0.16      &  -1.72\,$\pm$\,0.10                                           & -1.63\,$\pm$\,0.04                     &  848\,$\pm$\,159     \\ 
 NGC\,2482             & 07:55:07                        & -24:16:07                         &  8.9\,$\pm$\,0.1       & 3.69\,$\pm$\,0.72  & 3.68\,$\pm$\,0.47   &  8.52\,$\pm$\,1.37   & 10.45\,$\pm$\,0.25    & 0.08\,$\pm$\,0.06     & 8.70\,$\pm$\,0.10    &  0.00\,$\pm$\,0.17      &  -4.67\,$\pm$\,0.08                                           &  2.26\,$\pm$\,0.06                     &  116\,$\pm$\, 27     \\ 
 NGC\,2489             & 07:56:16                        & -30:03:49                         &  9.1\,$\pm$\,0.1       & 1.08\,$\pm$\,0.18  & 2.27\,$\pm$\,0.48   &  9.58\,$\pm$\,3.41   & 11.24\,$\pm$\,0.25    & 0.53\,$\pm$\,0.07     & 8.55\,$\pm$\,0.10    &  0.00\,$\pm$\,0.17      &  -2.52\,$\pm$\,0.07                                           &  2.17\,$\pm$\,0.06                     &   96\,$\pm$\, 31     \\ 
 NGC\,2506             & 08:00:05                        & -10:46:31                         & 10.4\,$\pm$\,0.2       & 2.97\,$\pm$\,0.45  & 6.39\,$\pm$\,1.22   & 27.60\,$\pm$\,8.86   & 12.45\,$\pm$\,0.20    & 0.07\,$\pm$\,0.05     & 9.35\,$\pm$\,0.05    & -0.22\,$\pm$\,0.28      &  -2.56\,$\pm$\,0.03                                           &  3.97\,$\pm$\,0.05                     &  913\,$\pm$\,262     \\ 
 NGC\,2509             & 08:00:48                        & -19:03:07                         &  9.7\,$\pm$\,0.2       & 1.70\,$\pm$\,0.34  & 3.86\,$\pm$\,0.90   & 17.51\,$\pm$\,6.90   & 11.84\,$\pm$\,0.20    & 0.07\,$\pm$\,0.04     & 9.20\,$\pm$\,0.10    &  0.22\,$\pm$\,0.10      &  -2.72\,$\pm$\,0.03                                           &  0.81\,$\pm$\,0.06                     &  240\,$\pm$\, 85     \\ 
 NGC\,2548             & 08:13:29                        & -05:41:34                         &  8.7\,$\pm$\,0.1       & 3.61\,$\pm$\,0.59  & 5.25\,$\pm$\,0.85   & 16.25\,$\pm$\,4.07   &  9.32\,$\pm$\,0.25    & 0.06\,$\pm$\,0.06     & 8.70\,$\pm$\,0.10    &  0.05\,$\pm$\,0.15      &  -1.29\,$\pm$\,0.12                                           &  1.04\,$\pm$\,0.15                     &  284\,$\pm$\, 71     \\ 
 NGC\,2571             & 08:18:59                        & -29:45:20                         &  8.8\,$\pm$\,0.1       & 2.57\,$\pm$\,0.63  & 3.40\,$\pm$\,0.56   &  9.78\,$\pm$\,2.08   & 10.51\,$\pm$\,0.25    & 0.14\,$\pm$\,0.07     & 7.40\,$\pm$\,0.15    &  0.05\,$\pm$\,0.20      &  -4.93\,$\pm$\,0.06                                           &  4.29\,$\pm$\,0.06                     &  103\,$\pm$\, 27     \\ 
 NGC\,2627             & 08:37:16                        & -29:56:19                         &  8.9\,$\pm$\,0.1       & 1.55\,$\pm$\,0.33  & 3.71\,$\pm$\,0.63   & 17.52\,$\pm$\,4.24   & 11.05\,$\pm$\,0.15    & 0.07\,$\pm$\,0.05     & 9.35\,$\pm$\,0.10    &  0.00\,$\pm$\,0.17      &  -2.37\,$\pm$\,0.06                                           &  2.90\,$\pm$\,0.07                     &  231\,$\pm$\, 61     \\ 
 NGC\,2635             & 08:38:26                        & -34:46:00                         & 10.0\,$\pm$\,0.2       & 3.13\,$\pm$\,0.93  & 4.45\,$\pm$\,0.68   & 13.55\,$\pm$\,1.67   & 13.00\,$\pm$\,0.25    & 0.37\,$\pm$\,0.06     & 8.85\,$\pm$\,0.10    & -0.32\,$\pm$\,0.24      &  -2.49\,$\pm$\,0.05                                           &  2.24\,$\pm$\,0.03                     &  250\,$\pm$\, 61     \\ 
 IC\,2395              & 08:42:26                        & -48:08:57                         &  8.3\,$\pm$\,0.1       & 1.10\,$\pm$\,0.14  & 2.00\,$\pm$\,0.15   &  7.44\,$\pm$\,0.14   &  9.18\,$\pm$\,0.30    & 0.16\,$\pm$\,0.10     & 7.10\,$\pm$\,0.20    &  0.14\,$\pm$\,0.20      &  -4.46\,$\pm$\,0.10                                           &  3.33\,$\pm$\,0.11                     &   40\,$\pm$\,  5     \\ 
 NGC\,2658             & 08:43:28                        & -32:39:49                         &  9.7\,$\pm$\,0.2       & 2.43\,$\pm$\,0.58  & 4.38\,$\pm$\,0.63   & 16.23\,$\pm$\,2.43   & 12.62\,$\pm$\,0.20    & 0.35\,$\pm$\,0.06     & 8.80\,$\pm$\,0.10    & -0.32\,$\pm$\,0.24      &  -2.48\,$\pm$\,0.04                                           &  2.28\,$\pm$\,0.05                     &  323\,$\pm$\, 71     \\ 
 Ruprecht\,68          & 08:44:37                        & -35:53:41                         &  9.1\,$\pm$\,0.2       & 2.31\,$\pm$\,0.59  & 4.76\,$\pm$\,0.91   & 19.80\,$\pm$\,5.21   & 12.04\,$\pm$\,0.25    & 0.43\,$\pm$\,0.05     & 9.25\,$\pm$\,0.10    & -0.13\,$\pm$\,0.16      &  -2.61\,$\pm$\,0.06                                           &  5.66\,$\pm$\,0.05                     &  353\,$\pm$\,103     \\ 
 NGC\,2670             & 08:45:34                        & -48:48:52                         &  8.4\,$\pm$\,0.1       & 2.31\,$\pm$\,0.40  & 2.73\,$\pm$\,0.42   &  7.16\,$\pm$\,1.67   & 10.84\,$\pm$\,0.30    & 0.46\,$\pm$\,0.10     & 7.90\,$\pm$\,0.10    &  0.28\,$\pm$\,0.18      &  -5.36\,$\pm$\,0.08                                           &  3.68\,$\pm$\,0.08                     &   87\,$\pm$\, 21     \\ 
 NGC\,2669             & 08:46:30                        & -52:56:34                         &  8.3\,$\pm$\,0.1       & 1.00\,$\pm$\,0.16  & 2.04\,$\pm$\,0.47   &  8.34\,$\pm$\,3.32   & 10.23\,$\pm$\,0.30    & 0.26\,$\pm$\,0.06     & 8.05\,$\pm$\,0.20    &  0.10\,$\pm$\,0.18      &  -4.11\,$\pm$\,0.08                                           &  4.69\,$\pm$\,0.07                     &   51\,$\pm$\, 18     \\ 
 NGC\,2818             & 09:16:08                        & -36:37:29                         &  9.1\,$\pm$\,0.2       & 2.65\,$\pm$\,0.66  & 5.70\,$\pm$\,1.06   & 24.57\,$\pm$\,6.34   & 12.27\,$\pm$\,0.25    & 0.15\,$\pm$\,0.04     & 9.05\,$\pm$\,0.10    &  0.00\,$\pm$\,0.11      &  -4.42\,$\pm$\,0.07                                           &  4.55\,$\pm$\,0.06                     &  424\,$\pm$\,121     \\ 
 NGC\,2849             & 09:19:25                        & -40:31:01                         &  9.9\,$\pm$\,0.3       & 1.71\,$\pm$\,0.24  & 4.23\,$\pm$\,0.80   & 20.66\,$\pm$\,6.68   & 13.45\,$\pm$\,0.25    & 0.40\,$\pm$\,0.06     & 9.10\,$\pm$\,0.10    & -0.06\,$\pm$\,0.13      &  -4.25\,$\pm$\,0.06                                           &  3.55\,$\pm$\,0.06                     &  282\,$\pm$\, 83     \\ 
 IC\,2488              & 09:27:24                        & -56:58:29                         &  8.2\,$\pm$\,0.1       & 3.21\,$\pm$\,0.43  & 4.57\,$\pm$\,0.67   & 13.91\,$\pm$\,3.23   & 10.47\,$\pm$\,0.25    & 0.32\,$\pm$\,0.07     & 8.15\,$\pm$\,0.15    &  0.00\,$\pm$\,0.17      &  -7.74\,$\pm$\,0.09                                           &  5.75\,$\pm$\,0.09                     &  234\,$\pm$\, 53     \\ 
 Trumpler\,12          & 10:06:29                        & -60:17:31                         &  8.1\,$\pm$\,0.2       & 2.37\,$\pm$\,0.46  & 4.28\,$\pm$\,0.73   & 15.87\,$\pm$\,4.10   & 12.48\,$\pm$\,0.30    & 0.22\,$\pm$\,0.06     & 8.90\,$\pm$\,0.10    &  0.05\,$\pm$\,0.15      &  -6.89\,$\pm$\,0.05                                           &  4.46\,$\pm$\,0.06                     &  257\,$\pm$\, 67     \\ 
 NGC\,3293             & 10:35:47                        & -58:14:28                         &  7.9\,$\pm$\,0.1       & 1.36\,$\pm$\,0.26  & 3.07\,$\pm$\,0.48   & 13.81\,$\pm$\,3.11   & 11.74\,$\pm$\,0.25    & 0.33\,$\pm$\,0.07     & 7.10\,$\pm$\,0.15    &  0.00\,$\pm$\,0.23      &  -7.63\,$\pm$\,0.14                                           &  3.36\,$\pm$\,0.11                     &  162\,$\pm$\, 39     \\ 
 Melotte\,101          & 10:42:04                        & -65:05:48                         &  7.8\,$\pm$\,0.2       & 4.60\,$\pm$\,0.94  & 4.04\,$\pm$\,0.43   &  8.54\,$\pm$\,0.81   & 11.40\,$\pm$\,0.30    & 0.39\,$\pm$\,0.08     & 8.35\,$\pm$\,0.20    &  0.05\,$\pm$\,0.20      &  -6.33\,$\pm$\,0.07                                           &  3.49\,$\pm$\,0.09                     &  247\,$\pm$\, 41     \\ 
 Ruprecht\,91          & 10:47:39                        & -57:27:39                         &  8.0\,$\pm$\,0.1       & 2.94\,$\pm$\,0.44  & 4.70\,$\pm$\,1.02   & 15.74\,$\pm$\,5.82   & 10.00\,$\pm$\,0.25    & 0.10\,$\pm$\,0.07     & 8.00\,$\pm$\,0.15    &  0.05\,$\pm$\,0.15      & -11.21\,$\pm$\,0.07                                           &  2.45\,$\pm$\,0.11                     &  173\,$\pm$\, 58     \\ 
 NGC\,3496             & 10:59:35                        & -60:20:05                         &  7.8\,$\pm$\,0.2       & 2.36\,$\pm$\,0.50  & 4.00\,$\pm$\,0.93   & 14.05\,$\pm$\,5.34   & 11.65\,$\pm$\,0.35    & 0.43\,$\pm$\,0.07     & 8.85\,$\pm$\,0.10    &  0.18\,$\pm$\,0.15      &  -7.42\,$\pm$\,0.08                                           &  2.74\,$\pm$\,0.11                     &  320\,$\pm$\,112     \\ 
 Trumpler\,19          & 11:14:25                        & -57:33:25                         &  7.7\,$\pm$\,0.1       & 2.67\,$\pm$\,0.48  & 4.43\,$\pm$\,0.73   & 15.30\,$\pm$\,3.84   & 11.75\,$\pm$\,0.15    & 0.20\,$\pm$\,0.05     & 9.60\,$\pm$\,0.10    &  0.05\,$\pm$\,0.15      &  -1.66\,$\pm$\,0.06                                           & -1.22\,$\pm$\,0.06                     &  419\,$\pm$\,106     \\ 
 Alessi Teutsch\,8     & 12:02:19                        & -60:52:54                         &  7.9\,$\pm$\,0.1       & 3.31\,$\pm$\,0.55  & 4.50\,$\pm$\,0.66   & 13.21\,$\pm$\,2.90   &  9.83\,$\pm$\,0.25    & 0.22\,$\pm$\,0.07     & 8.20\,$\pm$\,0.15    &  0.05\,$\pm$\,0.20      &  -6.64\,$\pm$\,0.08                                           &  1.67\,$\pm$\,0.10                     &  178\,$\pm$\, 41     \\

\hline

\end{tabular}

\end{minipage}
\end{sideways}
\label{tab:investig_sample}
\end{table*}

\begin{table*}
\centering
\tiny
%\rotcaption{ Central coordinates, Galactocentric distances, structural and fundamental parameters, mean proper motion components and half-light relaxation times ($t_{rh}$) for the studied sample. }
\contcaption{          } 
\begin{sideways}
\begin{minipage}{225mm}

\begin{tabular}{lcrrccrrccrrrr}

 Cluster               & $RA$                            & $DEC$                           & R$_\textrm{G}^{(*)}$    & $r_c$              & $r_{h}$             & $r_t$                & $(m-M)_0$             & $E(B-V)$              & log\,$t$             & $[Fe/H]^{(**)}$         & $\langle\mu_{\alpha}\,\textrm{cos}\,\delta\rangle^{\dag}$    & $\langle\mu_{\delta}\rangle^{\dag}$     & $t_{rh}$   \\    
                       &($h$:$m$:$s$)                    & ($^{\circ}$:$'$:$''$)           &  (kpc)                  &  (pc)              &     (pc)            &  (pc)                & (mag)                 & (mag)                 & (dex)                & (dex)                   & (mas yr$^{-1}$)                                              & (mas yr$^{-1}$)                         & (Myr)      \\

\hline                                                                                                                                                                                                                                                                                                                                                                                                                              
                                                                                                                     
 NGC\,4103             & 12:06:31                        & -61:15:46                       & 7.6\,$\pm$\,0.2         & 2.0\,$\pm$\,0.3    &4.9\,$\pm$\,1.2     &23.3\,$\pm$\,10.1     &11.3\,$\pm$\,0.3      &0.38\,$\pm$\,0.08     &7.45\,$\pm$\,0.30    & 0.10\,$\pm$\,0.23      &-6.19\,$\pm$\,0.14                                           &  0.21\,$\pm$\,0.14                   &  322\,$\pm$\,121     \\ 
 NGC\,4349             & 12:24:23                        & -61:51:29                       & 7.6\,$\pm$\,0.2         & 3.7\,$\pm$\,0.7    &3.8\,$\pm$\,0.4     & 8.8\,$\pm$\, 1.2     &11.0\,$\pm$\,0.2      &0.40\,$\pm$\,0.05     &8.60\,$\pm$\,0.15    &-0.06\,$\pm$\,0.20      &-7.84\,$\pm$\,0.10                                           & -0.26\,$\pm$\,0.10                   &  252\,$\pm$\, 44     \\ 
 Trumpler\,20          & 12:39:28                        & -60:38:40                       & 7.1\,$\pm$\,0.3         & 4.0\,$\pm$\,0.5    &8.2\,$\pm$\,1.0     &34.4\,$\pm$\, 6.7     &12.5\,$\pm$\,0.3      &0.45\,$\pm$\,0.06     &9.15\,$\pm$\,0.10    & 0.05\,$\pm$\,0.15      &-7.11\,$\pm$\,0.10                                           &  0.20\,$\pm$\,0.09                   & 1544\,$\pm$\,292     \\ 
 NGC\,4609             & 12:42:22                        & -62:58:54                       & 7.6\,$\pm$\,0.2         & 1.9\,$\pm$\,0.4    &3.4\,$\pm$\,0.8     &12.2\,$\pm$\, 4.4     &10.6\,$\pm$\,0.3      &0.43\,$\pm$\,0.07     &7.90\,$\pm$\,0.20    & 0.05\,$\pm$\,0.25      &-4.90\,$\pm$\,0.06                                           & -1.01\,$\pm$\,0.09                   &  132\,$\pm$\, 45     \\ 
 UBC\,290              & 12:47:30                        & -59:24:40                       & 7.5\,$\pm$\,0.2         & 5.3\,$\pm$\,0.6    &7.5\,$\pm$\,1.0     &22.7\,$\pm$\, 4.8     &11.0\,$\pm$\,0.2      &0.27\,$\pm$\,0.07     &7.85\,$\pm$\,0.15    & 0.22\,$\pm$\,0.14      &-5.93\,$\pm$\,0.06                                           & -0.23\,$\pm$\,0.07                   &  404\,$\pm$\, 86     \\ 
 NGC\,4755             & 12:53:41                        & -60:22:43                       & 7.4\,$\pm$\,0.2         & 1.4\,$\pm$\,0.3    &2.9\,$\pm$\,0.5     &12.0\,$\pm$\, 2.6     &11.4\,$\pm$\,0.2      &0.44\,$\pm$\,0.07     &7.20\,$\pm$\,0.15    & 0.05\,$\pm$\,0.20      &-4.71\,$\pm$\,0.12                                           & -1.06\,$\pm$\,0.13                   &  159\,$\pm$\, 38     \\ 
 NGC\,4852             & 13:00:05                        & -59:35:22                       & 7.6\,$\pm$\,0.1         & 2.1\,$\pm$\,0.5    &2.7\,$\pm$\,0.4     & 7.5\,$\pm$\, 1.6     &10.4\,$\pm$\,0.2      &0.44\,$\pm$\,0.08     &8.15\,$\pm$\,0.25    & 0.18\,$\pm$\,0.19      &-8.15\,$\pm$\,0.08                                           & -1.15\,$\pm$\,0.07                   &   83\,$\pm$\, 21     \\ 
 Collinder\,272        & 13:30:15                        & -61:19:08                       & 7.2\,$\pm$\,0.2         & 2.2\,$\pm$\,0.4    &4.7\,$\pm$\,1.0     &20.5\,$\pm$\, 7.0     &11.4\,$\pm$\,0.3      &0.54\,$\pm$\,0.08     &7.60\,$\pm$\,0.25    &-0.06\,$\pm$\,0.26      &-3.47\,$\pm$\,0.09                                           & -1.77\,$\pm$\,0.11                   &  275\,$\pm$\, 88     \\ 
 Pismis\,18            & 13:36:57                        & -62:05:20                       & 7.1\,$\pm$\,0.3         & 1.0\,$\pm$\,0.3    &1.7\,$\pm$\,0.3     & 6.0\,$\pm$\, 0.9     &11.7\,$\pm$\,0.3      &0.70\,$\pm$\,0.08     &8.85\,$\pm$\,0.10    & 0.05\,$\pm$\,0.25      &-5.70\,$\pm$\,0.07                                           & -2.27\,$\pm$\,0.06                   &   62\,$\pm$\, 15     \\ 
 Collinder\,277        & 13:47:55                        & -66:04:49                       & 7.4\,$\pm$\,0.1         & 2.0\,$\pm$\,0.3    &3.2\,$\pm$\,0.6     &10.6\,$\pm$\, 3.2     &10.8\,$\pm$\,0.2      &0.21\,$\pm$\,0.05     &9.00\,$\pm$\,0.10    & 0.18\,$\pm$\,0.11      &-9.25\,$\pm$\,0.08                                           & -4.11\,$\pm$\,0.06                   &  135\,$\pm$\, 40     \\ 
 NGC\,5381             & 14:00:47                        & -59:34:25                       & 7.0\,$\pm$\,0.3         & 3.0\,$\pm$\,0.6    &4.0\,$\pm$\,0.5     &11.7\,$\pm$\, 1.5     &11.6\,$\pm$\,0.3      &0.52\,$\pm$\,0.07     &8.75\,$\pm$\,0.10    & 0.18\,$\pm$\,0.15      &-6.07\,$\pm$\,0.07                                           & -2.92\,$\pm$\,0.07                   &  285\,$\pm$\, 50     \\ 
 NGC\,5822             & 15:04:16                        & -54:21:13                       & 7.7\,$\pm$\,0.1         & 2.7\,$\pm$\,0.4    &4.8\,$\pm$\,0.8     &17.4\,$\pm$\, 4.8     & 9.3\,$\pm$\,0.2      &0.15\,$\pm$\,0.05     &9.05\,$\pm$\,0.10    &-0.06\,$\pm$\,0.20      &-7.48\,$\pm$\,0.13                                           & -5.50\,$\pm$\,0.13                   &  306\,$\pm$\, 81     \\ 
 NGC\,5823             & 15:05:26                        & -55:37:19                       & 7.0\,$\pm$\,0.2         & 2.2\,$\pm$\,0.4    &3.5\,$\pm$\,0.7     &11.3\,$\pm$\, 3.6     &11.1\,$\pm$\,0.3      &0.70\,$\pm$\,0.07     &8.20\,$\pm$\,0.15    & 0.00\,$\pm$\,2.29      &-3.69\,$\pm$\,0.11                                           & -2.41\,$\pm$\,0.09                   &  206\,$\pm$\, 62     \\ 
 NGC\,5925             & 15:27:16                        & -54:30:18                       & 7.2\,$\pm$\,0.2         & 2.6\,$\pm$\,0.5    &3.9\,$\pm$\,0.6     &12.3\,$\pm$\, 2.9     &10.5\,$\pm$\,0.2      &0.52\,$\pm$\,0.05     &8.70\,$\pm$\,0.10    & 0.14\,$\pm$\,0.16      &-4.35\,$\pm$\,0.08                                           & -5.12\,$\pm$\,0.08                   &  203\,$\pm$\, 49     \\ 
 NGC\,6025             & 16:03:03                        & -60:26:48                       & 7.6\,$\pm$\,0.1         & 2.0\,$\pm$\,0.3    &2.8\,$\pm$\,0.5     & 8.2\,$\pm$\, 2.6     & 9.4\,$\pm$\,0.3      &0.20\,$\pm$\,0.07     &8.20\,$\pm$\,0.30    & 0.14\,$\pm$\,0.20      &-2.98\,$\pm$\,0.13                                           & -3.07\,$\pm$\,0.14                   &   88\,$\pm$\, 26     \\ 
 NGC\,6067             & 16:13:11                        & -54:14:21                       & 6.7\,$\pm$\,0.2         & 2.2\,$\pm$\,0.3    &4.8\,$\pm$\,0.8     &20.7\,$\pm$\, 5.3     &11.3\,$\pm$\,0.3      &0.39\,$\pm$\,0.06     &8.10\,$\pm$\,0.10    & 0.10\,$\pm$\,0.14      &-1.93\,$\pm$\,0.08                                           & -2.58\,$\pm$\,0.11                   &  453\,$\pm$\,111     \\ 
 NGC\,6087             & 16:18:57                        & -57:56:19                       & 7.5\,$\pm$\,0.1         & 1.5\,$\pm$\,0.3    &1.7\,$\pm$\,0.3     & 4.5\,$\pm$\, 1.3     & 9.8\,$\pm$\,0.2      &0.24\,$\pm$\,0.06     &7.90\,$\pm$\,0.15    & 0.05\,$\pm$\,0.15      &-1.62\,$\pm$\,0.11                                           & -2.41\,$\pm$\,0.11                   &   37\,$\pm$\, 11     \\ 
 NGC\,6152             & 16:32:54                        & -52:39:36                       & 7.0\,$\pm$\,0.2         & 3.1\,$\pm$\,0.7    &3.5\,$\pm$\,0.5     & 8.8\,$\pm$\, 1.7     &10.8\,$\pm$\,0.2      &0.36\,$\pm$\,0.05     &8.40\,$\pm$\,0.15    & 0.10\,$\pm$\,0.14      &-2.58\,$\pm$\,0.05                                           & -4.98\,$\pm$\,0.06                   &  146\,$\pm$\, 35     \\ 
 NGC\,6204             & 16:46:09                        & -47:02:07                       & 7.2\,$\pm$\,0.2         & 0.7\,$\pm$\,0.1    &1.7\,$\pm$\,0.4     & 8.0\,$\pm$\, 3.0     &10.2\,$\pm$\,0.3      &0.48\,$\pm$\,0.12     &8.10\,$\pm$\,0.35    & 0.10\,$\pm$\,0.23      &-0.73\,$\pm$\,0.06                                           & -0.61\,$\pm$\,0.05                   &   44\,$\pm$\, 15     \\ 
 NGC\,6208             & 16:49:19                        & -53:42:08                       & 7.3\,$\pm$\,0.1         & 1.5\,$\pm$\,0.2    &2.6\,$\pm$\,0.5     & 9.0\,$\pm$\, 2.9     &10.2\,$\pm$\,0.1      &0.26\,$\pm$\,0.04     &9.25\,$\pm$\,0.10    & 0.00\,$\pm$\,0.11      &-1.01\,$\pm$\,0.10                                           & -1.53\,$\pm$\,0.09                   &  113\,$\pm$\, 34     \\ 
 NGC\,6281             & 17:04:37                        & -37:56:53                       & 7.8\,$\pm$\,0.1         & 1.1\,$\pm$\,0.2    &1.4\,$\pm$\,0.2     & 3.9\,$\pm$\, 0.6     & 8.4\,$\pm$\,0.2      &0.20\,$\pm$\,0.06     &8.50\,$\pm$\,0.15    & 0.00\,$\pm$\,0.17      &-1.87\,$\pm$\,0.20                                           & -4.03\,$\pm$\,0.20                   &   30\,$\pm$\,  6     \\ 
 Trumpler\,25          & 17:24:33                        & -38:59:58                       & 6.4\,$\pm$\,0.4         & 2.0\,$\pm$\,0.4    &4.2\,$\pm$\,1.0     &18.3\,$\pm$\, 7.3     &11.3\,$\pm$\,0.4      &1.10\,$\pm$\,0.15     &8.05\,$\pm$\,0.20    & 0.18\,$\pm$\,0.19      & 0.33\,$\pm$\,0.05                                           & -2.11\,$\pm$\,0.11                   &  414\,$\pm$\,145     \\ 
 Trumpler\,29          & 17:41:31                        & -40:10:45                       & 6.9\,$\pm$\,0.2         & 3.1\,$\pm$\,0.6    &3.7\,$\pm$\,0.6     &10.0\,$\pm$\, 2.4     &10.7\,$\pm$\,0.2      &0.30\,$\pm$\,0.08     &7.75\,$\pm$\,0.20    & 0.18\,$\pm$\,0.15      & 0.46\,$\pm$\,0.07                                           & -2.34\,$\pm$\,0.08                   &  140\,$\pm$\, 36     \\ 
 NGC\,6416             & 17:43:58                        & -32:20:59                       & 7.2\,$\pm$\,0.2         & 3.7\,$\pm$\,0.8    &3.3\,$\pm$\,0.5     & 7.2\,$\pm$\, 1.4     &10.0\,$\pm$\,0.2      &0.32\,$\pm$\,0.06     &8.50\,$\pm$\,0.15    & 0.05\,$\pm$\,0.20      &-1.95\,$\pm$\,0.06                                           & -2.36\,$\pm$\,0.09                   &  128\,$\pm$\, 32     \\ 
 NGC\,6494             & 17:56:59                        & -18:57:43                       & 7.6\,$\pm$\,0.1         & 2.7\,$\pm$\,0.4    &3.9\,$\pm$\,0.8     &12.0\,$\pm$\, 4.1     & 9.1\,$\pm$\,0.3      &0.46\,$\pm$\,0.07     &8.55\,$\pm$\,0.10    &-0.06\,$\pm$\,0.20      & 0.31\,$\pm$\,0.15                                           & -1.83\,$\pm$\,0.15                   &  237\,$\pm$\, 73     \\ 
 NGC\,6531             & 18:04:09                        & -22:31:13                       & 7.2\,$\pm$\,0.2         & 1.4\,$\pm$\,0.3    &1.5\,$\pm$\,0.1     & 3.5\,$\pm$\, 0.3     &10.1\,$\pm$\,0.3      &0.32\,$\pm$\,0.10     &7.05\,$\pm$\,0.20    & 0.14\,$\pm$\,0.20      & 0.54\,$\pm$\,0.12                                           & -1.45\,$\pm$\,0.09                   &   27\,$\pm$\,  5     \\ 
 NGC\,6568             & 18:12:53                        & -21:36:27                       & 7.3\,$\pm$\,0.2         & 2.9\,$\pm$\,0.7    &2.4\,$\pm$\,0.4     & 5.0\,$\pm$\, 0.9     & 9.9\,$\pm$\,0.2      &0.23\,$\pm$\,0.06     &8.90\,$\pm$\,0.10    & 0.14\,$\pm$\,0.16      & 0.56\,$\pm$\,0.08                                           & -1.41\,$\pm$\,0.08                   &   73\,$\pm$\, 18     \\ 
 NGC\,6645             & 18:32:40                        & -16:54:50                       & 6.9\,$\pm$\,0.2         & 2.1\,$\pm$\,0.4    &3.1\,$\pm$\,0.5     & 9.9\,$\pm$\, 2.6     &10.8\,$\pm$\,0.2      &0.41\,$\pm$\,0.06     &8.80\,$\pm$\,0.10    &-0.13\,$\pm$\,0.23      & 1.32\,$\pm$\,0.06                                           & -0.65\,$\pm$\,0.09                   &  164\,$\pm$\, 43     \\ 
 IC\,4756              & 18:38:45                        &  05:28:30                       & 7.9\,$\pm$\,0.1         & 3.5\,$\pm$\,0.6    &4.4\,$\pm$\,0.6     &12.2\,$\pm$\, 2.6     & 8.2\,$\pm$\,0.2      &0.22\,$\pm$\,0.07     &9.00\,$\pm$\,0.10    &-0.06\,$\pm$\,0.20      & 1.26\,$\pm$\,0.20                                           & -4.98\,$\pm$\,0.22                   &  214\,$\pm$\, 49     \\ 
 NGC\,6705             & 18:51:01                        &  -6:17:55                       & 6.6\,$\pm$\,0.2         & 2.3\,$\pm$\,0.4    &5.0\,$\pm$\,0.8     &22.5\,$\pm$\, 5.2     &11.4\,$\pm$\,0.2      &0.47\,$\pm$\,0.06     &8.45\,$\pm$\,0.10    & 0.00\,$\pm$\,0.23      &-1.55\,$\pm$\,0.15                                           & -4.16\,$\pm$\,0.18                   &  615\,$\pm$\,149     \\ 
 NGC\,6709             & 18:51:22                        &  10:21:47                       & 7.5\,$\pm$\,0.1         & 1.8\,$\pm$\,0.2    &3.1\,$\pm$\,0.5     &11.2\,$\pm$\, 2.6     &10.0\,$\pm$\,0.2      &0.34\,$\pm$\,0.06     &8.20\,$\pm$\,0.15    & 0.05\,$\pm$\,0.20      & 1.44\,$\pm$\,0.09                                           & -3.54\,$\pm$\,0.07                   &  109\,$\pm$\, 24     \\ 
 NGC\,6728             & 18:58:52                        & -08:58:14                       & 6.8\,$\pm$\,0.2         & 2.6\,$\pm$\,0.4    &3.4\,$\pm$\,0.4     & 9.8\,$\pm$\, 1.8     &11.1\,$\pm$\,0.2      &0.28\,$\pm$\,0.06     &8.75\,$\pm$\,0.10    & 0.05\,$\pm$\,0.15      & 1.32\,$\pm$\,0.05                                           & -1.79\,$\pm$\,0.05                   &  143\,$\pm$\, 30     \\ 
 NGC\,6793             & 19:23:12                        &  22:08:17                       & 7.9\,$\pm$\,0.1         & 0.9\,$\pm$\,0.1    &1.7\,$\pm$\,0.4     & 6.6\,$\pm$\, 2.6     & 8.7\,$\pm$\,0.2      &0.31\,$\pm$\,0.08     &8.65\,$\pm$\,0.20    & 0.22\,$\pm$\,0.17      & 3.79\,$\pm$\,0.16                                           &  3.55\,$\pm$\,0.14                   &   38\,$\pm$\, 13     \\ 
 IC\,1311              & 20:10:46                        &  41:10:52                       & 8.9\,$\pm$\,0.5         & 3.1\,$\pm$\,0.7    &6.5\,$\pm$\,1.0     &27.2\,$\pm$\, 5.3     &13.8\,$\pm$\,0.3      &0.55\,$\pm$\,0.06     &9.15\,$\pm$\,0.10    &-0.13\,$\pm$\,0.16      &-3.30\,$\pm$\,0.07                                           & -5.09\,$\pm$\,0.08                   &  868\,$\pm$\,201     \\ 
 NGC\,6939             & 20:31:47                        &  60:40:04                       & 8.6\,$\pm$\,0.1         & 2.5\,$\pm$\,0.5    &4.7\,$\pm$\,0.8     &17.9\,$\pm$\, 4.4     &11.1\,$\pm$\,0.3      &0.40\,$\pm$\,0.07     &9.20\,$\pm$\,0.10    & 0.00\,$\pm$\,0.23      &-1.82\,$\pm$\,0.09                                           & -5.46\,$\pm$\,0.09                   &  422\,$\pm$\,106     \\ 
 NGC\,6940             & 20:34:30                        &  28:16:48                       & 8.0\,$\pm$\,0.1         & 2.8\,$\pm$\,0.5    &3.8\,$\pm$\,0.5     &11.3\,$\pm$\, 1.7     & 9.7\,$\pm$\,0.2      &0.19\,$\pm$\,0.06     &9.05\,$\pm$\,0.10    & 0.00\,$\pm$\,0.17      &-1.96\,$\pm$\,0.11                                           & -9.43\,$\pm$\,0.12                   &  200\,$\pm$\, 38     \\ 
 NGC\,7082             & 21:28:46                        &  47:06:55                       & 8.3\,$\pm$\,0.1         & 2.5\,$\pm$\,0.4    &4.1\,$\pm$\,0.7     &13.7\,$\pm$\, 4.0     &10.4\,$\pm$\,0.3      &0.33\,$\pm$\,0.08     &8.00\,$\pm$\,0.15    & 0.00\,$\pm$\,0.23      &-0.29\,$\pm$\,0.06                                           & -1.18\,$\pm$\,0.07                   &  169\,$\pm$\, 46     \\ 
 NGC\,7209             & 22:05:04                        &  46:30:22                       & 8.4\,$\pm$\,0.1         & 2.5\,$\pm$\,0.5    &5.8\,$\pm$\,1.4     &27.1\,$\pm$\,11.8     &10.2\,$\pm$\,0.3      &0.20\,$\pm$\,0.07     &8.70\,$\pm$\,0.15    & 0.00\,$\pm$\,0.23      & 2.30\,$\pm$\,0.10                                           &  0.25\,$\pm$\,0.09                   &  321\,$\pm$\,122     \\ 
 NGC\,7243             & 22:14:60                        &  49:51:54                       & 8.4\,$\pm$\,0.1         & 3.6\,$\pm$\,0.5    &4.0\,$\pm$\,0.4     &10.0\,$\pm$\, 1.2     & 9.6\,$\pm$\,0.2      &0.27\,$\pm$\,0.06     &8.10\,$\pm$\,0.15    & 0.05\,$\pm$\,0.20      & 0.43\,$\pm$\,0.10                                           & -2.89\,$\pm$\,0.11                   &  151\,$\pm$\, 24     \\ 
 LP\,1800              & 22:31:24                        &  58:02:59                       & 8.6\,$\pm$\,0.1         & 3.3\,$\pm$\,0.7    &3.5\,$\pm$\,0.5     & 8.6\,$\pm$\, 1.6     &10.3\,$\pm$\,0.3      &0.56\,$\pm$\,0.08     &8.45\,$\pm$\,0.15    &-0.13\,$\pm$\,0.23      &-4.55\,$\pm$\,0.06                                           & -2.70\,$\pm$\,0.08                   &  128\,$\pm$\, 28     \\ 
 Berkeley\,98          & 22:42:42                        &  52:24:34                       & 9.5\,$\pm$\,0.2         & 1.7\,$\pm$\,0.4    &4.3\,$\pm$\,1.0     &21.6\,$\pm$\, 8.6     &12.5\,$\pm$\,0.2      &0.22\,$\pm$\,0.05     &9.50\,$\pm$\,0.10    & 0.05\,$\pm$\,0.15      &-1.35\,$\pm$\,0.05                                           & -3.26\,$\pm$\,0.03                   &  279\,$\pm$\,104     \\

\hline

\end{tabular}

$Obs.:$ The $RA$ and $DEC$ coordinates here correspond to the redetermined central coordinates, as outlined in Section~\ref{sec:struct_params}. \\
%$Obs.\,(2)$: This table presents the results for 73 OCs; see Appendix~\ref{sec:appendix_astroph_params} for the remaining 41 objects and additional parameters. \\
$^{(*)}$  For the Sun, it is assumed $R_{G,\odot}$\,=\,8.23\,$\pm$\,0.12\,kpc \citep{Leung:2023}. \\                                                                                                                                                                                                                                                                                                                                                                                                                  
$^{(**)}$ Obtained from isochrone fits, using as initial guesses the mean spectroscopic $[Fe/H]$ values for member stars, whenever available (see Section~\ref{sec:decontam_CMD_analysis}). \\                                                                                                                                                                                                                                                                                                                                                                                                                 
$^{\dag}$ The numbers after the ``$\pm$"\,signal correspond to the intrinsic (i.e., corrected for measurement uncertainties) dispersions derived from the member stars data. \\ 
\end{minipage}
\end{sideways}
%\label{tab:investig_sample_compl}
\end{table*}

\begin{table*}
\tiny
\caption{ Cluster total Mass ($M_{\textrm{clu}}$), number of stars ($N_{\textrm{clu}}$), Jacobi radius ($R_J$), initial mass estimate ($M_{\textrm{ini}}$), dissolution time ($t_{\textrm{95}}$) and limiting mass ($M_{\textrm{lim}}$) for the studied sample (parameters derived in Section~\ref{sec:analysis}).  } 

\begin{tabular}{lrrrrrc}

 Cluster            &  $M_{\textrm{clu}}$      & $N_{\textrm{clu}}$      &      $R_J$            &  $M_{\textrm{ini}}$      &   $t_{\textrm{95}}$     & $M_{\textrm{lim}}^{(*)}$  \\    
                    &  ($M_{\odot}$)           &                         &      (pc)             &  ($M_{\odot}$)           &      (Gyr)              &   ($M_{\odot}$)           \\     
										                                                                                                                                           
\hline                                                                                                                                                         
                                                                                                                                                                                                                  
 NGC\,103           &   1205\,$\pm$\,56        &    2807\,$\pm$\,169     & 10.6\,$\pm$\,1.5      &   1761\,$\pm$\,181       &   0.96\,$\pm$\,0.06     &   0.9            \\
 NGC\,436           &   1179\,$\pm$\,56        &    2625\,$\pm$\,161     & 11.9\,$\pm$\,1.0      &   1510\,$\pm$\,81        &   1.10\,$\pm$\,0.04     &   0.9            \\
 NGC\,457           &   3172\,$\pm$\,104       &    6531\,$\pm$\,276     & 16.5\,$\pm$\,1.0      &   3531\,$\pm$\,95        &   1.86\,$\pm$\,0.06     &   0.9            \\
 NGC\,581           &   1552\,$\pm$\,108       &    3287\,$\pm$\,294     & 11.8\,$\pm$\,0.6      &   1801\,$\pm$\,83        &   1.06\,$\pm$\,0.03     &   0.8            \\
 Trumpler\,2        &    420\,$\pm$\,24        &     983\,$\pm$\,64      &  6.9\,$\pm$\,0.3      &    656\,$\pm$\,45        &   0.50\,$\pm$\,0.02     &   0.5            \\
 NGC\,1039          &    720\,$\pm$\,25        &    1726\,$\pm$\,54      &  8.6\,$\pm$\,0.4      &   1262\,$\pm$\,243       &   0.79\,$\pm$\,0.09     &   0.3            \\
 NGC\,1193          &   2582\,$\pm$\,142       &    7481\,$\pm$\,489     & 24.0\,$\pm$\,3.3      &  10172\,$\pm$\,1202      &   8.51\,$\pm$\,0.61     &   0.9            \\
 Trumpler\,3        &    389\,$\pm$\,22        &     928\,$\pm$\,59      &  6.7\,$\pm$\,0.5      &    509\,$\pm$\,30        &   0.42\,$\pm$\,0.01     &   0.4            \\
 NGC\,1245          &   4296\,$\pm$\,134       &   11044\,$\pm$\,432     & 23.2\,$\pm$\,2.1      &   7938\,$\pm$\,406       &   4.70\,$\pm$\,0.23     &   0.8            \\
 NGC\,1528          &    893\,$\pm$\,35        &    2172\,$\pm$\,107     &  9.1\,$\pm$\,0.4      &   2194\,$\pm$\,265       &   1.06\,$\pm$\,0.08     &   0.6            \\
 NGC\,1664          &    906\,$\pm$\,35        &    2274\,$\pm$\,110     &  9.2\,$\pm$\,1.0      &   2832\,$\pm$\,281       &   1.19\,$\pm$\,0.07     &   0.6            \\
 Berkeley\,14       &   3066\,$\pm$\,178       &    8439\,$\pm$\,574     & 16.7\,$\pm$\,3.3      &  13533\,$\pm$\,1736      &   3.90\,$\pm$\,0.29     &   1.0            \\
 NGC\,1778          &    730\,$\pm$\,35        &    1685\,$\pm$\,103     &  9.0\,$\pm$\,0.4      &   1127\,$\pm$\,70        &   0.77\,$\pm$\,0.03     &   0.7            \\
 NGC\,1798          &   2939\,$\pm$\,129       &    7780\,$\pm$\,417     & 20.6\,$\pm$\,2.5      &   6583\,$\pm$\,463       &   4.08\,$\pm$\,0.21     &   1.0            \\
 Berkeley\,17       &   3346\,$\pm$\,174       &    9783\,$\pm$\,601     & 17.1\,$\pm$\,5.2      &  52773\,$\pm$\,9603      &   8.26\,$\pm$\,0.86     &   0.9            \\
 Berkeley\,70       &   3433\,$\pm$\,163       &    9448\,$\pm$\,535     & 20.4\,$\pm$\,3.1      &  11141\,$\pm$\,1317      &   4.84\,$\pm$\,0.37     &   1.0            \\
 NGC\,1907          &   1628\,$\pm$\,76        &    4132\,$\pm$\,246     & 11.7\,$\pm$\,0.6      &   4428\,$\pm$\,392       &   1.73\,$\pm$\,0.09     &   0.7            \\
 NGC\,1912          &   2040\,$\pm$\,54        &    4847\,$\pm$\,160     & 12.1\,$\pm$\,0.7      &   3253\,$\pm$\,191       &   1.34\,$\pm$\,0.05     &   0.6            \\
 NGC\,1960          &    772\,$\pm$\,43        &    1620\,$\pm$\,109     &  8.9\,$\pm$\,0.4      &    875\,$\pm$\,30        &   0.64\,$\pm$\,0.01     &   0.6            \\
 NGC\,2194          &   7040\,$\pm$\,158       &   18235\,$\pm$\,536     & 21.5\,$\pm$\,1.8      &  11470\,$\pm$\,431       &   3.64\,$\pm$\,0.13     &   0.9            \\
 NGC\,2192          &   1548\,$\pm$\,85        &    4110\,$\pm$\,279     & 17.5\,$\pm$\,0.9      &   3380\,$\pm$\,136       &   3.77\,$\pm$\,0.11     &   0.8            \\
 NGC\,2236          &   2820\,$\pm$\,107       &    7204\,$\pm$\,342     & 14.8\,$\pm$\,1.4      &   6611\,$\pm$\,490       &   2.33\,$\pm$\,0.11     &   0.8            \\
 Trumpler\,5        &  26339\,$\pm$\,391       &   73840\,$\pm$\,1326    & 32.3\,$\pm$\,2.7      &  57718\,$\pm$\,3544      &   8.95\,$\pm$\,0.39     &   0.9            \\
 NGC\,2266          &   2007\,$\pm$\,89        &    5174\,$\pm$\,289     & 20.0\,$\pm$\,1.7      &   3842\,$\pm$\,219       &   3.79\,$\pm$\,0.18     &   0.8            \\
 Collinder\,115     &   1445\,$\pm$\,55        &    3520\,$\pm$\,172     & 11.3\,$\pm$\,0.5      &   1834\,$\pm$\,92        &   1.03\,$\pm$\,0.03     &   0.7            \\
 NGC\,2286          &    890\,$\pm$\,63        &    2306\,$\pm$\,210     &  9.7\,$\pm$\,1.1      &   2626\,$\pm$\,384       &   1.26\,$\pm$\,0.11     &   0.7            \\
 NGC\,2281          &    448\,$\pm$\,19        &    1085\,$\pm$\,44      &  7.5\,$\pm$\,0.5      &   1391\,$\pm$\,286       &   0.84\,$\pm$\,0.10     &   0.3            \\
 LP\,930            &   1022\,$\pm$\,85        &    2709\,$\pm$\,270     &  9.9\,$\pm$\,2.1      &   6841\,$\pm$\,998       &   2.03\,$\pm$\,0.17     &   0.8            \\
 NGC\,2301          &    995\,$\pm$\,33        &    2516\,$\pm$\,98      &  9.3\,$\pm$\,0.4      &   1670\,$\pm$\,155       &   0.88\,$\pm$\,0.05     &   0.4            \\
 NGC\,2309          &   1262\,$\pm$\,51        &    3210\,$\pm$\,171     & 10.9\,$\pm$\,1.2      &   2548\,$\pm$\,242       &   1.24\,$\pm$\,0.07     &   0.8            \\
 Berkeley\,32       &   3898\,$\pm$\,180       &   11297\,$\pm$\,619     & 19.4\,$\pm$\,5.7      &  26885\,$\pm$\,4019      &   6.43\,$\pm$\,0.55     &   0.7            \\
 Tombaugh\,1        &   2332\,$\pm$\,81        &    6166\,$\pm$\,273     & 16.0\,$\pm$\,2.5      &   6173\,$\pm$\,539       &   2.88\,$\pm$\,0.17     &   0.7            \\
 NGC\,2324          &   4271\,$\pm$\,100       &   10855\,$\pm$\,329     & 20.6\,$\pm$\,1.6      &   6874\,$\pm$\,258       &   3.33\,$\pm$\,0.14     &   0.8            \\
 NGC\,2335          &    731\,$\pm$\,37        &    1697\,$\pm$\,111     &  8.1\,$\pm$\,0.3      &   1137\,$\pm$\,116       &   0.75\,$\pm$\,0.05     &   0.7            \\
 NGC\,2354          &   1044\,$\pm$\,73        &    2779\,$\pm$\,237     & 10.2\,$\pm$\,0.7      &   6092\,$\pm$\,1283      &   2.14\,$\pm$\,0.26     &   0.6            \\
 NGC\,2355          &   1439\,$\pm$\,59        &    3828\,$\pm$\,201     & 14.6\,$\pm$\,1.3      &   4137\,$\pm$\,387       &   2.64\,$\pm$\,0.15     &   0.6            \\
 Haffner\,5         &   2294\,$\pm$\,146       &    6562\,$\pm$\,494     & 12.7\,$\pm$\,3.7      &  37339\,$\pm$\,6658      &   5.18\,$\pm$\,0.54     &   0.6            \\
 Melotte\,66        &   8674\,$\pm$\,228       &   24722\,$\pm$\,784     & 31.4\,$\pm$\,6.4      &  21555\,$\pm$\,1811      &  10.57\,$\pm$\,0.82     &   0.9            \\
 NGC\,2396          &    425\,$\pm$\,24        &    1023\,$\pm$\,73      &  6.8\,$\pm$\,0.3      &    790\,$\pm$\,146       &   0.61\,$\pm$\,0.07     &   0.6            \\
 NGC\,2414          &   1581\,$\pm$\,101       &    3105\,$\pm$\,267     & 13.9\,$\pm$\,2.4      &   1687\,$\pm$\,85        &   1.24\,$\pm$\,0.06     &   1.2            \\
 NGC\,2423          &   1201\,$\pm$\,52        &    3114\,$\pm$\,174     &  9.7\,$\pm$\,1.1      &   6153\,$\pm$\,802       &   1.79\,$\pm$\,0.14     &   0.4            \\
 NGC\,2425          &   2120\,$\pm$\,114       &    5825\,$\pm$\,379     & 13.2\,$\pm$\,2.4      &  13638\,$\pm$\,2146      &   3.55\,$\pm$\,0.31     &   0.8            \\
 Melotte\,72        &   1119\,$\pm$\,61        &    2922\,$\pm$\,203     & 11.8\,$\pm$\,2.2      &   4020\,$\pm$\,449       &   1.95\,$\pm$\,0.13     &   0.7            \\
 NGC\,2420          &   2667\,$\pm$\,86        &    7376\,$\pm$\,301     & 21.0\,$\pm$\,3.9      &   7666\,$\pm$\,455       &   5.43\,$\pm$\,0.27     &   0.6            \\
 NGC\,2428          &    662\,$\pm$\,41        &    1649\,$\pm$\,133     &  8.2\,$\pm$\,0.7      &   2364\,$\pm$\,259       &   1.08\,$\pm$\,0.07     &   0.6            \\
 NGC\,2439          &   3440\,$\pm$\,110       &    7065\,$\pm$\,294     & 18.2\,$\pm$\,1.3      &   3788\,$\pm$\,102       &   2.21\,$\pm$\,0.09     &   1.0            \\
 Haffner\,13        &    279\,$\pm$\,18        &     638\,$\pm$\,30      &  5.9\,$\pm$\,0.3      &    353\,$\pm$\,20        &   0.33\,$\pm$\,0.01     &   0.3            \\
 NGC\,2437          &   5714\,$\pm$\,87        &   14024\,$\pm$\,269     & 17.5\,$\pm$\,1.4      &   9164\,$\pm$\,331       &   2.57\,$\pm$\,0.06     &   0.6            \\
 NGC\,2447          &   1472\,$\pm$\,40        &    3636\,$\pm$\,117     & 10.4\,$\pm$\,0.7      &   3909\,$\pm$\,333       &   1.39\,$\pm$\,0.07     &   0.5            \\
 Berkeley\,39       &   6932\,$\pm$\,274       &   20275\,$\pm$\,941     & 30.2\,$\pm$\,5.4      &  24704\,$\pm$\,2733      &  11.61\,$\pm$\,0.82     &   0.8            \\
 NGC\,2482          &    373\,$\pm$\,36        &     914\,$\pm$\,112     &  6.7\,$\pm$\,0.5      &   1458\,$\pm$\,173       &   0.81\,$\pm$\,0.06     &   0.6            \\
 NGC\,2489          &   1793\,$\pm$\,60        &    4330\,$\pm$\,188     & 11.3\,$\pm$\,0.8      &   3316\,$\pm$\,181       &   1.31\,$\pm$\,0.04     &   0.8            \\
 NGC\,2506          &   8421\,$\pm$\,157       &   23116\,$\pm$\,545     & 29.2\,$\pm$\,5.4      &  17595\,$\pm$\,681       &   7.64\,$\pm$\,0.43     &   0.7            \\
 NGC\,2509          &   2060\,$\pm$\,93        &    5372\,$\pm$\,304     & 14.5\,$\pm$\,1.7      &   7130\,$\pm$\,746       &   2.84\,$\pm$\,0.18     &   0.7            \\
 NGC\,2548          &   1040\,$\pm$\,46        &    2578\,$\pm$\,149     & 10.4\,$\pm$\,0.9      &   2421\,$\pm$\,187       &   1.28\,$\pm$\,0.06     &   0.4            \\
 NGC\,2571          &    436\,$\pm$\,28        &    1004\,$\pm$\,77      &  6.6\,$\pm$\,0.3      &    516\,$\pm$\,22        &   0.45\,$\pm$\,0.01     &   0.6            \\
 NGC\,2627          &   1975\,$\pm$\,103       &    5397\,$\pm$\,344     & 13.0\,$\pm$\,1.0      &  10677\,$\pm$\,1415      &   3.13\,$\pm$\,0.24     &   0.6            \\
 NGC\,2635          &   1375\,$\pm$\,79        &    3494\,$\pm$\,252     & 13.3\,$\pm$\,1.9      &   3126\,$\pm$\,271       &   1.91\,$\pm$\,0.13     &   0.9            \\
 IC\,2395           &    278\,$\pm$\,18        &     645\,$\pm$\,33      &  5.9\,$\pm$\,0.3      &    310\,$\pm$\,19        &   0.30\,$\pm$\,0.01     &   0.3            \\
 NGC\,2658          &   2922\,$\pm$\,82        &    7419\,$\pm$\,269     & 17.8\,$\pm$\,2.8      &   5171\,$\pm$\,249       &   2.77\,$\pm$\,0.13     &   0.8            \\
 Ruprecht\,68       &   2312\,$\pm$\,92        &    6245\,$\pm$\,311     & 14.0\,$\pm$\,1.0      &   8986\,$\pm$\,1047      &   2.96\,$\pm$\,0.20     &   0.8            \\
 NGC\,2670          &    734\,$\pm$\,39        &    1658\,$\pm$\,111     &  8.3\,$\pm$\,0.6      &   1013\,$\pm$\,39        &   0.63\,$\pm$\,0.02     &   0.7            \\
 NGC\,2669          &    536\,$\pm$\,27        &    1244\,$\pm$\,73      &  7.5\,$\pm$\,0.5      &    834\,$\pm$\,71        &   0.57\,$\pm$\,0.03     &   0.6            \\
 NGC\,2818          &   2029\,$\pm$\,83        &    5240\,$\pm$\,271     & 16.1\,$\pm$\,2.0      &   4881\,$\pm$\,389       &   2.87\,$\pm$\,0.16     &   0.8            \\
 NGC\,2849          &   2203\,$\pm$\,122       &    5766\,$\pm$\,385     & 17.4\,$\pm$\,1.0      &   4838\,$\pm$\,338       &   3.72\,$\pm$\,0.17     &   1.0            \\
 IC\,2488           &   1312\,$\pm$\,47        &    2999\,$\pm$\,133     &  9.9\,$\pm$\,0.7      &   1945\,$\pm$\,101       &   0.90\,$\pm$\,0.03     &   0.6            \\
 Trumpler\,12       &   1733\,$\pm$\,61        &    4395\,$\pm$\,200     & 11.6\,$\pm$\,2.1      &   4858\,$\pm$\,449       &   1.67\,$\pm$\,0.10     &   0.8            \\
 NGC\,3293          &   3459\,$\pm$\,122       &    6843\,$\pm$\,312     & 13.1\,$\pm$\,0.5      &   3698\,$\pm$\,152       &   1.26\,$\pm$\,0.03     &   0.8            \\
 Melotte\,101       &   2260\,$\pm$\,65        &    5413\,$\pm$\,200     & 12.4\,$\pm$\,1.2      &   3452\,$\pm$\,243       &   1.38\,$\pm$\,0.07     &   0.8            \\
 Ruprecht\,91       &    504\,$\pm$\,27        &    1138\,$\pm$\,70      &  6.9\,$\pm$\,0.4      &    787\,$\pm$\,52        &   0.50\,$\pm$\,0.02     &   0.5            \\
 NGC\,3496          &   4280\,$\pm$\,92        &   10674\,$\pm$\,298     & 13.6\,$\pm$\,1.3      &   9301\,$\pm$\,622       &   2.04\,$\pm$\,0.08     &   0.8            \\
 Trumpler\,19       &   4394\,$\pm$\,159       &   12364\,$\pm$\,543     & 13.9\,$\pm$\,4.0      &  42532\,$\pm$\,6895      &   4.58\,$\pm$\,0.43     &   0.7            \\
 Alessi Teutsch\,8  &    652\,$\pm$\,31        &    1489\,$\pm$\,86      &  7.4\,$\pm$\,0.6      &   1149\,$\pm$\,93        &   0.60\,$\pm$\,0.03     &   0.5            \\  
 NGC\,4103          &   2973\,$\pm$\,149       &    6237\,$\pm$\,404     & 12.1\,$\pm$\,0.6      &   3409\,$\pm$\,151       &   1.15\,$\pm$\,0.03     &   0.8            \\ 
 NGC\,4349          &   3046\,$\pm$\,71        &    7463\,$\pm$\,223     & 12.1\,$\pm$\,0.6      &   5579\,$\pm$\,436       &   1.52\,$\pm$\,0.07     &   0.7            \\
 Trumpler\,20       &  13728\,$\pm$\,196       &   35857\,$\pm$\,652     & 20.3\,$\pm$\,1.3      &  29970\,$\pm$\,1925      &   4.25\,$\pm$\,0.18     &   0.9            \\
 NGC\,4609          &    968\,$\pm$\,42        &    2172\,$\pm$\,117     &  8.2\,$\pm$\,0.8      &   1340\,$\pm$\,82        &   0.62\,$\pm$\,0.02     &   0.7            \\
 UBC\,290           &    820\,$\pm$\,40        &    1812\,$\pm$\,112     &  8.0\,$\pm$\,0.6      &   1113\,$\pm$\,52        &   0.60\,$\pm$\,0.02     &   0.7            \\ 
 NGC\,4755          &   4217\,$\pm$\,128       &    8478\,$\pm$\,336     & 13.7\,$\pm$\,0.6      &   4599\,$\pm$\,167       &   1.40\,$\pm$\,0.04     &   0.8            \\
 NGC\,4852          &    678\,$\pm$\,36        &    1550\,$\pm$\,104     &  7.0\,$\pm$\,0.3      &   1120\,$\pm$\,134       &   0.62\,$\pm$\,0.04     &   0.7            \\
 Collinder\,272     &   1980\,$\pm$\,66        &    4322\,$\pm$\,180     & 10.2\,$\pm$\,0.7      &   2371\,$\pm$\,97        &   0.87\,$\pm$\,0.03     &   0.8            \\
 Pismis\,18         &   1798\,$\pm$\,68        &    4492\,$\pm$\,219     &  9.7\,$\pm$\,0.7      &   5437\,$\pm$\,524       &   1.38\,$\pm$\,0.08     &   0.9            \\ 
 Collinder\,277     &   1005\,$\pm$\,67        &    2580\,$\pm$\,215     &  8.4\,$\pm$\,1.0      &   5108\,$\pm$\,667       &   1.42\,$\pm$\,0.11     &   0.7            \\
 NGC\,5381          &   3004\,$\pm$\,85        &    7546\,$\pm$\,278     & 11.7\,$\pm$\,0.9      &   6549\,$\pm$\,448       &   1.58\,$\pm$\,0.07     &   0.9            \\
 NGC\,5822          &   1686\,$\pm$\,73        &    4364\,$\pm$\,240     & 10.0\,$\pm$\,0.7      &   7093\,$\pm$\,833       &   1.76\,$\pm$\,0.12     &   0.4            \\
 NGC\,5823          &   2896\,$\pm$\,82        &    6661\,$\pm$\,245     & 11.3\,$\pm$\,1.8      &   4215\,$\pm$\,202       &   1.13\,$\pm$\,0.05     &   0.8            \\  
 NGC\,5925          &   1474\,$\pm$\,50        &    3598\,$\pm$\,159     &  9.2\,$\pm$\,0.8      &   3786\,$\pm$\,318       &   1.14\,$\pm$\,0.06     &   0.7            \\
 NGC\,6025          &    663\,$\pm$\,30        &    1521\,$\pm$\, 79     &  7.3\,$\pm$\,1.0      &   1183\,$\pm$\,201       &   0.58\,$\pm$\,0.06     &   0.5            \\
 NGC\,6067          &   6623\,$\pm$\,191       &   15090\,$\pm$\,556     & 14.9\,$\pm$\,0.8      &   8641\,$\pm$\,205       &   1.84\,$\pm$\,0.04     &   0.8            \\
 NGC\,6087          &    463\,$\pm$\,31        &    1006\,$\pm$\, 81     &  6.5\,$\pm$\,0.7      &    695\,$\pm$\,45        &   0.43\,$\pm$\,0.02     &   0.6            \\
 NGC\,6152          &   1054\,$\pm$\,64        &    2480\,$\pm$\,195     &  8.2\,$\pm$\,0.7      &   2072\,$\pm$\,199       &   0.80\,$\pm$\,0.05     &   0.7            \\
 NGC\,6204          &    735\,$\pm$\,36        &    1753\,$\pm$\,110     &  6.8\,$\pm$\,0.3      &   1181\,$\pm$\,193       &   0.59\,$\pm$\,0.06     &   0.7            \\
 NGC\,6208          &   1376\,$\pm$\,80        &    3675\,$\pm$\,264     &  9.4\,$\pm$\,0.9      &  10291\,$\pm$\,1558      &   2.19\,$\pm$\,0.20     &   0.6            \\
 NGC\,6281          &    507\,$\pm$\,23        &    1192\,$\pm$\, 57     &  6.8\,$\pm$\,0.3      &   1355\,$\pm$\,183       &   0.67\,$\pm$\,0.05     &   0.3            \\
 Trumpler\,25       &   8233\,$\pm$\,163       &   18683\,$\pm$\,483     & 15.6\,$\pm$\,1.1      &  10508\,$\pm$\,345       &   2.00\,$\pm$\,0.06     &   1.0            \\
 Trumpler\,29       &    748\,$\pm$\,37        &    1697\,$\pm$\,105     &  7.7\,$\pm$\,0.4      &    982\,$\pm$\,53        &   0.55\,$\pm$\,0.02     &   0.7            \\
 NGC\,6416          &    798\,$\pm$\,57        &    1931\,$\pm$\,175     &  7.5\,$\pm$\,0.5      &   1905\,$\pm$\,232       &   0.76\,$\pm$\,0.06     &   0.6            \\
 NGC\,6494          &   2204\,$\pm$\,76        &    5395\,$\pm$\,241     & 11.0\,$\pm$\,0.4      &   4103\,$\pm$\,227       &   1.29\,$\pm$\,0.04     &   0.5            \\ 
 NGC\,6531          &    485\,$\pm$\,34        &     999\,$\pm$\, 83     &  6.4\,$\pm$\,0.4      &    503\,$\pm$\,31        &   0.35\,$\pm$\,0.01     &   0.5            \\
 NGC\,6568          &    555\,$\pm$\,39        &    1391\,$\pm$\,128     &  6.6\,$\pm$\,0.9      &   3494\,$\pm$\,501       &   1.03\,$\pm$\,0.09     &   0.6            \\
 NGC\,6645          &   1967\,$\pm$\,83        &    4936\,$\pm$\,267     & 10.1\,$\pm$\,0.7      &   5234\,$\pm$\,453       &   1.39\,$\pm$\,0.07     &   0.7            \\
 IC\,4756           &    927\,$\pm$\,48        &    2351\,$\pm$\,158     &  8.3\,$\pm$\,0.7      &   4769\,$\pm$\,624       &   1.41\,$\pm$\,0.11     &   0.3            \\
 NGC\,6705          &  10500\,$\pm$\,214       &   25112\,$\pm$\,650     & 17.2\,$\pm$\,1.9      &  15157\,$\pm$\,418       &   2.44\,$\pm$\,0.06     &   0.8            \\

\hline      
                                                                                                                                                                                                                                                                                                                                                                                                                                                                                                                                                                                                                                                                                                                                                                                                                                                                                                                                                                    
\end{tabular}

\label{tab:masses_and_other_params}
				
\end{table*}

\begin{table*}
\tiny
\contcaption{          } 

\begin{tabular}{lrrrrrc}

 Cluster            &  $M_{\textrm{clu}}$      & $N_{\textrm{clu}}$      &      $R_J$          &  $M_{\textrm{ini}}$   &   $t_{\textrm{95}}$   & $M_{\textrm{lim}}^{(*)}$  \\    
                    &  ($M_{\odot}$)           &                         &      (pc)           &  ($M_{\odot}$)        &      (Gyr)            &   ($M_{\odot}$)           \\     
                                                                                                                                                                                                                                            
\hline                                                                                                                 
                                                                                                                                                                              
 NGC\,6709          &    720\,$\pm$\, 33       &    1665\,$\pm$\, 93     &  7.7\,$\pm$\,0.3    &  1229\,$\pm$\,93      & 0.65\,$\pm$\,0.03     &   0.6          \\
 NGC\,6728          &    945\,$\pm$\, 62       &    2323\,$\pm$\,193     &  8.5\,$\pm$\,0.6    &  2877\,$\pm$\,288     & 1.08\,$\pm$\,0.07     &   0.6          \\
 NGC\,6793          &    394\,$\pm$\, 20       &    1017\,$\pm$\, 59     &  6.2\,$\pm$\,1.0    &  1615\,$\pm$\,400     & 0.70\,$\pm$\,0.10     &   0.5          \\
 IC\,1311           &   7831\,$\pm$\,289       &   20511\,$\pm$\,918     & 24.7\,$\pm$\,3.9    & 15210\,$\pm$\,1143    & 5.38\,$\pm$\,0.50     &   1.2          \\
 NGC\,6939          &   4022\,$\pm$\,116       &   10655\,$\pm$\,387     & 18.8\,$\pm$\,2.0    & 10124\,$\pm$\,832     & 3.90\,$\pm$\,0.23     &   0.7          \\
 NGC\,6940          &   1328\,$\pm$\, 53       &    3427\,$\pm$\,177     &  9.8\,$\pm$\,1.0    &  5906\,$\pm$\,723     & 1.71\,$\pm$\,0.12     &   0.5          \\
 NGC\,7082          &    861\,$\pm$\, 36       &    1975\,$\pm$\,100     &  8.6\,$\pm$\,0.4    &  1233\,$\pm$\,62      & 0.69\,$\pm$\,0.02     &   0.6          \\
 NGC\,7209          &   1027\,$\pm$\, 38       &    2488\,$\pm$\,114     &  9.5\,$\pm$\,1.1    &  2634\,$\pm$\,337     & 1.14\,$\pm$\,0.09     &   0.6          \\
 NGC\,7243          &    674\,$\pm$\, 31       &    1549\,$\pm$\, 83     &  7.5\,$\pm$\,0.2    &  1045\,$\pm$\,66      & 0.66\,$\pm$\,0.03     &   0.5          \\
 LP\,1800           &    652\,$\pm$\, 33       &    1544\,$\pm$\,101     &  7.8\,$\pm$\,0.5    &  1434\,$\pm$\,157     & 0.75\,$\pm$\,0.05     &   0.6          \\
 Berkeley\,98       &   1756\,$\pm$\,133       &    4848\,$\pm$\,432     & 14.4\,$\pm$\,2.6    & 11568\,$\pm$\,1727    & 4.10\,$\pm$\,0.35     &   0.8          \\

\hline

\multicolumn{7}{l}{$^{(*)}$ Limiting mass corresponding to $G$\,=\,19\,mag (Section~\ref{sec:sample_data}).}

\end{tabular}

%\label{tab:masses_and_other_params_compl}

\end{table*}

%%%%%%%%%%%%%%%%%%%%%%
\section{Supplementary figures - main plots}
%%%%%%%%%%%%%%%%%%%%%%

This Appendix shows the main results (RDP, CMD, $\varpi\times G\,$mag plot, VPD and skymap; figs. B1-B113) for 113 investigated OCs, not shown in the main text.

%%%%%%%%%%%%%%%%%%%%%%%%%%%%%%%%%%%
\section{Supplementary figures - Spectroscopic plots}
\label{appendix_spect}
%%%%%%%%%%%%%%%%%%%%%%%%%%%%%%%%%%%

This Appendix shows the set of spectroscopic data (whenever available) for the member stars of the investigated OCs. In each case (figs. C1-C113), the plots are: $[Fe/H]$\,$\times$\,$G_{\textrm{mag}}$, $V_{\textrm{rad}}$\,$\times$\,$G_{\textrm{mag}}$ and the spectroscopic \textit{Hertzsprung-Russell} diagram (log\,$g$\,$\times$\,log\,$T_{\textrm{eff}}$).

%%%%%%%%%%%%%%%%%%%%%%%%%%%%%%%%%%%
\section{Supplementary figures - Mass functions}
\label{appendix_MFs}
%%%%%%%%%%%%%%%%%%%%%%%%%%%%%%%%%%%

This Appendix shows the observed mass functions (MF) for 113 investigated OCs (figs. D1-D10). The MF for the OC NGC\,6940 is shown in Figure~\ref{fig:MF} of the main text.

\bsp

\label{lastpage}

\end{document}